\documentclass[twocolumn]{aastex631}

\usepackage[T1]{fontenc}
\usepackage{graphicx}	% Including figure files
\usepackage{amsmath}	% Advanced maths commands
\usepackage{amssymb}	% Extra maths symbols
\usepackage{hyperref}
\usepackage{makecell}
\usepackage{color} 		% Colors for text commenting
\usepackage{comment}    % Allows for commenting out blocks of text
\usepackage{verbatim}

\newcommand{\simgeq}{\; \raisebox{-0.4ex}{\small$\stackrel
{{\textstyle>}}{\sim}$}\;}

\definecolor{BA}{rgb}{1,0.6,0 }

\usepackage{mathrsfs}

\begin{document}

\label{firstpage}

\title{
Wind-confined jet collimation revealed by the acceleration-phase photosphere of GRB 220426A
}

\correspondingauthor{Felix Ryde}
\email{fryde@kth.se}

\author[0000-0002-9769-8016]{Felix Ryde}
\affiliation{Department of Physics, KTH Royal Institute of Technology, \\
and The Oskar Klein Centre, SE-10691 Stockholm, Sweden}

\author[0000-0001-8667-0889]{Asaf Pe'er}
\affiliation{Department of Physics, Bar-Ilan University, Ramat-Gan 52900, Israel}

\begin{abstract}

We analyze the prompt emission of the exceptionally bright GRB 220426A observed by {\it Fermi}/GBM, whose time-resolved spectra are among the narrowest measured in any GRB. Here we show that observations during the first $\sim 5$~s are consistent with the signal being emitted while the jet was still in the initial radiation-dominated acceleration phase. The time-resolved spectra allow the effective launch radius, $r_0$, to be inferred with unusual precision. We find $r_0 \sim \mathrm{few} \times 10^{10}\,\mathrm{cm}$, increasing linearly with time. These findings match the theoretical expectation for the recollimation shock, implying that this GRB shows the first clear evidence for the existence and evolution of a recollimation shock. Using this interpretation, 
the linear increase observed is sustained over a period longer than expected from a pure jet breakout. Therefore, the jet collimation must have persisted even after breakout. We suggest that the collimation was maintained by a finite, dense, wind-like circumburst medium, which would reproduce the observed behavior of the prompt emission. We conclude that late-stage progenitor mass-loss can shape the earliest prompt emission and that the photospheric emission  provides a way to probe both the jet collimation and the innermost region of the circumburst medium (CBM) surrounding  the progenitor, independently of
the constraints from interacting supernovae.

\end{abstract}

\section{Introduction}

The earliest prompt emission from a gamma-ray burst (GRB) can carry information about the dynamics of the relativistic jet as it emerges from its progenitor environment.
Numerical simulations of GRB jets drilling through their progenitor stars show that stellar confinement drives recollimation shocks 
that inhibit free expansion and delay efficient bulk acceleration until the jet approaches breakout from the progenitor core, at radii  $\sim10^{10}\ \mathrm{cm}$ \citep[e.g.,][]{Thompson2007,Bromberg2016,Gottlieb2019}. In this scenario, a substantial fraction of the dynamical shaping of the outflow might occur after breakout, so that interaction with the immediate circumburst medium (CBM) can further regulate collimation, baryon loading, and the effective acceleration history \citep[e.g., ][]{Matzner2003, Bromberg11}. 

Observational evidence that such jet-environment interactions occur is growing. Possible examples  are the low-luminosity GRBs who have a significantly lower luminosity and softer prompt emission compared to typical long GRBs \citep{Gehrels+06, Campana2006}. Low-luminosity GRBs have been interpreted as being due to relativistic jets that are choked, where the jet fails to fully emerge from an extended enveloppe or dense CBM and instead inflates a hot cocoon that powers the observed high-energy emission \citep[e.g.,][]{Nakar2015, DuffellHo2020, SuzukiMaeda2024, Hamidani2025}.  Related models involving the interaction of engine-driven outflows with extended CBM have also recently been proposed for even softer and less luminous X-ray transients, appearing as X-ray flashes \citep[e.g.,][]{Srinivasaragavan2026}. Moreover, recent studies of MeV spectral features during the prompt emission,  suggest that some GRBs are surrounded by a dense, likely non-steady CBM \citep{Oganesyan2026}. Further observational evidence for a jet-CBM interaction was provided by \citet{Ryde2022}, who interpreted late prompt pulses (at $\sim$100 s post-trigger) as an external-shock encounter with a pre-existing Wolf-Rayet nebular shell at $ \sim10^{18}$ cm. \citet{PeerRyde2024} later provided the theoretical framework for such interaction-driven delayed emission, identifying highly efficient synchrotron emission from the jet encounter with the wind-termination shock. Such matter could also correspond to ejection due to activity of the progenitor star around  1000 years prior to the GRB.

In addition to this, recent observational progress in detecting and analysing the earliest part of the light curve of supernovae in the optical wavelenghts %, but also IR and X-rays.
 suggest that many progenitors of supernovae have to be surrounded by  significant and dense matter structures. In particular, highly stripped supernovae types, which are typically associated with GRBs, show in many cases indications of fast,  massive, and detatched shells lying close to the progenitor star \cite[e.g., ][]{Smith17}.  Prominent examples are given by SN2021csp \citep{Perley2022} and SN2022oqm \citep{Irani2024} which are interpreted as having preexisting eruptive shells with velocities up to 4000 km/s. This matter needs to lie at radii even smaller than the shells studied by \citet{PeerRyde2024}, with many cases indicating sizes of around $10^{13} - 10^{15}$ cm. Matter lying at such small radii could be the results of eruptive mass loss during the late stages of stellar evolution,
 %would reflect instable activity of the progenitor,
 at times much closer to the SN/GRB explosion, on a time-scale of less than a year.  
Since the GRB jet needs to drills through such a CBM, there might be interaction signatures even on the very earliest prompt emission, at times much closer to the GRB trigger compared to the pulses studied in \citet{PeerRyde2024}. Therefore,  analysis of the initial prompt-phase signal can probe the innermost CBM and, and thereby the most recent pre-explosion mass-loss activity of the progenitor.

The observational method to infer properties of the dynamics of the jet from the prompt GRB emission depends on the energy content in the jet, that in most cases has to be assumed.
The energy in a GRB jet outflow can be carried by multiple components, including radiation, matter (baryons and leptons), and magnetic fields. The relative importance of these components varies both along the jet and between individual bursts.  When radiation dominates the energy content, the jet dynamics gets simplified and  the interpretation of the observations become less uncertain. First, the unique spectral shape expected from such a emission unambigously identifies the energy content as radiation-dominated \citep{Ryde2017}. Second, the physics of the acceleration phase, which is the radiation-dominated phase, provides the least uncertainties in the analysis of the data. In particular, the radius at which the acceleration begins, $r_0$, is given uniquely from the observations \citep{Peer2007}. %This fact is essential since $r_0$ is a fundamental quantity of the fireball model of GRBs. 
In contrast to the radiation-dominated, acceleration phase, the coasting phase is kinetic-energy-dominated, which means that the observed photospheric component will be affected by adiabatic cooling and part of the kinetic luminosity can potentially be dissipated in internal shocks. Both effects introduce additional uncertainty in the inferred initial flow parameters \citep{Peer2007}. The cleanest constraints therefore come from intervals of GRB emission that remain radiation dominated.

%The challenge is that such intervals are observationally rare and typically weak \citep{Ryde2004}. 

We are faced with two challenges when wanting to study the radiation-dominated phase in GRBs, and thereby taking  advantage of its simplicity.  First, identification of pure radiation-dominated flow, namely flows in which the photons escape during the early acceleration phase, are very rare. 
These are identified by their very narrow spectra, originating from the photosphere \citep[see eq. 2 in ][]{Ryde2017}. The spectra are narrower than that expected if the photons decouple the plasma during the  coasting phase, the so-called "non-dissipative photosphere" \citep[NDP; ][]{Acuner2019}. Fitting pure radiation-dominated and NDP spectra with the commonly used empirical \citet{Band1993}-model, the narrowness is translated to the requirement that the photon index of the low-energy power-law is $\alpha \simgeq -0.4$ for NDP \citep{Acuner2020} and $\alpha \sim 0.6$ for pure radiation-dominated photosphere \citep{Acuner2019}. In that work, \citet{Acuner2019} further deduced that not more than a few per cent of bursts have such narrow spectra.

The second challenge in the analysis of spectra from the acceleration phase lies in the small expected change of spectral width as the flow transitions from the pure radiation-dominated to the coasting phases \citep{Ryde2017}. The model parameter describing the width of the observed spectrum (and thereby the transition between dynamical phases) is the ratio $\eta/\eta_* \equiv  ({r_{\rm ph}}/{r_{\rm s}})^{-3/4}$, where $r_{\rm ph}$ and $r_{\rm s}$ are the photosphere and the saturation radii, and $\eta$ and $\eta_*$ are the dimensionless enthalpy of the flow and the critical enthalpy (see Appendix \ref{sec:AppRD} below, Eq. \ref{eq:rphrs}).  Even small variations in the observed spectral shape translate into substantial changes in the fitted ratio $\eta/\eta_*$.
Therefore, spectra with large statistical noise and poorly determined widths lead to poorly constrained $\eta/\eta_*$-values, which then translates into poorly constrained derived properties of the flow (see \S \ref{sec:quantaties}). This limitation is illustrated by the poorly constrained values found so far in the few attempts made to measure this ratio. For example, \citet{Ryde2017} measured  $\eta/\eta_*$ in two temporal intervals during the acceleration phase of GRB~100507, although with large associated uncertainties. Therefore, in order to make a detailed study of the radiation-dominated phase in GRBs, one needs to identify bursts with the narrowest spectra, but at the same time require a great brightness to reduce the noise level.

In this paper, we reanalyze GRB 220426A, which meets these requirements. It is an exceptionally bright burst and indeed has very narrow spectra, similar to the bursts presented in \citet{Ryde2004}. The observed spectra unambigeously identifies the first 5 seconds of the prompt phase  to be radiation dominated.  Its brightness allows the evolution of $r_0$  to be followed with great precision, and we find that $r_0$ grows linearly with time  (Section 2).  Interpreting $r_0$ as set by a recollimation shock in the outflow, we conclude that the jet must have traversed a very dense circumburst medium in order to remain collimated (Section 3).  We discuss the consequences of this interpretation in Section 4 and we conclude in Section 5.

\section{Analysis of radiation-dominated GRB 220426A}
\label{sec:analysis}

\subsection{A uniquely suitable burst to study}

The requirements discussed in the Introduction are met by GRB 220426A, which was both very bright and had very narrow time-resolved spectra, as observed by the {\it Fermi Gamma-ray Space Telescope} and its GBM detector \citep{Malacaria2022}.  The narrowness of the spectra is manifested by the very hard spectral power-law index, $\alpha$, below the peak energy.  Indeed, the maximal, time-resolved $\alpha$-parameter of the Band fits is unusually large: $\alpha_{\rm max} = 0.7\pm 0.1$; \citep[][, see also \citet{Song2022, Wang2022}]{Deng2022}.  This value is well within the expectation of the pure radiation-dominated photosphere spectra ($\eta/\eta_* \gg 1$), which is $\alpha \sim 0.6$ \citep[see][]{Acuner2019}. 
The uniqueness of such a $\alpha_{\rm max}$-value is realized by comparing to Figure 5 in \citet{Acuner2019} and figure 3 in \citet{Acuner2020}. GRB 220426A therefore adds to the small sample of bursts with very narrow spectra, consistent with being produced in the acceleration-phase. Moreover, due to its unique brightness, it is so far the only burst for which a precise determination of $\eta/\eta_*$ is made possible. Previous  analyses of this burst have employed other spectral models, such as a single Planck function, a multi-colour blackbody, or the Band function. However, those fits do not constrain $\eta/\eta_*$, which is the parameter needed here.
We will thus reanalyse this burst using the methodology in  \citet{Ryde2017}.
%Our spectral modelling is performed 

For the spectral analysis, we chose the data ranging between  8 keV and 40 MeV from the  {\it Fermi Gamma-ray space telescope}  including sodium iodide (NaI1, NaI2) and bismuth germanate (BGO1) detectors \citep{Meegan2009}. To perform time-resolved spectral fits we
followed the standard analysis procedure for the data handling, K-edge effects, spectral-fitting method, as well as background selections and modeling  \citep[e.g., ][]{Goldstein2012, Yu2018}. In our particular case we selected the intervals $[-63, -3]$ s and $[100, 250]$ s, relative to the GBM trigger time for the background modeling. The analysis is carried out with a joint-likelihood analysis using the {\tt Multi-Mission Maximum Likelihood (3ML)} software \citep{Vianello2015}.

\subsection{The prompt emission}
\label{sec:prompt}

\begin{figure}
    \centering
    \includegraphics[width = \columnwidth]{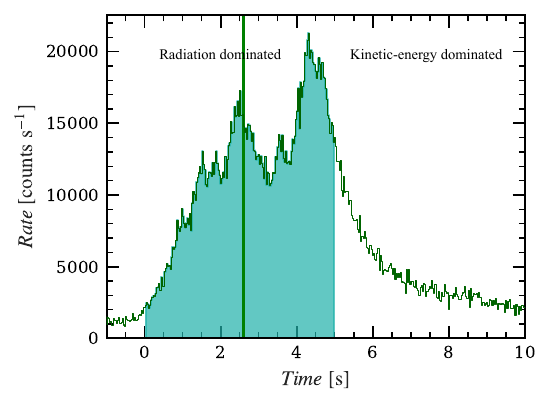}
    \caption{Light curve of GRB 220426A, observed by the GBM  onboard the {\it{Fermi Gamma-ray Space Telescope}} (using the NaI1 detector).  
    During most of its duration (green shaded area) the emission is consistent with being emitted during the radiation-dominated phase of the flow. The green verticle line corresponds to the break in the temperature decay (Fig. \ref{fig:observations_kT}). 
    }
    \label{fig:lc}
\end{figure}

Figure \ref{fig:lc} shows the light curve of GRB 220426A as the  count rate detected in the GBM NaI1 detector, over the energy range 8 keV to 1 MeV.
The light curve consists of a single emission episode, with a $T_{90} = 5.6 \pm 0.4$ s \citep{Malacaria2022}, on top of which there are two main pulses which peak around 2.6 and 4.3 seconds. We also calculated $T_{95} \sim 8.1 \pm 0.2$ s, which considers the long emission tail\footnote{$T_{95}$ were calculated on the background subtracted, deconvolved photon light-curves, using NaI1 and NaI2, identifying the duration during which 95\% of the photon fluence was accumulated. We used the standard selection of 50-300 keV energy range, 64 ms resolution, and used the standarad COMP model for the deconvolution.}. During the first 5 seconds there is substantial variability, with a minimum variability time given by $t_{\rm FWHM} = 0.6\pm 0.2$ s, which is the full-width at half maximum of the shortest pulse \citep{Maccary}.

In the present study the light-curve is rebinned by requiring each timebin to have a significance of $S = 100$ \citep{Vianello2018}, which resulted in 20 timebins over the interval 0-6.7 s.   For each timebin the observed spectrum is fitted with a table model containing the numerical spectra derived in \citep[][see their Fig. 2]{Ryde2017}, for which the spectral width provides a measure of the location of the photosphere relative to the saturation radius. %(see further the  {\tt 3ML} manual). 
This table model will  be denoted "radiation-dominated phase photosphere" (RDP) below. The time-resolved count spectra of the 20 timebins analysed in GRB 220426A and their best-fitting RDP models are shown in Appendix \ref{AppCount}. The fitted values of the evaluated parameters are given  in Table 1: peak energy of the $F_{\nu}$-spectrum, $E_{\rm p}$, which is related to the temperature at the photosphere\footnote{Similarly to the Planck function, the peak energy of the RDP model is related to the temperature at the photosphere, $kT_{\rm ph}$. For instance, in the pure acceleration phase ($\eta/\eta_* >>1$) the relation is given by $E_{\rm p} = 3.66 \, kT_{\rm ph}$ \citep{Ryde2017}.},  and the ratio of enthalpies, $\eta / \eta_*$, which is related to the spectral width,  and finally the normalization, which is given by photons flux $N_{\gamma}$  (ph cm$^{-2}$ s$^{-1}$, over the energy range 1 keV - 30 MeV) and, equivalently, by the corresponding energy flux, $F_{\rm E}$ (keV cm$^{-2}$ s$^{-1}$, over the same energy range).  With an estimation of the  luminosity distance, $d_{\rm L}$, the isotropically equivalent luminosity can be found from $L = 4\pi d_L^2 F_{\rm E}$. 
%%%%%%%%%%%%%%%%%%%%%

\begin{figure*}
    \centering
        \includegraphics[width = 0.7\textwidth]{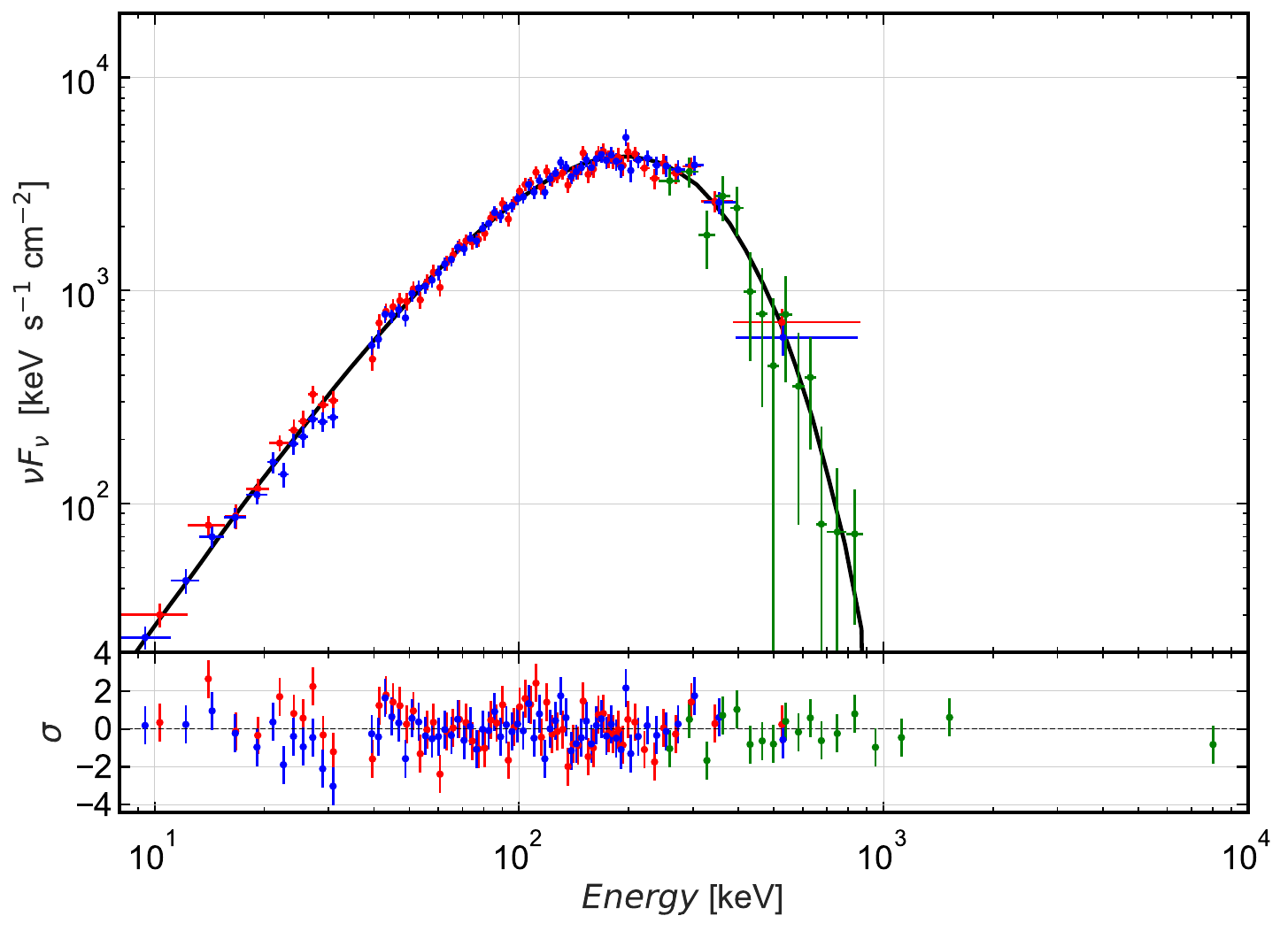}
        %{Figures/nuFnu_bn220426285_with_residuals_RP26_Fig4_1b.pdf}
        %{Figures/Peak_eta_2025_SNR150vFv_220426285_bin0_Lund2.pdf}
        \includegraphics[width = 0.7\textwidth]{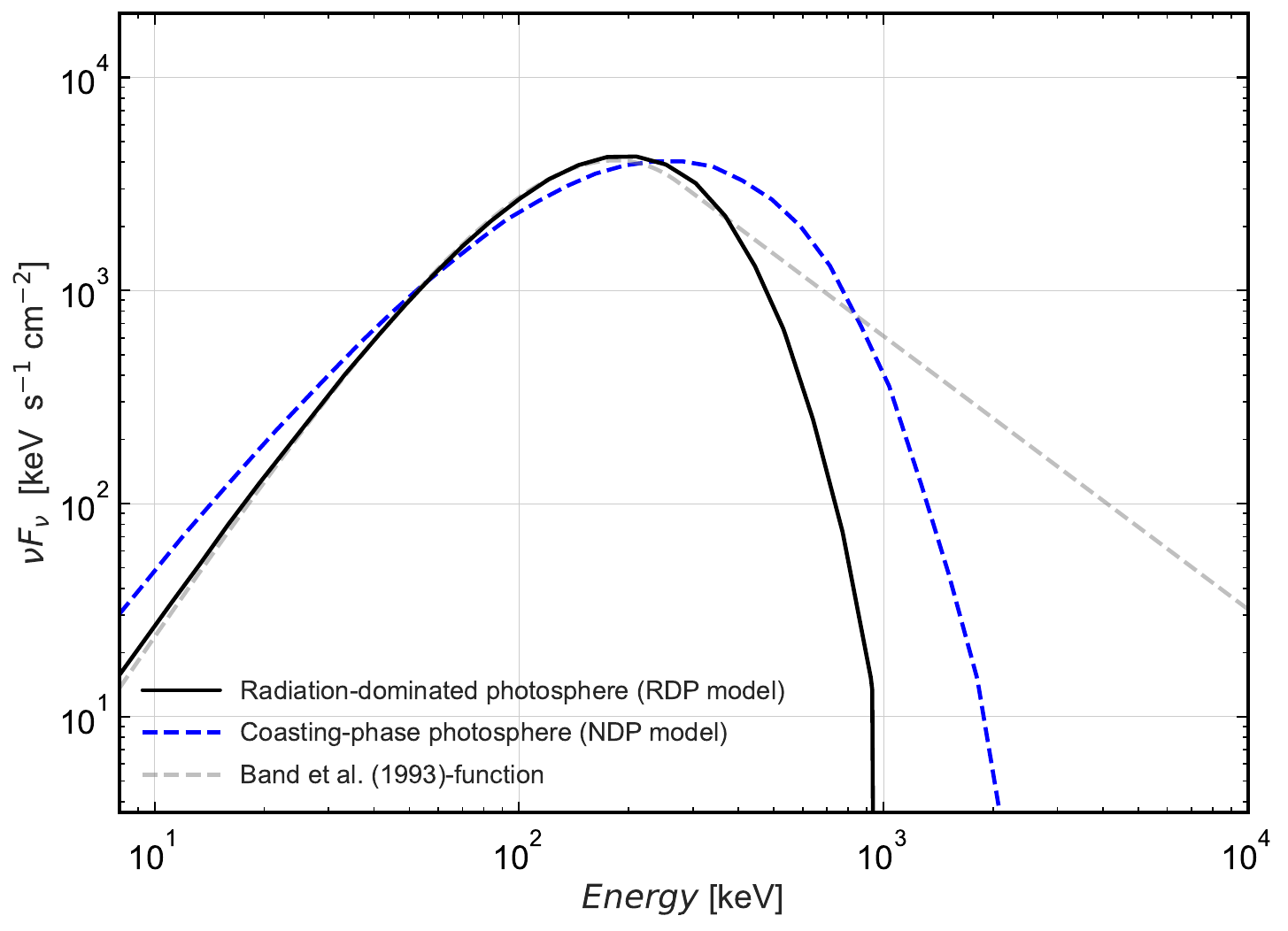}
        %Figures/nuFnu_bn220426285_with_residuals_RP26_Fig4_2b.pdf}
     %   \includegraphics[width = \columnwidth]{Figures/Peak_eta_2025_SNR150_ApJ_likelihood_220426285_bin0_Lund.pdf}
        %\includegraphics[width = \columnwidth]{Figures/spectra2.png}
        %\includegraphics[width = \columnwidth]{Figures/likelihood2.png}
        %\includegraphics[width = \columnwidth]{Figures/spectra2.png}
        %\includegraphics[width = \columnwidth]{Figures/likelihood2.png}
    \caption{Time-resolved spectrum  of GRB220426A, at the trigger time. Upper panel: Observation by the {\it Fermi Gamma-Ray Space Telescope} detectors NaI1, NaI2 (blue and red), and BGO1 (green), together with residuals between the data and the model. The black line shows the best spectral fit using the radiation-dominated phase photosphere model \citep[RDP; ][]{Ryde2017}.   
    Lower panel: The RDP model fit (black line) shown together with the corresponding fit using the NDP-spectrum, assuming the emission originates in the coasting phase \citep[blue dashed line][]{Acuner2019}, and the fit using the empirical Band function \citep[grey dashed line][]{Band1993}.}
    \label{fig:spectra}
\end{figure*}

\begin{figure*}
    \centering
        \includegraphics[width = 0.7\textwidth]{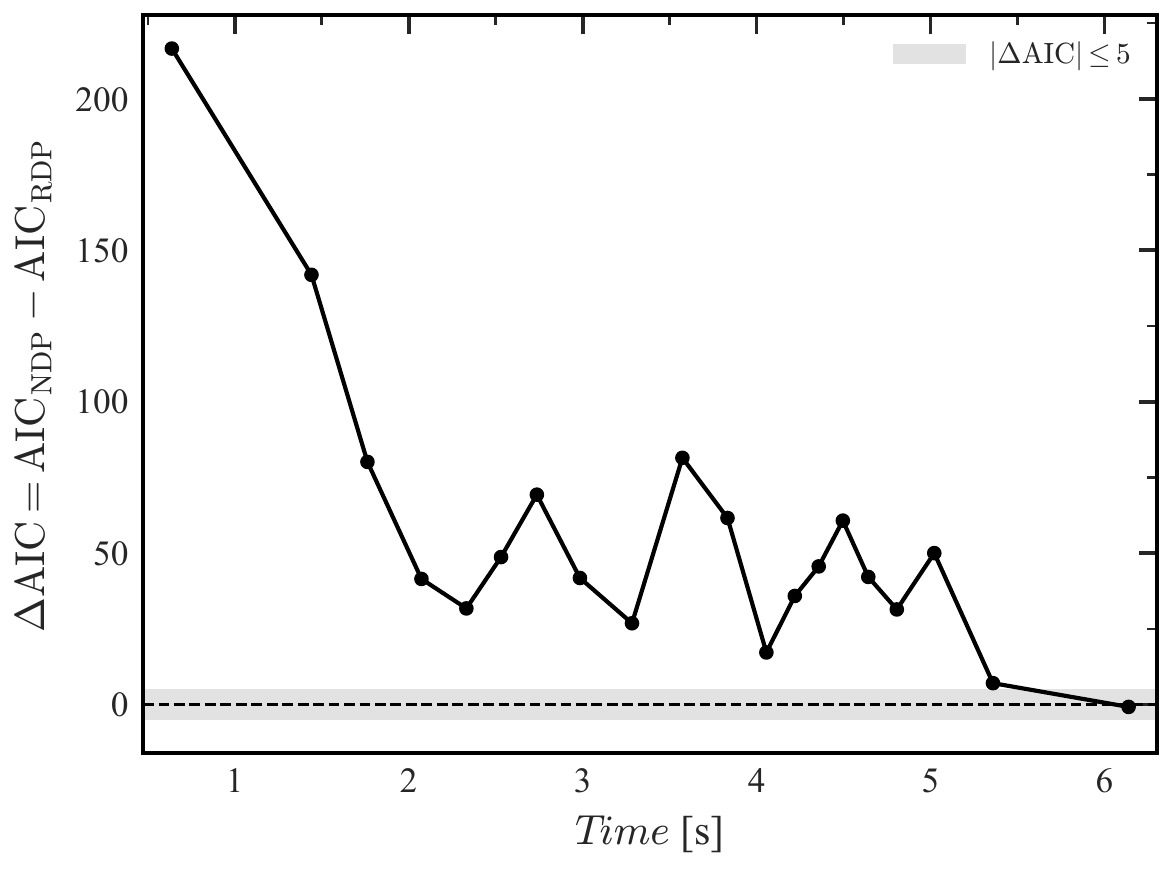}
    \caption{Difference in AIC between the NDP and RDP models as a function of time, where positive values favor the RDP model. The gray region marks $|\Delta\mathrm{AIC}|\leq5$, where the model preference is weak or inconclusive.}
    \label{Fig:AIC}
\end{figure*}

Figure \ref{fig:spectra} shows one of the narrowest time-resolved spectra, occurring right after the trigger. It is shown as $\nu F_{\nu}$-spectra (given by $E^2 N_{\rm E}$, where $E$ is the photon energy and $N_{\rm E}$ is the photon flux).  The upper panel shows the best fit of the RDP model as the solid lines and the associated data points from three of the {\it Fermi} detectors. Below the spectrum there is a panel showing the residuals between the data and the model.  The residuals are randomly distributed around zero with no obvious systematic trends, indicating statistically acceptable fits. All the spectra in the 20 timebins shown in App \ref{AppCount} show similar residual distributions. Note
that in Figure \ref{fig:spectra}, since the data points in a $\nu F_{\nu}$ spectrum are forward-folded through the  GBM response matrix, they will depend on the model used, and thus the plotted data points should not be used when comparing models. However, the residuals are based on the folded, count data and therefore correctly represent the comparison between the RDP models and the data, indicating the goodness-of-fit. In addition to the RDP model, in the bottom panel in Figure \ref{fig:spectra}, we also show the best-fit model spectra found when using the NDP model (the dashed, blue line) and the empirical Band function
(grey dashed line \citet{Band1993}). The NDP model only has two free parameters, the peak energy and the normalisation, and is expected from a photosphere occurring in the coasting phase ($\eta/\eta_* << 1$). Since the Band function is not physical model it is only shown for comparison, but not used in the model comparison. The clear difference between the RDP and the NDP spectral shapes illustrates why the width parameter $\eta/\eta_*$ is so well determined, being clearly larger than unity.   This fact is also illustrated in Figure \ref{Fig:AIC}, where we have plotted the difference in the Akaike Information Criterion (AIC) between the NDP and the RDP models for each analyzed time bin ($\Delta\mathrm{AIC}\equiv\mathrm{AIC}_{\rm NDP}-\mathrm{AIC}_{\rm RDP}$). 
The AIC is a relative model-selection criterion that apart from the maximum likelihood also takes into account a penalty for the number of fitted parameters. The $\Delta\mathrm{AIC}$ was shown by \citet[][their Fig. 2]{Acuner2020} to be a good model selection criteria. A lower AIC indicates greater support among models fitted to the same data and therefore positive value $\Delta\mathrm{AIC}>0$ favors the RDP model. The gray region marks the threshold $|\Delta\mathrm{AIC}|\leq5$, within which  a strong model preference should not be claimed \citep{Acuner2020}.

\begin{figure*}
    \centering
        \includegraphics[width = 0.7\textwidth]{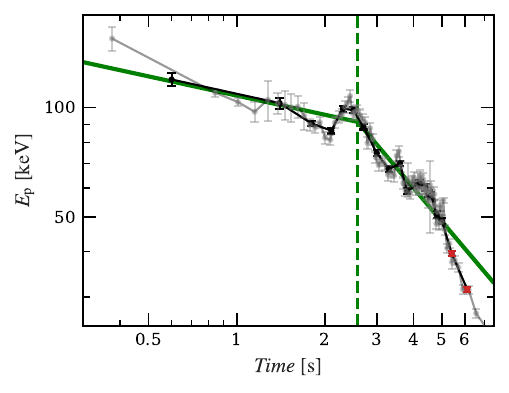}
    \caption{Temporal evolution of the properties of the photosphere in GRB 220426A: the observed peak-energy ($\sim$ temperature), showing the characteristic break. The break-time is shown by the dark-green, dashed line and is at $t_{\rm b} = 2.59 \pm 0.19$ s after the trigger. The 20 black databins have a significance of $S = 100$, while the grey points with $S=50$ are shown for comparison. The red points are beyond the studied interval, as the flow enters the coasting phase.}
    \label{fig:observations_kT}
\end{figure*}

\begin{figure*}
    \centering
        \includegraphics[width = 0.7\textwidth]{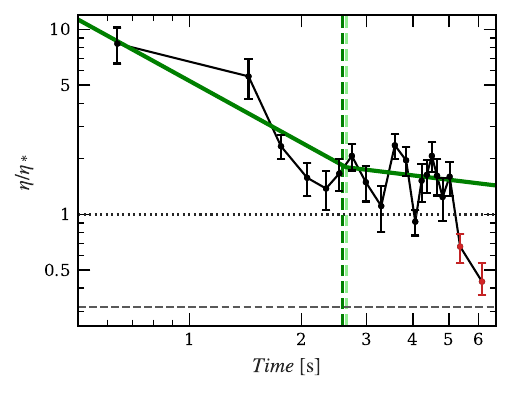}
    \caption{Temporal evolution of the properties of the photosphere in GRB 220426A: The observed width of the spectra, measured by $\eta/\eta_*$. The brightness of the burst makes the parameter well-determined. The break-time is shown by the light-green, dashed line, which coincides with the break-time of the the temperature decay (dark-green dashed line, see Fig. \ref{fig:observations_kT}). %Note the change in time interval compared to Fig. \ref{fig:observations_kT}
    }
    \label{fig:observations_eta}
\end{figure*}

%%%%%%%%%%%%%%
Figures \ref{fig:observations_kT} and \ref{fig:observations_eta} show the evolution of two of the RDP model parameters of GRB 220426A: the peak energy, $E_{\rm p}$ (in units of keV)  and the spectral-width parameter, $\eta/ \eta_*$.  The peak energy shows a monotonic cooling behaviour, following a broken power-law behaviour. We fitted the evolution using the broken power-law function in \citet{Ryde1999}, using a sharp break. The power-law slopes\footnote{Note that the values of the power-law indices depend on the exact value of the trigger time.} before and after the break time are $s_1 = -0.18 \pm 0.05$ and $s_2 = -0.96 \pm 0.10 $. This type of cooling beaviour is typical seen in thermal GRB pulses \citep[e.g., ][]{Ryde2004, Ryde2005, RydePeer2009, Iyyani2013}. %In this particular case the best fit requires two break in the power law, at times $t_1$ and $t_2$. 
The break time {occurs} at $t_{\rm b} = 2.59 \pm 0.19$ s and is also indicated by the green line in Figure \ref{fig:lc}.  Inspecting where the break time occurs in the light curve it corresponds to the first of two major peaks, before which the flux  increases. 
For comparison, the corresponding $E_{\rm p}$-values for a higher time resolution is also shown in Figure \ref{fig:observations_kT}  in grey (here $S = 50$ was used to bin the data). The coarser time binning is motivated by the need to constrain the $\eta/\eta_*$ parameter, which is more sensitive to the statistical significance of the data.

Figure \ref{fig:observations_eta} shows the evolution of  the $\eta / \eta_*$-parameter.  Due to the brightness of the bursts, this plot shows the most detailed and well determined $\eta / \eta_*$-evolution observed to date. The plot shows a monotonic decrease, starting off at very large values $\eta / \eta_* \sim 10$. Only for the last two time bins (marked by red color in Fig. \ref{fig:observations_eta}) do the spectra reach  values of $\eta$ which are below the critical value $\eta_*$, which is defined as the limit for saturation in \S \ref{sec:relations}. We note that, in reality, the transition to the coasting phase is gradual in $\eta / \eta_*$, and a fully developped NDP, has  $\eta / \eta_* << 1$.  On the other hand, for values smaller than $\eta / \eta_* \sim 0.316$ the spectra become indistinguishable from a NDP \citep{Ryde2017}, and therefore $\eta / \eta_* = 0.316$ is the smallest value used in the fits (shown by the  dashed line in Fig. \ref{fig:observations_eta}). Nevertheless, in the following, we will consider the emission to be truly in the acceleration phase when $\eta / \eta_* >1$, which thus occurs during period $0-5$ s, corresponding the green shaded area in Figure \ref{fig:lc}.  Here again we fit a broken power-law, which has the indices $-1.11\pm 0.14$ and $-0.2\pm0.7$, before and after the break time. The break time ($2.6\pm 1.0$ s) coincides with the break time of the the temperature decay.

Finally, in order to compare with empirical model fits, in Appendix \ref{sec:AppBand},  we present in Table \ref{Tab:App} the low-energy power-law index, $\alpha$, from fits using the  \citet{Band1993}-function \cite[see, e.g., ][]{Deng2022}. In particular, we show that the parameters, $\eta/\eta_*$ and $\alpha$ from the spectral analyses are closely related to each other.

\subsection{Evolution of derived properties}
\label{sec:properties}

The basic scalings of the dynamical evolution of the fireball in the radiation-dominated phase are summarized in Appendix \ref{sec:AppRD} \citep[see also, e.g., ][]{Meszaros2006, Hascoet2013, peer2015}. % in \S \ref{sec:quantaties} 
We combine these with the time-resolved observed quantities in order to make estimations of the flow properties in GRB220426A and their evolution.
Unfortunately, the redshift is not measured for GRB 220426A. By comparing the position of the properties of the prompt phase to that of other observed GRBs in the  $E_{\rm p} - E_{\rm iso}$ diagram, a broad range of redshift $z=0.2 -5$ are consistent with the current data %, with a center value at $z \sim 1.4$ 
 \citep{Wang2022}.
 %(their figure 4a).  
Since its value is highly uncertain, we use the fiducial redshift $z = 1$.

\begin{figure*}
    \centering
         \includegraphics[width = 0.7\textwidth]{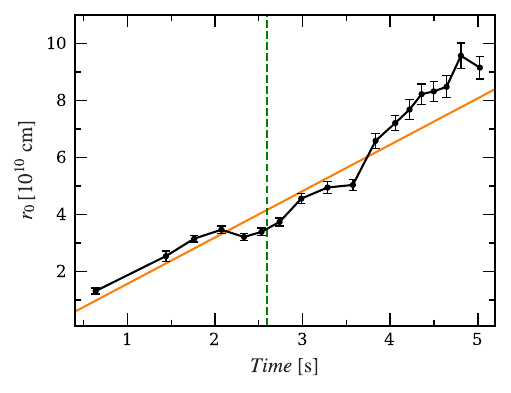}
         \caption{Linear evolution of $r_0$ during the radiation-dominated phase (Pearson $r=0.97$). The best fit relation is shown by the orange line, whose linear proportionality constant is $k = 1.63 \times 10^{10}$ cm/s. Redshift $z=1$ is assumed.} %(see further \S \ref{sec:redshift_dep}).}
    \label{fig:r0}
\end{figure*}

From the observed energy flux, ${F}_{\rm E}$, and temperature, ${T}$, we can derive the normalization of the thermal component, $\mathscr{R}$, from equation (\ref{eq:R}). The values of $\mathscr{R}$ inserted in equation (\ref{eq:r0}) gives $r_0$, which is the effective radius  at which the flow starts to accelerate freely. The evolution of $r_0$ over the radiation-dominated interval 0-5 s is shown in Figure \ref{fig:r0}. We find that the size of $r_0$ is few $\times 10^{10}$ cm and it increases linearly with time, $r_0(t) = k\, t$, which is shown by the orange line, for which $k = 1.63\times10^{10}$ cm/s. Indeed, we find a strong positive Pearson correlation between $r_0$ and $t$, with a correlation coefficient of $r=0.97$. The large values of $r_0$ that we find are consistent with the values found in the analysis in \citet{Song2022} and \citet{Deng2022}, even though their spectral modelling were different. 

\begin{figure*}
    \centering
        \includegraphics[width = 0.7\textwidth]{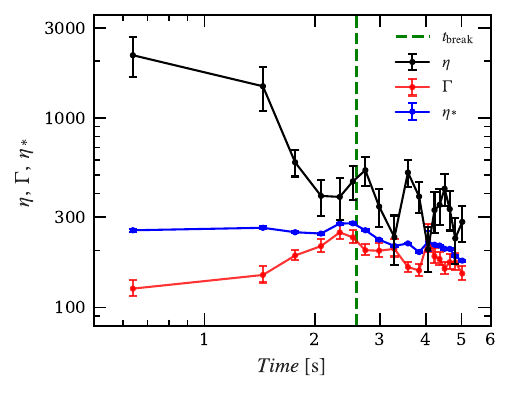}
         \caption{Derived properties of the flow. The black points show the dimensionless enthalpy, $\eta$, the blue points show the critical enthalpy, $\eta_*$, and the red points show the bulk Lorentz factor, $\Gamma$. The break time is from the temperature decay shown in Fig. \ref{fig:observations_kT}. Redshift $z=1$ is assumed.}
         %Right-hand panel: Time evolution of the photosphere radii}
    \label{fig:gammas}
\end{figure*}

With the measured values of $L$ and the derived values of $r_0$, the critical enthalpy, $\eta_*$ can be found from equation (\ref{eq:eta*}).  The combination of $\eta_*$ and the measurements of $\eta/\eta_*$ gives the Lorentz factor, $\Gamma$ from equation (\ref{eq:Gamma}), and the enthalpy, $\eta$, from equation (\ref{eq:eta}). The three parameters $\eta/\eta_*$, $\eta$, and $\Gamma$ and their evolutions are shown in Figure \ref{fig:gammas}.  %The other cases are show in Figure.
Even though the $\eta$-parameter starts off very large, with values reaching $\eta \sim 1000$, the Lorentz factor, $\Gamma$ remains modest with an averaged value of $\Gamma = 177$. The reason is that the Lorentz factor is only a fraction of $\eta$ during the acceleration phase, since the emission decouples from the flow before it has saturated at the maximal value. 

\begin{figure*}
    \centering
        \includegraphics[width = 0.7\textwidth]{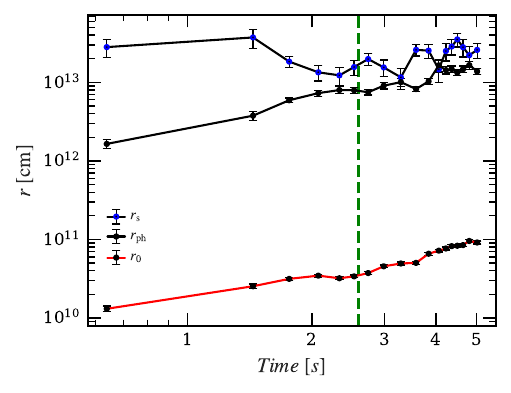}
         \caption{Evolution of derived fireball radii.  The effective launch radius, $r_0$, in red, the photospheric radius, $r_{\rm ph}$, in black, and the saturation radius, $r_{\rm s}$, in blue. The green dashed line is the break time of the temperarure decay. Redshift $z=1$ is assumed.}
         %Right-hand panel: Time evolution of the photosphere radii}
    \label{fig:radii_0sph}
\end{figure*}

Finally, the saturation radius, $r_{\rm s}$, and the photosphere radius, $r_{\rm ph}$, can then be found through equations (\ref{eq:ph6}) and (\ref{eq:rs2}) in appendix \ref{sec:relations}. These two radii are shown together with $r_0$ in Figure \ref{fig:radii_0sph}. All the derived parameters, $r_0$, $r_{\rm s}$, $r_{\rm ph}$, $\eta$, $\eta_*$, and $\Gamma$ are also given in Table \ref{tab:physical_parameters}.

\section{A collimating dense circumburst shell}

The analysis of the prompt, thermal episode in GRB 220426A gives three main results, which are all shown in Figure \ref{fig:r0}: (i) the typical value of $r_0$ is large and lies around a few $\times 10^{10}$ cm, (ii) this large value is maintained throughout the episode, on a timescale of 5 seconds in the observer frame, (iii) $r_0$ has a linear increase with time during this period.

 Indeed, the large value of $r_0$ is the main cause of the strong thermal emission. The reason is that the saturation radius ($r_{\rm s}$;  Eq. \ref{eq:rs}) also increases as $r_0$ increases, and consequently the ratio between the photosphere and saturation radii, $r_{\rm ph}/r_{\rm s}$ decreases (for the same opacity and Lorentz factor of the flow) leading to stronger radiative efficiency \citep[e.g., ][]{Meszaros&Rees2000}. Similarly large values have been derived in other bursts in which a thermal component dominates \citep[e.g., ][]{Ghirlanda2013, peer2015, Samulesson2022}.

\subsection{Recollimation shock at $r_0$ and its linear evolution}
\label{sec:linear}

%%%%%%%%%%
The results (i) and (ii) suggest a persistent jet nozzle, naturally associated with the scale of the progenitor star. Long-GRB progenitors are commonly modeled as compact Wolf--Rayet (WR) stars with characteristic radii of order $\sim 10^{10}$--$10^{11}\,\mathrm{cm}$ ($\sim 0.2$-a few $R_\odot$) 
\citep[e.g.,][]{Crowther2007, Tchekhovskoy2015, Sander2019}.
%Observed WO stars, in particular, are extremely compact with atmosphere analyses yielding $R_\star \simeq 0.24$--$0.44\,R_\odot$ \citep[]{Sander2019}.
During jet propagation through such a star, simulations show that a flow nozzle is formed by a strong recollimation shock and that its location migrates outward and can reach $z\sim 10^{10}\,\mathrm{cm}$ by breakout \citep[e.g., ][]{BrombergTchekhovskoy2016}. 

For the recollimation shock to be persistent even after breakout from the star, a pressure from the outside of the jet is needed. Several suggestions for the origin of such a pressure have been given. \citet{Thompson2007} suggested that $r_0 \sim 10^{10}$ cm could be supported by  a shear layer surviving the emergence of the jet. Similarily, a mildly-relativistic cocoon accompanying the jet breakout (i.e. the stellar cocoon) has been discussed by, e.g.,  \citet{Ioka11}, \citet{Nakar2017} and \citet{Salafia2020}. 
Such structures can contribute to the post-breakout collimation, 
however, if the confinement is provided only by a stellar cocoon or shear layer, then such a period is short. The reason is that the drop in boundary pressure is rapidly communicated back to the recollimation region by a rarefaction wave on a timescale of $t = R/c_{\rm s} \sim  0.6 $ s for $R \sim 10^{10}$ cm, where  $c_{\rm s} = c/\sqrt{3}$ is the speed of sound, and $c$ is the speed of light. The pressure decline at the site of the recollimation shock is, therefore, expected to be on a time-scale that is much shorter than the duration of the observed thermal episode.
%The cocoon can thus delay the disappearance of the recollimation shock by providing a finite post-breakout pressure reservoir. The onset of the pressure decline, however, is still communicated to the recollimation region on the causal timescale $t\sim R/c_{\rm s}$ and the cocoon mainly makes the subsequent weakening more gradual.

Result (iii), namely the linear increase of $r_0$  over the full radiation-dominated episode, suggests a different possibility. Linear increase in length-scales in the jet-cocoon system has been argued to be the result of the cocoon being embedded in a wind-like medium \citep[e.g., ][]{Bromberg11, Harrison18} and has also been indicated in numerical simulations \citep[e.g., ][]{Mizuta09, Hamidani2021}. In appendix \ref{sec:AppRC}, we show that this is also for the case for the position of the recollimation shock, $r_{\rm rc}$, as long as the cocoon pressure is maintained. In a wind-like medium, with  the wind density $\rho_{\rm w} = A r^{-2}$, then its position is given by 
\begin{equation}
r_{\rm cs}(t) \sim \sqrt{\frac{L_{\rm iso} \theta_{\rm j}^2}{4\pi \epsilon_{\rm j} c A }} \, t
\label{eq:zrcM}
\end{equation}
(see eq. \ref{eq:zrc}),  where %$L_{\rm j}$ is the jet luminosity, and jet 
$\theta_{\rm j}$ is the jet opening angle and $\epsilon_{\rm j}$ is the radiative efficiency, which has a typical value of $\epsilon_{\rm j}\sim 0.6$ for the acceleration phase\footnote{$\epsilon_{\rm j} = 1-r_{\rm ph}/r_{\rm s} = 1 - (\eta/\eta_*)^{-4/3}$, where the last step used Eq. (\ref{eq:rphrs}). In Figure \ref{fig:observations_eta} the value of $\eta/\eta_* \sim 2$ on average. Therefore, $\epsilon_{\rm j} \sim 1 - 2^{-4/3} = 0.6$}. The time is measured in the rest frame of the central-engine, and the linear relation is maintained in the observer frame as well (see App. \ref{sec:AppRC}). 
%Due to cosmological redshift, the time in the central-engine frame is related to the observed time by $t = t^{\rm obs}/(1+z)$. 
Equation (\ref{eq:zrcM}) therefore gives an explicit
linear scaling of the recollimation-shock position with
observer time, $r_{\rm cs}\propto t_{\rm obs}$.
Weak variations in the jet or wind
properties, during the observed interval, would enter through the coefficient,
$\theta_{\rm j}(L_{\rm iso}/\epsilon_{\rm j}A)^{1/2}$, and
would only show up as deviations from a strictly linear evolution, rather than removing the basic linear scaling.

{A finite wind-like shell can therefore explain a linear increase of the position of the recollimations shock. 
In addition, following breakout from the CBM, a jet-cleared polar channel is created and  the resulting drop in lateral confining pressure is communicated back to the recollimation region on a timescale longer than the subsecond timescale discussed above.
The effect of breakout will only become noticeable at the recollimation shock at $r_{\rm cs} \sim10^{10}\ {\rm cm}$ once the rarefaction wave reaches it, after a time $t\sim r_{\rm CBM}/c_{\rm s}$. %, where $R_{\rm CBM}\gg r_{\rm cs}$ has been assumed. 
The duration of the prompt emission in GRB~220426A, $\Delta t_{\rm obs}\sim10\ {\rm s}$ (Fig.~\ref{fig:lc}), can therefore be used to give a lower limit\footnote{This lower limit assumes that the disappearance of the photospheric emission, and hence the end of the observed light curve, is mainly caused by the loss of collimation and the resulting decrease in $r_0$. However, the emission may also disappear when $\eta/\eta_* \ll 1$, since the photospheric radiation then suffers substantial adiabatic losses (see Fig. \ref{fig:observations_eta}).} on the radial extent of the confining medium, $r_{\rm CBM} \simgeq c_{\rm s} \Delta t_{\rm obs}/(1 + z) \sim 10^{11}$ cm.

 The value of $r_0$ that is measured from the photopsheric emission will be related to the position of this recollimation shock, $r_{\rm cs}$. The reason is that after passing through $r_{\rm cs}$, the jet will remain approximately cylindrical, while it is confined by the surrounding cocoon.  When the confinement weakens at $\sim r_{\rm CBM}$, the flow starts a renewed, approximately conical radiation-dominated expansion. The radial streamlines of this flow extrapolate to a shifted virtual focus at $r_{\rm f} = r_{\rm CBM} - r_{\rm i}$,  where $r_{\rm i}$ is the radial distance between the virtual focus and the base of the renewed conical expansion \citep{Gottlieb2019}.
%where $R_{\rm i}$ is the distance below $R_{\rm CBM}$ to the focus point. 
 The subsequent evolution can approximately be described as a fireball in the shifted coordinate $R = r - r_{\rm f}$, with the acceleration beginning\footnote{Here an analogue fireball notation of $R_0$ could be used.}  at $R = r_{\rm i}$. 
 If the jet  re-expands with an opening angle similar to that of the incident jet, geometrical arguments give  $r_{\rm i} \sim r_{\rm cs}$ independent of the length of the collimation region \citep{Gottlieb2019}. 
 The measured value of $r_0$ is then identified with $r_0 = r_{\rm i} \sim r_{\rm cs}$ and is therefore expected to trace the linear increase of the recollimation shock. % assuming $\Gamma_{\rm i} \sim  \Gamma_0$}.
}

%SUMMARY
%Because the linear trend remains well beyond $t \sim R/c_{\rm s}$, the observations suggest that the jet collimation is maintained by a circumburst wind-like medium encountered after breakout from the progenitor star. Such a medium should then exert a pressure large enough to support the collimation of the jet in order to preserve the recollimation shock over the whole observed thermal episode. 

\subsection{Collimation requirement}
\label{sec:52}

For a circumburst wind medium to be able to collimate
the jet, it has to be very dense. The cocoon pressure
produced by the jet--CBM interaction must be large enough to maintain
an oblique reconfinement shock in the outflow. 
The pressure  of the shocked material in the cocoon is of the order of
the ambient ram pressure of the moving sidewalls
\begin{equation}
p_{\rm c}\sim {1 \over 4\pi r^2}{\dot M\over v_{\rm w}}v_{\rm c}^2 .
\end{equation}
where we have used equation (\ref{eq:cocoonpressure}) and defined the mass-loading as $\dot M/v_{\rm w}=4\pi A$, with the normalization of the wind profile $A$ defined in equation (\ref{eq:wind}).  Here, $\dot M$ is the mass injection rate
of the wind and $v_w$ is the wind velocity, and $v_{\rm c}$ is the lateral expansion speed of the shocked cocoon into the surrounding medium.

This pressure should be compared with the upstream transverse ram pressure of the jet given by equation (\ref{eq:Pperp}), 
\begin{equation}
p_{\perp, {\rm j}}(r) \sim  \frac{L_{\rm iso}}{4\pi\epsilon_{\rm j} c r^2} \, \theta_{\rm j}^2. 
\end{equation}
The condition $p_{\rm c}\gtrsim p_{\perp, {\rm j}}$ therefore gives the
characteristic mass-loading required for the shocked medium to collimate the jet,
\begin{equation}
{\dot M\over v_{\rm w}}
\gtrsim
{L_{\rm iso}\theta_{\rm j}^2\over
\epsilon_{\rm j}c^3 \beta_{\rm c}^2}
= 6.9 \times 10^{20} \, \mathrm{g/cm}.
\label{eq:massloading_collimation}
\end{equation}
Note that this is a local pressure-balance condition and the $r^{-2}$ scalings of $p_{\rm c}$ and $p_{\perp, {\rm j}}$ cancel in the wind case.
The numerical value is calculated assuming  measured value $L_{\rm iso}=10^{53}~{\rm erg~s^{-1}}$ (assuming $z=1$), mean value $\epsilon_j=0.6$, an adopted fiducial opening angle $\theta_{\rm j}=0.1$ \citep[e.g.,][]{Lloyd2020}, and taking $\beta_{\rm c} =0.3$ to be mildly relativistic \citep[see, e.g., ][]{Hamidani2021, Hamidani2025}.% and $\chi=1$.

This inferred mass loadings correspond to very dense winds, several orders of magnitude above those typically assumed for persistent Wolf–Rayet winds \citep{Nugis2000}. Indeed, assuming a continuous and persistent wind with such large mass-loading, and estimating the radius, $r$, as above which the optical depth of the wind reaches unity
\begin{equation}
\tau(r) =
\int_{r}^\infty
{\sigma_T\over 4\pi m_p r^2}
{\dot M\over v_{\rm w}}\,dr
=
{\sigma_T\over 4\pi m_p}
{\dot M\over v_{\rm w}}
{1\over r}
=1,
\end{equation}
gives a few $\times 10^{19}\,{\rm cm}$. Therefore, the
collimating medium cannot be an ordinary persistent wind
extending to large radii. It must instead be finite, for
example a dense shell produced by a short episode of
eruptive mass loss shortly before the GRB. This is in line with the discussion in \S \ref{sec:linear} where the jet breaks out of a finite dense shell.

Using the above estimates of $r_{\rm CMB} \sim 10^{11}$ cm and the mass loading of $\dot{M} /v_{\rm w} = 6.9 \times 10^{20} \, \mathrm{g/cm}$ (Eq. \ref{eq:massloading_collimation}), the mass contained  in such a shell is ($\dot{M} /v_{\rm w}) \, r_{\rm CBM} \sim$  a few $\%$ of a solar mass.

%external pressure will cease as the cocoon pressure decreases due to the jet break-out. 

Since the burst redshift is unknown, we present in Table~\ref{tab:derived_values_small} the results for four fiducial values, 
$z=[0.2, 0.5, 1, 1.4]$, corresponding to luminosity distances of $d_{\rm L} = [0.31; 0.90; 2.1; 3.2] \times 10^{28}$ cm. The table lists the inferred isotropic luminosity 
$L$, the Lorentz factor, $\Gamma$, the linear proportionality constant $k$, the mass loading, $\dot{M}/v$ %, and the jet angle $\theta$
for each assumed redshift. For smaller $z$, all four inferred quantities decrease.

The ability of a dense, finite shell surrounding the progenitor star to collimate a relativistic jet has been demonstrated by numerical simulations \citep[e.g.,][]{SuzukiMaeda2024, Hamidani2025}. As the jet propagates through the CBM, it inflates a shocked, mildly relativistic ejecta that forms a wind-driven cocoon. This newly formed wind cocoon merges with the original stellar cocoon, and the resulting confining pressure on the jet produces recollimation shocks along the outflow \citep{SuzukiMaeda2024}.

\subsection{Requirement for a successful jet}
\label{sec:requirement}

In order for the jet not to be choked by the dense medium and thereby fail to emerge (and thus not being observed as a GRB), an additional requirement on the properties of the shell is set \citep[e.g.,][]{Bromberg11, DuffellHo2020}:  A useful sufficient condition for avoiding strong deceleration of the jet head is that the jet energy flux, which is given by $L_{\rm j} / (\pi  \theta_{\rm j}^2 r^2)$,  exceeds the ambient rest-mass energy flux, $\rho_{\rm w} c^3 = Ac^3\, r^{-2}$. This condition, therefore, gives an rough estimate of the upper limit on the mass-loading [wind $A = \dot{M}/(4\pi v_{\rm w})$] as
\begin{equation}
\frac{\dot{M}}{v_{\rm w}} = \frac{ 4 L_{\rm j} }{\theta_{\rm j}^2 c^3} = \frac{L_{\rm
iso} }{ c^3 \epsilon_{\rm j}} = 6.2 \times 10^{21} \, \mathrm{g/cm},
\label{eq:Amax}
\end{equation}
where $z=1$ is assumed. In the more general case, whether a jet is choked also depends on the engine duration relative to the breakout time through the finite confining region.

Summarising, when a radiation-dominated GRB spectrum is observed  on a timescale longer than $r/c_{\rm s}$, and with a linearly increasing value of $r_0$, then the two conditions expressed in equations (\ref{eq:massloading_collimation}) and (\ref{eq:Amax}) can be combined. Together they give an estimated range of possible mass-loading of the wind:
\begin{equation}
\frac{L_{\rm iso} }{c^3\epsilon_{\rm j}} \, {\theta_{\rm j}^2\over
\chi \beta_{\rm c}^2}   \leq \frac{\dot{M}}{v_{\rm w}} \leq \frac{L_{\rm iso} }{c^3\epsilon_{\rm j}} 
\label{eq:requirement}
\end{equation}
Here, the lower limit is set by characteristic mass-loading for pressure balance
(collimation requirement) and the upper limit by the successful jet requirement.   Thus, for there to be an allowed range, the requirement on the opening angle is  $\theta_{\rm j}  \lesssim   \chi^{1/2}\beta_{\rm c} \sim 0.3$.

\section{Discussion}

\subsection{The observed cooling behaviour tracing the central engine}

The observed variability in GRB220426A is given by the intrinsic, central engine variability and not by high-latitude smoothing. The reason is that the angular time-scale due to high-latitude emission is small for the flow properties derived above.  For emission at radius $r$, curvature delays across the relativistic beaming cone occur on a few angular timescales \citep[e.g., ][]{PeerRyde2011, Hascoet2013}
\begin{equation}
t_{\rm ang}\simeq \frac{(1+z)r}{2c\Gamma^{2}}.
\label{eq:7}
\end{equation}
With $r\sim 10^{13}\ {\rm cm}$ and $\Gamma\simeq 200$, equation (\ref{eq:7}) gives that $t_{\rm ang}\approx 4\times 10^{-3}(1+z)\ {\rm s}$, showing that high-latitude emission only has an effect on the variation in the light curve on a very short time-scale. %, and that the observed variability is expected to  be longer than this time-scale. 
As mentioned in \S \ref{sec:prompt}, \citet{Maccary} measured $t_{\rm FWHM} = 0.6\pm 0.2$ s in GRB 220426A. However, the relevant observed time-scale to be compared to the high-latitude time-scale is shorter, and better captured by the pulse rise-time \citep{Camisasca, Maccary}, which is close to the timescale found by using MEPSA \citep[Multiple Excess Peak Search Algorithm; ][]{guidorzi2015}\footnote{The MEPSA detection timescale is the interval over which the significance of the narrowest detected peak is maximized and is found to track the pulse rise-time. For GBM, MEPSA was applied to 8–1000 keV background-subtracted TTE light curves following the binning and significance criteria of Maccary et al. (2025).}. For GRB 220426A the MEPSA times-scale is  $\sim 0.036$ s (private communication, Maccary 2026),  which is still longer than $t_{\rm ang}$.

Moreover, since the dominating part of the prompt emission in  GRB 220426A is from the photosphere occurring during the acceleration phase, the observer frame luminosity  and temperatures therefore correspond to the central engine values, $L_0$ and $T_0$. This is because the linear increase of the Lorentz factor in a thermal outflow, exactly compensates for the adiabatic losses \citep[e.g., ][]{Meszaros2006}.  The observed luminosity and temperature evolutions therefore directly trace the central engine evolution, for which   
\begin{equation}
T_0 = \left( \frac{L_0}{4 \pi r_0^2 c a \Gamma_0}  \right) ^{1/4} \propto L_0^{1/4}\, r_0^{-1/2}
\label{eq:T0}
\end{equation}
Since the temperature is given by the luminosity per unit area, the increase in emitting area ($\propto r^2_0$) causes the temperature to drop, even during the initial increase in luminosity (see Fig. \ref{fig:observations_kT} and also, e.g., \citet{Ryde2004}). This makes the pulse a hard-to-soft type, rather than of a tracking type \citep{Norris1986}. The latter type would require a weaker evolution $r_0$, or with $r_0$ remaining at a constant value.

During the first 5 seconds of GRB220426A, the observed emission and its variability therefore trace the central engine emission and its variability.
Later on, however, in the coasting phase the observed emission will be affected smoothing effects, by adiabatic cooling and by dissipation, which makes inferring the behaviour  of the central engine less straight-forward.

\subsection{Implications of mass-loading for photospheric emission and spectral evolution}
\label{sec:implications}

In this paper we have argued that the properties of the CBM surrounding the GRB progenitor has an important effect on the strength and appearance of the prompt, photospheric emission.

For typical GRB luminosities and Lorentz factors, a strong photospheric component generally requires an effective acceleration radius ($r_0$) substantially larger than the engine scale, often of order $r_0 \sim 10^{9}-10^{10}$ cm, although the exact threshold depends sensitively on $\eta$ \citep{Rees&Meszaros2005, Ryde2017}. For smaller values of  $r_0 \sim 10^{7}$ cm, the photosphere typically lies well above the saturation radius, causing the photospheric emission to be substantially weakened by adiabatic cooling, and thereby less probable to be observed. Thus, if an extended photospheric phase is observed in a burst, then a persistently large value of $r_0$ is required, which in turn suggests a sustained recollimation shock.

In \S\ref{sec:requirement} we argued that the mass loading required to maintain such a persistent photosphere depends on the jet luminosity. For very luminous bursts, such as GRB 220426A with $L_{\rm iso} \sim 10^{53}\ {\rm erg\ s^{-1}}$, an extremely large mass loading of order $10^{21}\ {\rm g\ cm^{-1}}$ is needed to sustain the collimation (with the exact value depending on the assumed redshift). Guided by the mass loadings inferred around interacting supernovae, such dense surroundings appear to lie at the tail end of the observed distribution and are therefore probably rare. This may explain why GRB 220426A is such an unusual case, with a very intense and unambiguous photospheric component. By contrast, for less luminous photospheric bursts, of which GRB 100507 is an example with $L_{\rm iso} \sim 5 \times 10^{50}\ {\rm erg\ s^{-1}}$ \citep{Ghirlanda2013, Ryde2017}, the required mass loading is lower by a factor of $\sim 200$, bringing it into a range more typical of supernova observations \citep[e.g.,][]{Das2024, Srinivasaragavan2025}.

For any given jet luminosity and Lorentz factor, at sufficiently low mass loading, the collimating effect of the wind is no longer strong enough to sustain the recollimation shock, which then weakens and eventually dissolves. The effective radius at which the jet acceleration begins, $r_0$, should then decrease, allowing newly injected material to accelerate from much smaller radii. The ratio of the photospheric to saturation radii, $r_{\rm ph}/r_{\rm s}=r_{\rm ph}/(\Gamma r_0)$, should as a consequence increase. Once this ratio exceeds unity, the flow is no longer radiation dominated at the photosphere, and the spectrum broadens into the characteristic NDP shape \citep{Beloborodov2011, Ryde2017, Acuner2019}. Indeed, observational analyses of photospheres forming above the saturation radius, using the method outlined by \citet{Peer2007}, infer typical values of $r_0 \sim 10^{8} - 10^{9}$ cm \citep{Ryde2010, peer2015}. Another consequence of the photosphere occurring in the coasting phase is that the radiative efficiency decreases due to enhanced adiabatic losses, which scale as $(r_{\rm ph}/r_{\rm s})^{-2/3}$. The spectra therefore both get broader and weaker as they enter the coasting phase. On the other hand, energy dissipation can naturally occur above the saturation radius, which may generate radiation-mediated shocks, partly increasing the radiative efficiency while broadening the spectrum even further \citep[e.g.,][]{Ito2013,Samuelsson2023,Wistemar2026,Wistemar2025}.

If the photospheric emission is severely degraded by adiabatic cooling and thereby not detected, the prompt emission may instead be dominated by other non-thermal emission mechanisms.  Such radiation can be produced by internal dissipation through internal shocks, but these are typically inefficient \citep{Kobayashi_etal_1997}. On the other hand, external dissipation, in which  the jet interacts with larger-scale circumburst material can produce strong and efficient synchrotron emission \citep{Ryde2022, PeerRyde2024}.

%The existence of dense CBM surrounding progenitors of GRBs, which is suggested above, could have additional effects on the prompt emission. \citet{Oganesyan2026} argue that dense CBM can modify the prompt MeV spectrum through external $\gamma - \gamma$ absorption, and use this to explain the MeV absorption feature in GRB190114C. Furthermore, \citet{Salafia2026} use a dense CBM to explain the pair-annihilation feature of GRB 221009A.

The dense CBM suggested above could also affect the prompt emission in other ways.
\citet{Oganesyan2026} showed that X-ray photons scattered back by electrons in a dense CBM can cause $\gamma - \gamma$ absorption of the outgoing MeV photons. %Using this model on the prompt emission of GRB190114C, they find that the spectral fit improves.
Moreover,  \citet{Salafia2026} argued that the MeV line in GRB221009A could be produced by electron-positron annihilation in a precursor blast wave moving through a dense CBM.

%The surrounding matter can therefore affect the observed $\gamma$-ray emission (the prompt phase) in several ways. Material at small radii $(\lesssim 10^{13}-10^{14}$ cm) will primarily affect the photospheric emission, occurring within tens of seconds of the trigger, as shown in this paper. 

\subsection{Properties of dense circumstellar shells}

{The mass loading inferred here, is of the order of
$\dot M/v_{\rm w}\sim10^{21}\ {\rm g\ cm^{-1}}$, is large,
but it is still within the upper end of values inferred for compact
circumstellar material around stripped-envelope supernovae. In
particular, broad-lined Type Ic SNe are observationally associated
with long GRBs, and early optical and X-ray light curves many times
require dense material at radii
$r_{\rm CSM}\sim10^{12}$--$10^{15}\ {\rm cm}$. For example,
\citet{Srinivasaragavan2025} compile cases with typical
$M_{\rm CSM}\sim0.1$--$0.5\,M_\odot$ and
$r_{\rm CSM}\sim0.1$--$25\times10^{13}\ {\rm cm}$. Interpreting
these as shell-equivalent mass loadings\footnote{For
a shell with $\rho_{\rm w}=Ar^{-2}$ between $r_{\rm in}$ and
$r_{\rm out}\gg r_{\rm in}$, the shell-equivalent mass loading is
$\dot M/v_{\rm w}=4\pi A\simeq M_{\rm CSM}/r_{\rm CSM}$.} gives
$\dot M/v_{\rm w}\sim M_{\rm CSM}/r_{\rm CSM}
\sim10^{18}$--$5\times10^{20}\ {\rm g\ cm^{-1}}$.
Similarly, \citet{Das2024} infer
$\dot M/v_{\rm w}\sim10^{18}$--$3\times10^{21}\,
{\rm g\ cm^{-1}}$ for SNe Ibc, with about ten per cent of the
sample showing evidence for dense compact shells in their optical
light curves.  For a freely expanding shell, the inferred radius corresponds to ejection only hours to about a day before the collapse (using $r \sim 10^{12}-10^{13}$ cm and shell velocity of $\sim 10^3-$ few $\times10^3\ {\rm km\,s^{-1}}$).

%If the material is a freely expanding shell, then
%$r\simeq v_{\rm sh}\Delta t_{\rm ej}$ implies ejection only hours
%to days before collapse for
%$r\sim10^{12}$--$10^{13}\ {\rm cm}$ and
%$ v_{\rm sh} \sim {\mathrm{few } }\times 10^3 \ {\rm km\ s^{-1}}$. 

The early light curves constrain the effective mass and radial extent of the extended material, but do not uniquely determine its physical origin. Possible origins include late binary mass transfer and circumbinary material \citep{ercolino, sun}, wave-driven pre-SN mass loss \citep{fuller}, jet/cocoon-shocked material \citep{nakarpiran}.  Moreover, there are examples of extreme pre-collapse
mass loss that can reach even larger mass per unit radius.
For instance, pulsational pair instability (PPI)
models can eject several solar masses, and in some cases
much more, with velocities of a few $10^3\,{\rm km\,s^{-1}}$
and radii $r_{\rm CSM}\sim10^{12}-10^{15}\,{\rm cm}$
before collapse \citep{Leung2019, Renzo2020}. Such compact,
massive shells can therefore reach shell-equivalent mass
loadings comparable to, or larger than, the values required here.
For example, material moving at $2500\,{\rm km\,s^{-1}}$
reaches $r_{\rm sh}\sim 2\times10^{12}\ {\rm cm}$
after a couple of hours. At such a radius,
$\dot M/v_{\rm w}=10^{21}\,{\rm g\,cm^{-1}}$
corresponds to
$M_{\rm sh}\simeq0.9\,M_\odot .$
The PPI channel has been proposed for some ultra-long GRBs \citep[e.g.,][]{Marchant20}.

The requirement in Equation~(\ref{eq:massloading_collimation}) is therefore
demanding, but not unphysical. It lies above the mass loadings
inferred for many stripped-envelope SN environments, but overlaps
the most extreme observational cases and can be produced by several
physically distinct channels that place dense material close to the
progenitor shortly before collapse.}

\subsection{Inclusion of magnetic fields}

{A dynamically important magnetic field will affect both the jet dynamics and the transverse pressure balance. Indeed, some models suggest that magnetic fields play an important role in the launching of GRB jets, since both black-hole--accretion-disc and magnetar central engines can naturally produce initially Poynting-flux-dominated outflows \citep{Lyutikov06, Giannios2008, ZY11}. Furthermore, 3D relativistic MHD simulations show that substantial magnetic dissipation can  occur while the jet is traversing the progenitor core, after which the jet  becomes predominantly thermal \citep{BrombergTchekhovskoy2016, Gottlieb2022}. %Similarly, GRMHD simulations of collapsar jets find strong magnetic dissipation around the collimation nozzle, \citep{Gottlieb2022}. Some magnetization may nevertheless survive to larger radii.

For GRB~220426A, the narrowness of the observed photospheric spectra suggests a thermally driven outflow, i.e., radiation-dominated accelerating flow. In such a flow, $\Gamma\propto r$, and the variation of the Doppler factor over the distribution of last-scattering radii is compensated by the corresponding evolution of the flow, producing the narrowest expected photospheric spectrum. In contrast, photospheric emission from gradually magnetically accelerated flows, for which  $\Gamma\propto r^{1/3}$ \citep{Drenkhahn02}, is expected to be broader \citep{Peer17}. The observed spectra therefore disfavor a flow  whose dynamics at the photosphere is dominated by magnetic acceleration. 

%A consequence of this fact is that any persistence of the jet collimation after breakout can 
%not be supported by an internal collimating, magnetic stress.
%This constraint does not, however, exclude a dynamically important magnetic component outside the photospheric spine.
While the data clearly disfavors a magnetized jet, it provides little information on the confining environment, which could be magnetized.  
A possible configuration is that the radiation-dominated jet is surrounded by a more strongly magnetized sheath. Such a structure is qualitatively consistent with MHD simulations in which toroidal magnetic fields are concentrated towards the outer parts of the jet, and even a subdominant magnetization can substantially suppress mixing between the jet and the cocoon \citep[e.g., ][]{Gottlieb2020}. 

For a predominantly toroidal magnetic field and a Lorentz boost along the poloidal direction, the transverse confining magnetic stress can be written as $p_{\rm conf} \sim {B'^2}/{8\pi}$, where $B'$ is the magnetic-field strength in the comoving frame \citep{SL2018}.
A magnetic sheath could therefore contribute significantly to the collimation, as soon as $p_{\rm conf}$ becomes comparable to the transverse pressure of the jet spine, $p_{{\rm j},\perp}$ (Equation~(\ref{eq:Pperp}). Equating the pressures leads to 
\begin{equation}
B' \sim  \left( \frac{2\, L_{\rm iso}}{\epsilon_{\rm j} c} \right)^{1/2} \, \frac{\theta_{\rm j}}{r}. 
\end{equation}
which corresponds to $B'$ of order $10^{10}$~G at $r \sim 10^{10}$ cm  for the parameters considered here. In magnetized GRB jet models, such a size of the field strength  is not implausible at these radii \citep[e.g., ][]{Komissarov2009, Bromberg2014}.

Such a magnetized sheath could therefore provide a confining stress, without the need of a dense wind. However, this alone is not enough to explain  the observed evolution of $r_0(t)$. Since $p_{{\rm j},\perp}\propto r^{-2}$, the approximately linear evolution $r_0\sim r_{\rm cs}\propto t$ requires the confining stress to decrease approximately as $p_{\rm conf}\propto t^{-2}$. A magnetized sheath could in principle satisfy this condition, for example if a predominantly toroidal field evolves as $B'\propto t^{-1}$. However, if such a scaling can be maintained depends on the dynamics, magnetic dissipation, and stability of the sheath. Detailed relativistic-MHD calculations would therefore be required to determine if magnetic confinement can reproduce both the inferred magnitude and temporal evolution of $r_0(t)$.}

%Interestingly, a predominantly toroidal magnetic sheath could also provide a pressure evolution similar to that required for the recollimation position to increase linearly with time, namely $p_{\rm B}(t)\propto t^{-2}$ [Eq. (\ref{eq:Pc2})]. For an approximately flux-frozen toroidal field, the lab-frame field scales as $B_\phi\propto R_{\rm sh}^{-1}$, where $R_{\rm sh}$ is the characteristic transverse size of the sheath. If the sheath participates in the approximately self-similar expansion inferred above, with $R_{\rm sh}\propto t$, this gives $B_\phi\propto t^{-1}$. 
%If the sheath has reached an approximately coasting phase in the region of interest, the comoving magnetic pressure evolves as $p_{\rm B}\sim \frac{B_\phi^2}{8\pi}\propto t^{-2}$. Therefore, under these specific assumptions, a collimating, magnetic sheath could explain the observed approximately linear evolution of the recollimation position.

\subsection{Comparison to previous work}

An increasing normalisation of the thermal  component (given by $\mathscr{R}$; Eq. \ref{eq:R}), as found above in GRB220426A, has been regularly observed in many bursts. This was, e.g., the case in the large sample of smooth pulses studied by \citet{RydePeer2009}. These pulses, however, differ from the temporal property of GRB220426A which has a much shorter variability time scale and is much brighter. The pulses in \citet{RydePeer2009} all had a variability time scale of the order of the pulse length. Another difference from GRB220426A, is that the bursts in \citet{RydePeer2009} are not purely thermal. Instead, they are all fitted with a hybrid model, consisting of a power-law spectral component in addition to the thermal component \citep[see also, e.g., ][]{Ryde2005, Battelino2007, Guiriec2011, Iyyani2013}.  Therefore, the photopshere for these bursts are interpreted to occur in the coasting phase.  An important consequence is that the  high-latitude time-scale becomes much longer %compared to the time-scale during the acceleration phase 
and can even determine the pulse duration.  This fact was used by \citet{PeerRyde2011} to explain the temporal and spectral behaviors of these bursts.
%Similar fits was confirmed later, but alternative interpretation exist including multiple breaks in the spectra.
%

Nevertheless, we note that the average slope found in \citet{RydePeer2009} for the observed parameter $\mathscr{R} (t) \propto t^{0.4}$, which can be translated to a temporal evolution of $r_{\rm cs} \propto t^{0.4}$ using equation (\ref{eq:r0}). Using the scaling of lengths in the jet-cocoon system found in \citet{Harrison18}, the exponent is expected to be  $({4+s})/{2(5-s)}$, where the density profile $\rho \propto r^{-s}$, which gives to $r_{\rm cs}\propto t^{2/5}$ in a uniform environment ($s=0$) and to $r_{\rm cs}\propto t$ in a wind–like medium ($s=2$). Therefore, the averaged temporal index of 0.4 corresponds to a uniform medium. 
%The broad distribution of temporal indices found in \citet{RydePeer2009}, which span from 0 to 1.2, 
The observed differences might  thus be interpreted as variation in the properties of the dense environments around the progenitor star.

Indeed, a steady wind profile is  not required for the material surrounding a GRB progenitor.  Observed interacting supernovae show that massive-star surroundings can differ significantly from the standard  wind profile, for which $s=2$.  For example, \citet{Lane2025} infer $s=1.4$ for SN2019vxm, suggesting a structured and asymmetric surrounding rather than a stationary wind. 
This H-rich event is not a direct analogue of a stripped-envelope GRB-SN, but it demonstrates that massive-star mass loss can produce non-wind-like density structures and therefore a plausible alternative for the medium confining a GRB jet.

Other interpretations of the observed spectral evolution seen in GRB220426A are possible. The gradual broadening could be due to changing superposition of temperatures from different regions of the photosphere, for instance, from a structured jet \citep{Lundman2013}. Indeed, \citet{Wang2022} modelled the time resolved spectra of GRB220426A with a multicolour blackbody. Moreover \citet{Song2022} suggested a hybrid outflow for which the Poynting flux is thermalized below the photosphere, and the magnetic field regulates the baryon loading and amount of energy that can be dissipated.   Similarily, \citet{Deng2022} suggested an increasing baryon-load and changes in the environment where the jet is launched.  Finally, \citet{Gupta2026} interpreted the spectral change as being due to multiple emission zones, with physically distinct regions. These explanations are for the broadening of the spectrum, but do not address the linearly increasing value of $r_0$ found here.

%%%%%%%%%%%%%%%%%%%%%%%%%%%%%%%%%%%%%%%%%%%%%%%%
\begin{deluxetable*}{ccccc|c}
\tabletypesize{\scriptsize}
\tablecaption{Time-resolved parameters with uncertainties from fits with the acceleration-phase photosphere model \citep[RDP;][]{Ryde2017}.\label{tab:fit_params_round}}
\tablehead{
\colhead{$t_{\rm mean}$} &
\colhead{$E_{\rm p}$} &
\colhead{$\eta/\eta_*$} &
\colhead{$N_{\gamma}$} &
\colhead{$F_{\rm E}$} &
\colhead{$\Delta\mathrm{AIC}$} \\
\colhead{s} &
\colhead{keV} &
\colhead{} &
\colhead{cm$^{-2}$ s$^{-1}$} &
\colhead{$10^{3}$\,keV cm$^{-2}$ s$^{-1}$} &
\colhead{}
}
\startdata
0.6 & $119.5\pm4.8$ & $8.4\pm1.9$ & $43.5\pm0.5$ & $5.4\pm0.1$ & $216.6$ \\
1.4 & $102.8\pm3.7$ & $5.6\pm1.4$ & $103.8\pm1.4$ & $11.7\pm0.2$ & $141.9$ \\
1.8 & $90.4\pm1.5$ & $2.3\pm0.4$ & $108.6\pm1.6$ & $11.7\pm0.2$ & $80.1$ \\
2.1 & $86.4\pm1.6$ & $1.6\pm0.3$ & $114.6\pm1.7$ & $12.1\pm0.3$ & $41.5$ \\
2.3 & $99.3\pm1.9$ & $1.4\pm0.3$ & $149.4\pm2.3$ & $18.2\pm0.4$ & $31.7$ \\
2.5 & $98.4\pm1.9$ & $1.7\pm0.3$ & $162.0\pm2.4$ & $19.4\pm0.5$ & $48.7$ \\
2.7 & $88.6\pm1.6$ & $2.1\pm0.4$ & $143.7\pm2.2$ & $15.3\pm0.3$ & $69.3$ \\
3.0 & $74.8\pm1.4$ & $1.5\pm0.3$ & $128.7\pm2.0$ & $11.8\pm0.2$ & $41.8$ \\
3.3 & $67.6\pm1.3$ & $1.1\pm0.3$ & $112.2\pm1.9$ & $9.4\pm0.2$ & $26.8$ \\
3.6 & $70.2\pm1.2$ & $2.4\pm0.4$ & $129.7\pm2.0$ & $10.9\pm0.2$ & $81.5$ \\
3.8 & $58.9\pm1.1$ & $2.0\pm0.3$ & $131.3\pm2.1$ & $9.4\pm0.2$ & $61.6$ \\
4.1 & $60.9\pm1.0$ & $0.9\pm0.1$ & $173.9\pm2.7$ & $13.2\pm0.3$ & $17.2$ \\
4.2 & $61.8\pm1.3$ & $1.5\pm0.3$ & $206.0\pm3.5$ & $15.6\pm0.3$ & $35.9$ \\
4.4 & $60.1\pm1.3$ & $1.6\pm0.4$ & $218.1\pm3.7$ & $16.0\pm0.3$ & $45.6$ \\
4.5 & $57.9\pm1.2$ & $2.1\pm0.4$ & $198.7\pm3.3$ & $13.8\pm0.3$ & $60.7$ \\
4.6 & $57.3\pm1.3$ & $1.6\pm0.4$ & $200.3\pm3.4$ & $14.0\pm0.3$ & $42.1$ \\
4.8 & $50.6\pm1.2$ & $1.2\pm0.3$ & $176.4\pm3.3$ & $11.1\pm0.2$ & $31.4$ \\
5.0 & $48.7\pm1.0$ & $1.6\pm0.3$ & $143.5\pm2.4$ & $8.6\pm0.2$ & $50.0$ \\
5.4 & $39.6\pm0.6$ & $0.7\pm0.1$ & $107.9\pm1.8$ & $5.4\pm0.1$ & $7.1$ \\
6.1 & $31.5\pm0.4$ & $0.4\pm0.1$ & $61.9\pm0.9$ & $2.47\pm0.03$ & $-0.8$ \\
\enddata
\tablecomments{Quoted uncertainties are $1\sigma$. Here $\Delta\mathrm{AIC}\equiv\mathrm{AIC}_{\rm NDP}-\mathrm{AIC}_{\rm RDP}$ and positive values favor the RDP model and negative values favor the NDP model.}
\end{deluxetable*}

\begin{deluxetable*}{ccccccc}
\tabletypesize{\scriptsize}
\tablecaption{Time-resolved parameters derived from the photospheric-model fits.\label{tab:physical_parameters}}
\tablehead{
\colhead{$t_{\rm mean}$} &
\colhead{$r_0$} &
\colhead{$r_{\rm s}$} &
\colhead{$r_{\rm ph}$} &
\colhead{$\eta$} &
\colhead{$\eta_*$} &
\colhead{$\Gamma$} \\
\colhead{s} &
\colhead{$10^{10}$ cm} &
\colhead{$10^{13}$ cm} &
\colhead{$10^{13}$ cm} &
\colhead{} &
\colhead{} &
\colhead{}
}
\startdata
0.6 & $1.31\pm0.11$ & $2.8\pm0.7$ & $0.16\pm0.02$ & $2150\pm520$ & $255.5\pm5.3$ & $126\pm12$ \\
1.4 & $2.54\pm0.18$ & $3.7\pm1.0$ & $0.38\pm0.05$ & $1470\pm390$ & $263.4\pm4.8$ & $148\pm15$ \\
1.8 & $3.15\pm0.11$ & $1.8\pm0.3$ & $0.59\pm0.04$ & $584\pm93$ & $249.6\pm2.6$ & $188\pm11$ \\
2.1 & $3.46\pm0.13$ & $1.3\pm0.3$ & $0.73\pm0.07$ & $388\pm84$ & $245.6\pm2.7$ & $211\pm17.$ \\
2.3 & $3.21\pm0.13$ & $1.2\pm0.3$ & $0.80\pm0.08$ & $383\pm97$ & $277.4\pm3.2$ & $249\pm23$ \\
2.5 & $3.39\pm0.14$ & $1.6\pm0.3$ & $0.79\pm0.07$ & $463\pm96$ & $278.0\pm3.2$ & $234\pm18$ \\
2.7 & $3.74\pm0.14$ & $2.0\pm0.4$ & $0.75\pm0.06$ & $530\pm93$ & $255.6\pm2.7$ & $200\pm13$ \\
3.0 & $4.55\pm0.18$ & $1.6\pm0.4$ & $0.91\pm0.08$ & $341\pm78$ & $227.9\pm2.5$ & $199\pm17$ \\
3.3 & $4.94\pm0.21$ & $1.2\pm0.4$ & $1.01\pm0.12$ & $235\pm69$ & $211.2\pm2.4$ & $204\pm22$ \\
3.6 & $5.04\pm0.19$ & $2.6\pm0.4$ & $0.82\pm0.06$ & $516\pm83$ & $217.8\pm2.3$ & $163\pm10$ \\
3.8 & $6.59\pm0.27$ & $2.5\pm0.5$ & $1.03\pm0.08$ & $385\pm73$ & $196.1\pm2.2$ & $157\pm11$ \\
4.1 & $7.2\pm0.3$ & $1.5\pm0.5$ & $1.65\pm0.29$ & $203\pm57$ & $222\pm26$ & $229\pm40$ \\
4.2 & $7.7\pm0.4$ & $2.5\pm0.6$ & $1.43\pm0.15$ & $326\pm80$ & $214.4\pm2.7$ & $186\pm17$ \\
4.4 & $8.2\pm0.4$ & $2.9\pm0.6$ & $1.48\pm0.14$ & $348\pm75$ & $212.3\pm2.6$ & $180\pm15$ \\
4.5 & $8.3\pm0.4$ & $3.5\pm0.7$ & $1.33\pm0.11$ & $424\pm82$ & $204.1\pm2.4$ & $160\pm12$ \\
4.6 & $8.5\pm0.4$ & $2.8\pm0.7$ & $1.47\pm0.15$ & $331\pm78$ & $203.8\pm2.5$ & $173\pm15$ \\
4.8 & $9.6\pm0.5$ & $2.2\pm0.6$ & $1.66\pm0.19$ & $232\pm64$ & $186.3\pm2.3$ & $173\pm18$ \\
5.0 & $9.1\pm0.4$ & $2.6\pm0.6$ & $1.38\pm0.13$ & $283\pm61$ & $176.7\pm2.1$ & $151\pm13$ \\
\enddata
\tablecomments{Quoted uncertainties are $1\sigma$.}
\end{deluxetable*}

%%%%%%%%%%%%%%%%%%%%%%%%%%%%%%%%%%%%%%%%%%%%%%%%%%%%%%%%%%

\begin{deluxetable*}{ccccc}
%\tabletypesize{\scriptsize}
\tablecaption{Dependence of the derived quantities on redshift. Values are averaged over the duration. The mass-loading parameter, $\dot{M}/v_{\rm w}$, is calculated from Equation~(\ref{eq:massloading_collimation}). \label{tab:derived_values_small}}
\tablehead{
\colhead{$z$} &
\colhead{$L$} &
\colhead{$\Gamma$} &
\colhead{$k$} &
\colhead{$\dot{M}/v_{\rm w}$} \\
\colhead{} &
\colhead{erg s$^{-1}$} &
\colhead{} &
\colhead{cm s$^{-1}$} &
\colhead{g cm$^{-1}$}
}
\startdata
0.2 & $2.3\times 10^{51}$ & 85  & $6.6\times 10^{9}$  & $1.6\times 10^{19}$\\
0.5 & $1.9\times 10^{52}$ & 124 & $1.2\times 10^{10}$ & $1.3\times 10^{20}$\\
1.0 & $1.0\times 10^{53}$ & 177 & $1.6\times 10^{10}$ & $6.9\times 10^{20}$\\
1.4 & $2.4\times 10^{53}$ & 215 & $1.7\times 10^{10}$ & $1.6\times 10^{21}$
\enddata
\end{deluxetable*}

\section{Conclusion}

In this paper we present a time-resolved spectroscopic study of the exceptionally bright GRB 220426A detected by \textit{Fermi}/GBM.
The dominant part of its prompt emission is unambiguously photospheric, produced while the outflow is still radiation-dominated and accelerating, i.e., prior to complete conversion of thermal energy to bulk kinetic energy. The time-resolved spectra are among the narrowest observed in GRBs, and bright enough to allow strong constraints on the flow conditions. 

Using this fact, we infer the jet effective launch radius to 
$r_0 \sim {\rm few \times 10^{10}}$ cm, pointing to an acceleration region that is not set by the central-engine scale, but by the scale at which the jet emerges from the progenitor core  region. We further find that the inferred $r_0$ follows a nearly linear increase in time, and provides evidence that the jet remains collimated for a duration exceeding the local dynamical time. We show that such behavior can be explained by a sustained external confinement, for instance,  by a dense, wind-like circumburst medium (CBM).  Using analytical scalings, we show that the measured $r_0(t)$ and luminosity require a finite, dense confining shell with a large mass-loading of the order of 
$\dot{M}/v_w \simeq 10^{21}\ \mathrm{g\ cm^{-1}}$ (depending on the assumed redshift). 

{Several transient events have been suggested to arise from interaction with circumburst material, including low-luminosity GRBs, some X-ray flashes, and late prompt synchrotron emission. A relativistic jet propagating through a dense external medium inflates a hot cocoon, which both collimates the jet and can weaken or even choke it. In the interpretation presented here, GRB 220426A may represent the successful, high-power end of the same general jet-CBM interaction phenomenon. Indeed, recent analytic work by \citet{Hamidani2025} similarly finds a continuum between successful, barely failed, and completely failed jets, determined by the properties of both the jet and the CBM. In contrast to the prompt X-ray emission in low-luminosity GRBs and some X-ray flashes, which has been suggested to originate from shock breakout of the cocoon, the dominant prompt component in GRB 220426A is consistent with emission from the photosphere of the jet itself. In this case, the CBM is sufficiently dense to collimate the jet, but not sufficiently extended or massive to choke it completely. The nearly linear evolution of the inferred $r_{0}$ therefore provides a direct probe of the recollimation region, rather than of the cocoon. Moreover, matter structures at even larger radii, such as Wolf--Rayet ring nebulae at $\sim 10^{18}$ cm, could instead power late-time $\gamma$-ray emission appearing on timescales of hundreds of seconds. All these different transient events may thus illustrate different outcomes of the same underlying interaction between a relativistic jet and dense material produced by pre-collapse mass-loss and winds. Prompt GRB observations therefore provide a way to probe the innermost CBM and the final mass-loss history of the progenitor, independently of the constraints from interacting supernovae.
}

\begin{acknowledgements}
We wish to thank Drs. Anjasha Gangopadhyay and Romain Maccary for useful discussions.  FR acknowledges support from the Swedish National Space Agency (2021-00180 and 2022–00205). This research has made use of data and/or software provided by the High Energy Astrophysics Science Archive Research Center (HEASARC), which is a service of the Astrophysics Science Division at NASA/GSFC.
\end{acknowledgements}

\appendix

\section{Scalings of radiation-dominated flows}
\label{sec:AppRD}

A photosphere that emerges while the outflow still is in the radiation-dominated phase will be referred to as a radiation-dominated phase photosphere (RDP).

\subsection{Relations describing the radiation-dominated flow}
\label{sec:relations}

During the the radiation-dominated phase, the Lorentz factor increases linearly with radius, $r$, as  
\begin{equation}
\Gamma = \frac{\Gamma_0}{r_0} \, r,
\label{eq:1}
\end{equation}
\noindent where ${\Gamma_0}$ and ${r_0}$ are the values of the Lorentz factor and radius at the base of the jet \citep{Meszaros&Rees2000}.

The ratio between the total, isotropic-equivalent luminosity and baryonic mass outflow rate
\begin{equation}
\eta = \frac{L}{\dot{M} c^2}.
\label{eq2}
\end{equation}
gives the dimensionless specific enthalpy density, i.e., the total energy (internal + kinetic) per baryon in units of rest-mass energy\footnote{For an adiabatic, radiation-dominated flow (i.e., thermal acceleration) $\eta$ (Eq. \ref{eq2}) is proportional with the specific entropy per baryon, differing by a constant $\Gamma_0 T_0$ \citep[e.g., ][]{Meszaros1993, Meszaros2006, Ryde2019}.}. 
The flow saturates to the coasting phase Lorentz factor $\Gamma = \eta$, when all the internal energy has been converted to kinetic energy.  The saturation radius occurs at
\begin{equation}
    r_{\rm s} = \eta \,  \frac{r_0}{\Gamma_0},
    \label{eq:rs}    
\end{equation}
\noindent where $\Gamma_0$ and $r_0$ is the Lorentz factor and the radius at which the fireball starts to be accelerated. 

We can define $\eta_*$ as a critical value of $\eta$, for a given $L$ and $r_0$, such that  if $\eta > \eta_*$ ($\eta < \eta_*$) ,  then $r_{\rm ph}$ < $r_{\rm s}$ ($r_{\rm ph} > r_{\rm s}$).  The expression for $\eta_*$ can be found by equating equation (\ref{eq:rs}) and
the photosphere radius, where the optical depth $\tau =1$,
\begin{equation}
    r_{\rm ph} = \frac{L \sigma_{\rm T} \kappa_\pm}{8 \pi m_p c^3 \Gamma^2 \eta},
    \label{eq:rph}    
\end{equation}
\noindent where $\kappa_\pm$ is the pair multiplicity and $m_{\rm p}$ is the proton mass and $c$ is the speed of light \citep{Peer2008}.  Solving for $\eta$ and using that a saturated jet has $\Gamma \sim \eta$, the expression becomes
\begin{equation}
\eta_* = \left( \frac{L \sigma_{\rm T} \kappa_\pm \Gamma_0}{8 \pi m_p c^3 r_0}\right)^{1/4},
\label{eq:eta*}
\end{equation}

In reality, both the end of the acceleration phase and photon decoupling occurs over more than one decade in radius, therefore, $\eta_*$ and $r_{\rm s}$, as well as parameters derived from them, should be treated as characteristic values. Similarly, the jet can be said to have saturated, when $\eta/\eta_*$ is around unity, however, the final Lorentz factor is only reached asymptotically (see further discussion below). 

The position of the photosphere relative to the saturation radius during the radiation-dominated phase (or equivalently the acceleration phase) can be found by inserting equation (\ref{eq:1}) with $r=r_{\rm ph}$ into equation (\ref{eq:rph}) to solve for $r_{\rm ph}$ and then dividing by $r_{\rm s}$ (eq. \ref{eq:rs}). Using the definition of $\eta_*$ (Eq. \ref{eq:eta*}) then gives that
\begin{equation}
\frac{r_{\rm ph}}{r_{\rm s}} = \left( \frac{\eta}{\eta_*} \right)^{-4/3}
\label{eq:rphrs}
\end{equation}

\subsection{Measured and derived quantities from a radiation-dominated phase photosphere}
\label{sec:quantaties}

 The (isotropic equivalent) $\gamma$-ray luminosity is related to the observed flux as $L_{\gamma} = 4\pi d_L^2 F_{\rm E}/\epsilon_\gamma$, where $F_{\rm E}$ is the observed $\gamma$-ray flux and $\epsilon_\gamma$ is the radiation efficiency, defined as $\epsilon_\gamma = L_\gamma / L_{\rm 0}$, where $L_{\rm 0}$ is the (isotropic equivalent) total injected power. If the emission spectrum during the acceleration phase is within the observed spectral band, then $\epsilon_\gamma \sim 1$.

The effective transverse size of the emitting region \citep{Peer2007} is given by 
%With $F_{\rm {BB}}(r_{\rm ph})$ and the temperature $T = \theta_{\rm u} \, mc^2 /k$ we can characterise the thermal emission by
\begin{equation}
 \mathscr{R} = \left( \frac{{F}_{\rm E}}{\sigma {T}_{}^4} \right)^{1/2} = \delta  \, \frac{r_{\rm ph}}{\Gamma},
\label{eq:R}
\end{equation}
where the constant $\delta = 1.06 \, (1+z)^2 \, d_{\rm L}^{-1} $ depends on the distance to the source. 

%\subsubsection{Base of the flow,  $r_{0}$}

Combining equations (\ref{eq:R}) with (\ref{eq:1}) gives
\begin{equation}
r_0 = \frac{\Gamma_0}{\delta} \, \mathscr{R},
\label{eq:r0}
\end{equation}
which gives a direct determination of the size of the base of the flow, $r_0$, from ${F}_{\rm BB}$ and ${T}$.

Furthermore, the value of $\eta$ can also be determined from observations. The shape of the spectrum released at the photosphere from a radiation-dominated jet will depend on the ratio $\eta/\eta_*$ \citep{Ryde2017}. Therefore, measuring the width of the spectrum gives an estimate of the value of $\eta$ through 
\begin{equation}
\eta = \left( \frac{\eta}{\eta_*}\right) \, \eta_*
\label{eq:eta}
\end{equation}
\noindent where $\eta_*$ can be estimated from equation (\ref{eq:eta*}) by using the measurements of luminosity, $L = L_{\rm 0} = L_{\gamma}/\epsilon_{\gamma} \sim L_{\gamma}$  and $r_0$ (eq. \ref{eq:r0}) with the use of $\mathscr{R}$ and $\Gamma_0 =1$ (eq. \ref{eq:R}).

The Lorentz factor can now be derived by combining the determined values of $\eta_*$ and $(\eta/\eta_*)$.
Combining equations (\ref{eq:1}) at $r=r_{\rm ph}$ with equation (6) in \citep{Meszaros&Rees2000}  %[equation (A7)] 
gives
\begin{equation}
\Gamma = \eta_* \left( \frac{\eta}{\eta_*} \right)^{-1/3}
\label{eq:Gamma}
\end{equation}

%\subsubsection{Photospheric radius, $r_{\rm ph}$}

Finally, a determination of the photopsheric radius is given by combining equation (\ref{eq:Gamma}) with equation (\ref{eq:1}) as 
\begin{equation}
r_{\rm ph} = \frac{r_0}{\Gamma_0}\, \Gamma,
\label{eq:ph6}
\end{equation}
and from Equations (\ref{eq:rs}), (\ref{eq:r0}), and (\ref{eq:eta}), the saturation radius is given by
\begin{equation}
r_{\rm s} = \delta^{-1} \mathscr{R}\left( \frac{\eta}{\eta_*} \right) \, \eta_*
\label{eq:rs2}
\end{equation}

\subsubsection{Summary of derived quantities}
For an observed spectrum from a radiation-dominated flow the following three quantities can be measured: ${F}_{\rm E}$,  $\mathscr{R}$, and width of the spectrum ($\eta/\eta_*$). Combined with measurements of luminosity distance, $d_{\rm L}$, and the redshift, $z$, of the burst,  we can thus derive
\begin{enumerate}
  \item radius at the base of the fireball, $r_0$ (from eq. \ref{eq:r0}),
  \item the critical enthalpy, $\eta_*$ (from eq. \ref{eq:eta*}),
  \item the enthalpy, $\eta$ (from eq. \ref{eq:eta}),
  \item  the Lorentz factor, $\Gamma$ (from eq. \ref{eq:Gamma}),
  \item the photospheric radius, $r_{\rm ph}$ (from eq. \ref{eq:ph6}),
  \item the saturation radius, $r_{\rm s}$ (eq. \ref{eq:rs2}). 
\end{enumerate}
The only unknowns are the initial Lorentz factor ($\Gamma_0$) at $r_0$, and the pair loading $\kappa_\pm$, both of which are assumed to be unity in the analysis in this paper.

Note that the relations introduced in this section, and used to infer the quantities above, are analytical scalings that are intended as order-of-magnitude estimates. They rely on the standard set of order-unity approximations commonly adopted in this literature.

\section{Interpretation of the linear growth of the position of the recollimtion shock}
\label{sec:AppRC}

From the analysis of the  observations of the thermal emission component in GRB 220426A we concluded that the recollimation shock has a linear variation in time. Here we interpret this evolution as a result of a dense, confining wind-like medium surrounding the progenitor star, resulting in a linear scaling given in equation (\ref{eq:zrc}) below.

In order to find the main dependances and temporal scalings of the recollimation shock, we follow the standard treatment of the cocoon-jet interaction, invoking a 1-zone model for the cocoon, which has a uniform pressure, $p_{\rm c}$ \citep{Bromberg11, Salafia2020}. The cocoon is assumed to be cylindrical in shape, with a lateral radius, $\tilde{r}_{\rm c}$.

We will first find the scaling of $\tilde{r}_{\rm c}$ and $p_{\rm c}$.  The sidewall of the cylinder forms a shock, which moves into the ambient medium with a velocity, $v_{\rm c}$. The ambient medium is assumed to be wind-like and have a radial dependence of the density following 
\begin{equation}
\rho_{\rm w}(r) = A \, r^{-2},
\label{eq:wind}
\end{equation}
where $r$ is the radial direction from the progenitor site, and a pressure which is assumed to be 
negligible.
Equating the cocoon pressure, $p_{\rm c}$, with the ambient ram pressure of the moving sidewalls
\begin{equation}
p_{\rm c} \sim v_{\rm c}^2 \, \rho_{\rm w}(r),
\label{eq:cocoonpressure}
\end{equation}
gives an estiate of $v_{\rm c}$.

A useful approximation is to use an effective density seen by the sidewall, which can be taken as the area-averged mean density $<\rho_{\rm w}>$ \citep{BegelmanCioffi89, Bromberg11}. Using this approximation for a wind-like medium (Eq. \ref{eq:wind}), one finds that, up to the first order, $<\rho_{\rm w}> \sim \rho_{\rm w}(r = \tilde{r}_{\rm c}) = A \, \tilde{r}_{\rm c}^{-2}$, where $\tilde{r}_c$ is the cocoon radius. Therefore,
\begin{equation}
v_{\rm c} = \frac{d\tilde{r}_{\rm c}}{dt} =\left({\frac{p_{\rm c}}{\rho_{\rm w}(\tilde{r}_{\rm c})}} \right)^{1/2}= \left({\frac{p_{\rm c}}{A}}\right)^{1/2} \, \tilde{r}_{\rm c}
\label{eq:drdt}
\end{equation}
where $t$ is the time measured in the rest-frame of the central-engine.

To find an expression for $p_{\rm c}$ in equation (\ref{eq:drdt}) we use the equation-of-state in the cocoon $p_{\rm c} = E_{\rm c}/3V_{\rm c}$, where the injected energy, $E_{\rm c}(t) \sim L_{\rm j} t$, with the jet luminosity $L_{\rm j}$ and the cocoon volume $V_{\rm c}(t) =  \pi \tilde{r}_{\rm c}^2 \, \beta_{\rm h} c t$, with $\beta_{\rm h}$ being the dimensionless velocity of the jet head, moving in the $r$-direction.
In the case of a wind-like medium (Eq. \ref{eq:wind}) $\beta_{\rm h}$ can be assumed to be constant, as the mass per unit-length that the jet encounters, $dM/dr \sim \rho(r) \, r^2$ is constant and if the jet-luminosity, $L_{\rm j}$ is also assumed to be constant, then $\beta_{\rm h}$ will be as well \citep[see also, ][]{Bromberg11, Sullivan24}. Therefore, 
\begin{equation}
p_{\rm c} \sim \frac{L_{\rm j}}{3\pi \tilde{r}_{\rm c}^2\beta_{\rm h}c}
\label{eq:Pc}
\end{equation}
Combining equations (\ref{eq:drdt}) and (\ref{eq:Pc}) then gives that 
\begin{equation}
\frac{d\tilde{r}_{\rm c}}{dt} = \left({\frac{L_{\rm j}}{3\pi \beta_{\rm h}c A}} \right)^{1/2}
\label{eq:drdt2}
\end{equation}
is a constant and therefore has a solution for the radius of the cocoon
\begin{equation}
\tilde{r}_{\rm c} (t) = \left({\frac{L_{\rm j}}{3\pi \beta_{\rm h}c A}}\right)^{1/2} \, t
\label{eq:rct}
\end{equation}
which scales linearly with time, $t$. Furthermore, using this scaling in equation (\ref{eq:Pc}) gives that the cocoon pressure scales as 
\begin{equation}
p_{\rm c} (t) \sim A t^{-2}
\label{eq:Pc2}
\end{equation}

With the scaling of $\tilde{r}_{\rm c}(t)$ and $p_{\rm c}(t)$ we can now find the scaling of the position of the recollimation shock, $r_{\rm rc}(t)$ by requiring that the collimation is dominated by oblique-shock deflection. As the jet moves through the cocoon, it is collimated by the cocoon pressure, $p_{\rm c}$. The upstream transverse ram pressure of the jet is, to order unity, 
\begin{equation}
p_{\perp, {\rm j}}(r) \sim \frac{L_{\rm j}}{\pi c r^2}   = \frac{L_{\rm iso}}{4\pi\epsilon_{\rm j} c r^2} \, \theta_{\rm j}^2 
\label{eq:Pperp}
\end{equation}
\citep[see, e.g.,][]{KomissarovFalle97, Bromberg07, Salafia2020}. 
In the last step we 
used that the jet luminosity is related to the measured isotropic-equivalent luminosity {$L_{\rm j}=
(f_{\rm b}/\epsilon_{\rm j})L_{\rm iso}$, where
$f_{\rm b}=\Omega_{\rm j}/4\pi=(1-\cos\theta_{\rm j})/2\simeq
\theta_{\rm j}^2/4$, and $\epsilon_{\rm j}$ is the gamma-ray radiative
efficiency of the jet and $\theta_{\rm j}$ is the opening-angle}.

The collimation shock position, $r_{\rm cs}$, is found from $p_{\perp, {\rm j}}(r_{\rm cs}) \sim p_{\rm c}(t)$ \citep{NalewajkoSikora09}.
Therefore, equating equations (\ref{eq:Pc2}) and (\ref{eq:Pperp})  then gives that 
\begin{equation}
r_{\rm cs}(t) \sim \sqrt{\frac{L_{\rm iso} \theta_{\rm j}^2}{4\pi \epsilon_{\rm j} c A}} \, t \equiv k_{\rm lab}\, t,
\label{eq:zrc}
\end{equation}
where $t = t_{\rm lab}$ is the time in the lab frame, i.e., the rest frame of the central engine, and the mass-loading is  $\dot{M}/v_{\rm w} = 4 \pi A $. Taking the cosmological time-dilation
and time arrival effects into account, the time measured by the observer is given by
\begin{equation}
\frac{t_{\rm obs}}{(1+z)} = t_{\rm lab} -\frac{r_{\rm rc}}{c}.
\end{equation}
Denoting the proportionality constant in Equation (\ref{eq:zrc}) with $k_{\rm lab}$  
the  equation can be rewritten as
\begin{equation}
r_{\rm cs}(t_{\rm obs}) \sim \left(\frac{k_{\rm lab}}{({1+z})(1 -k_{\rm lab}/c)} \right) \, t_{\rm obs}  \equiv k \, t_{\rm obs}.
%r_{\rm cs}(t_{\rm obs}) \sim \left(\frac{L_{\rm iso}}{c \epsilon_{\rm j}} \right)^{1/2} \left(\frac{\dot{M}}{v_{\rm w}}\right)^{-1/2} \, \frac{\theta_{\rm j}}{1+z} \; t_{\rm obs}
\label{eq:zrc2}
\end{equation}
Since we associate the derived value of $r_0$ as the recollimation shock at $r_{\rm rc}$ (\S \ref{sec:linear}),  the Equation (\ref{eq:zrc2}) shows the linear temporal scaling that we sought, with the first factor being the observed $k$. This is the trend that is observed for the full duration of the thermal episode in  GRB 220426A, as shown in Figure \ref{fig:r0}.  We note that for the measured value of $k$ and the assumed $z=1$ the light-travel-time compression is exactly compensated by cosmological time-dilation and $k=k_{\rm lab}$.

The numerical coefficient in Equation~(\ref{eq:zrc}) is uncertain at the factor-of-few level. This uncertainty is due to the dependence on the cocoon geometry, the density averaging, $\beta_{\rm h}$, the fraction of jet power transferred to the cocoon, and the obliquity of the reconfinement shock \citep[see, e.g.,][]{NalewajkoSikora09, Harrison18, Hamidani2021}.

\section{Comparison with Band parameters}
\label{sec:AppBand}

\begin{figure*}
    \centering
        \includegraphics[width = 0.7\columnwidth]{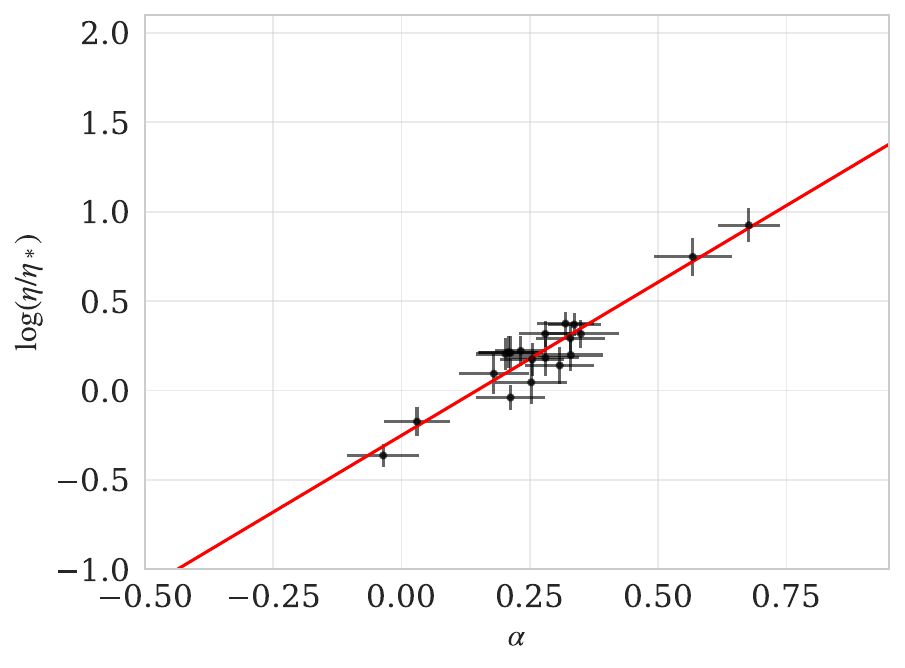}
         \caption{The spectral width $\eta/\eta_*$ as a function of the Band parameter $\alpha$, the photon index of the low-energy power-law of the Band function. The red line corresponds to the best fit which gives $\eta/\eta* = 0.56 * \exp(\alpha/0.25)$.}
    \label{fig:alpha}
\end{figure*}

The parameter $\eta/\eta_*$ of the photopshere model is directly linked to the width of the spectrum.
However, the low-energy power-law of the Band function, $\alpha$, can also be used to characterise the width of the spectrum\footnote{To do this, an implicit assumption is made that the high-energy power-law $\beta$ is related, such that larger $\alpha$ is accompanied by a smaller $\beta$ \citep{Yu2016}}.  In Table \ref{Tab:App} and in Figure \ref{fig:alpha}, we show the two parameters $\eta/\eta_*$ versus $\alpha$ for the time-resolved spectra in our analysis.  From the figure it is apparent that there is indeed a strong correlation between these parameters, in which larger values of $\alpha$ correspond to larger values of $\eta/\eta_*$. The best  fit to the correlation is given by an exponential function with 
\begin{equation}
\frac{\eta}{\eta_*} (\alpha) = 0.56 \; e^{\frac{\alpha}{0.25}}
\end{equation}
The linear relation shown in Figure \ref{fig:alpha} has a strong positive correlation  with Pearson correlation coefficient $r=0.95$ ($p \sim 1.0\times10^{-10}$).

\begin{deluxetable*}{ccc}
\tabletypesize{\scriptsize}
\tablecaption{Time-resolved values of the width parameter, $\eta/\eta_*$, and the low-energy index, $\alpha_{\rm Band}$ of the Band-function.\label{tab:eta_alpha}}
\tablehead{
\colhead{$t_{\rm mean}$} &
\colhead{$\eta/\eta_*$} &
\colhead{$\alpha_{\rm Band}$} \\
\colhead{s} &
\colhead{} &
\colhead{}
}
\startdata
0.6 & $8.4\pm1.9$ & $0.68\pm0.06$ \\
1.4 & $5.6\pm1.4$ & $0.57\pm0.08$ \\
1.8 & $2.3\pm0.4$ & $0.34\pm0.05$ \\
2.1 & $1.6\pm0.3$ & $0.33\pm0.06$ \\
2.3 & $1.4\pm0.3$ & $0.31\pm0.07$ \\
2.5 & $1.7\pm0.3$ & $0.23\pm0.05$ \\
2.7 & $2.1\pm0.4$ & $0.28\pm0.05$ \\
3.0 & $1.5\pm0.3$ & $0.25\pm0.06$ \\
3.3 & $1.1\pm0.3$ & $0.25\pm0.07$ \\
3.6 & $2.4\pm0.4$ & $0.32\pm0.06$ \\
3.8 & $2.0\pm0.3$ & $0.33\pm0.07$ \\
4.1 & $0.9\pm0.1$ & $0.21\pm0.07$ \\
4.2 & $1.5\pm0.3$ & $0.28\pm0.07$ \\
4.4 & $1.6\pm0.4$ & $0.21\pm0.06$ \\
4.5 & $2.1\pm0.4$ & $0.35\pm0.08$ \\
4.6 & $1.6\pm0.4$ & $0.21\pm0.06$ \\
4.8 & $1.2\pm0.3$ & $0.18\pm0.07$ \\
5.0 & $1.6\pm0.3$ & $0.20\pm0.06$ \\
5.4 & $0.7\pm0.1$ & $0.03\pm0.06$ \\
6.1 & $0.4\pm0.1$ & $-0.04\pm0.07$ \\
\enddata
\tablecomments{Quoted uncertainties are $1\sigma$.}
\label{Tab:App}
\end{deluxetable*}

\section{Time-resolved count spectra of GRB~220426A observed by {\it Fermi}}
\label{AppCount}

\begin{figure*}[p]
    \centering
    
    \gridline{
        \fig{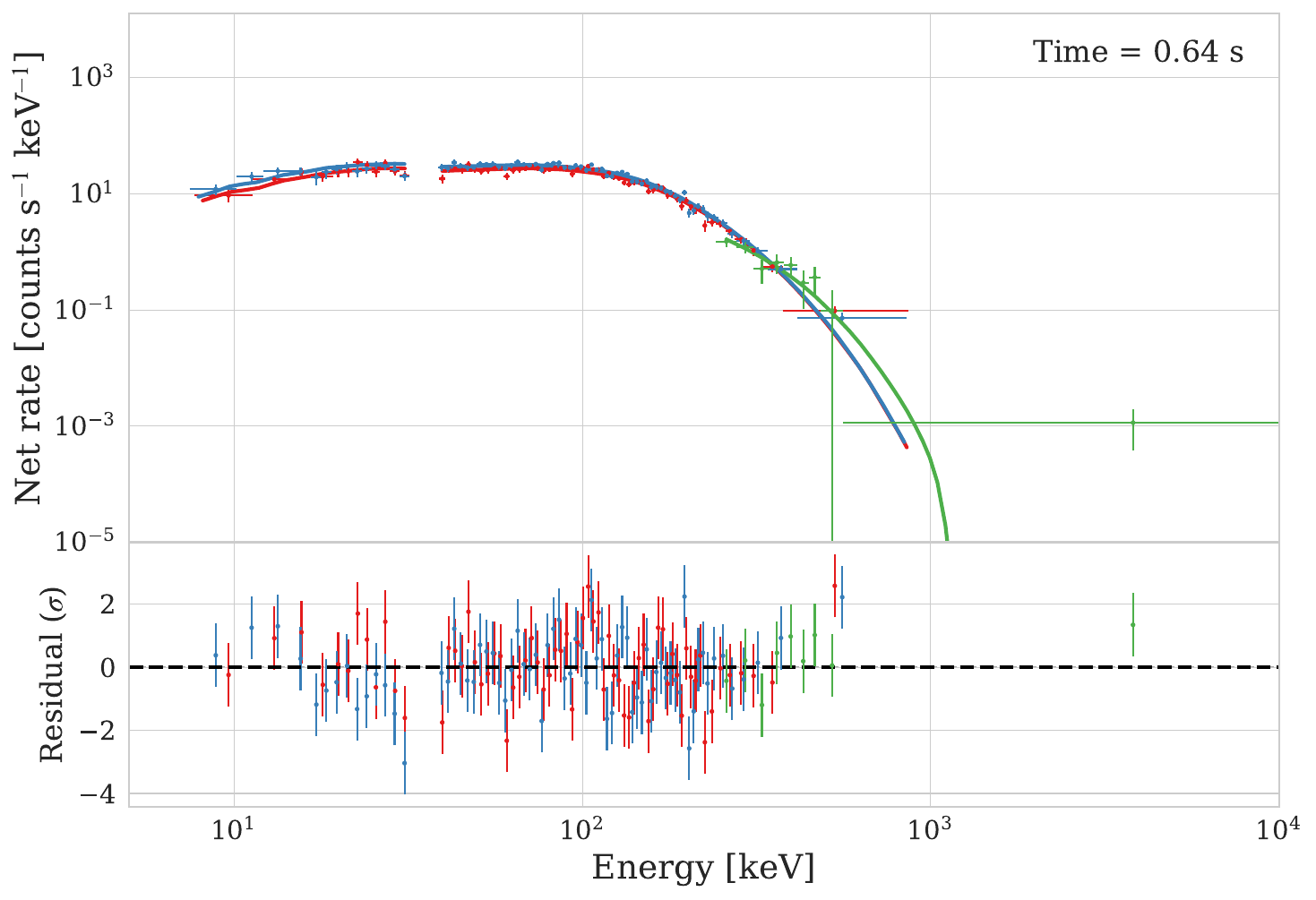}{0.31\textwidth}{(a)}
        \fig{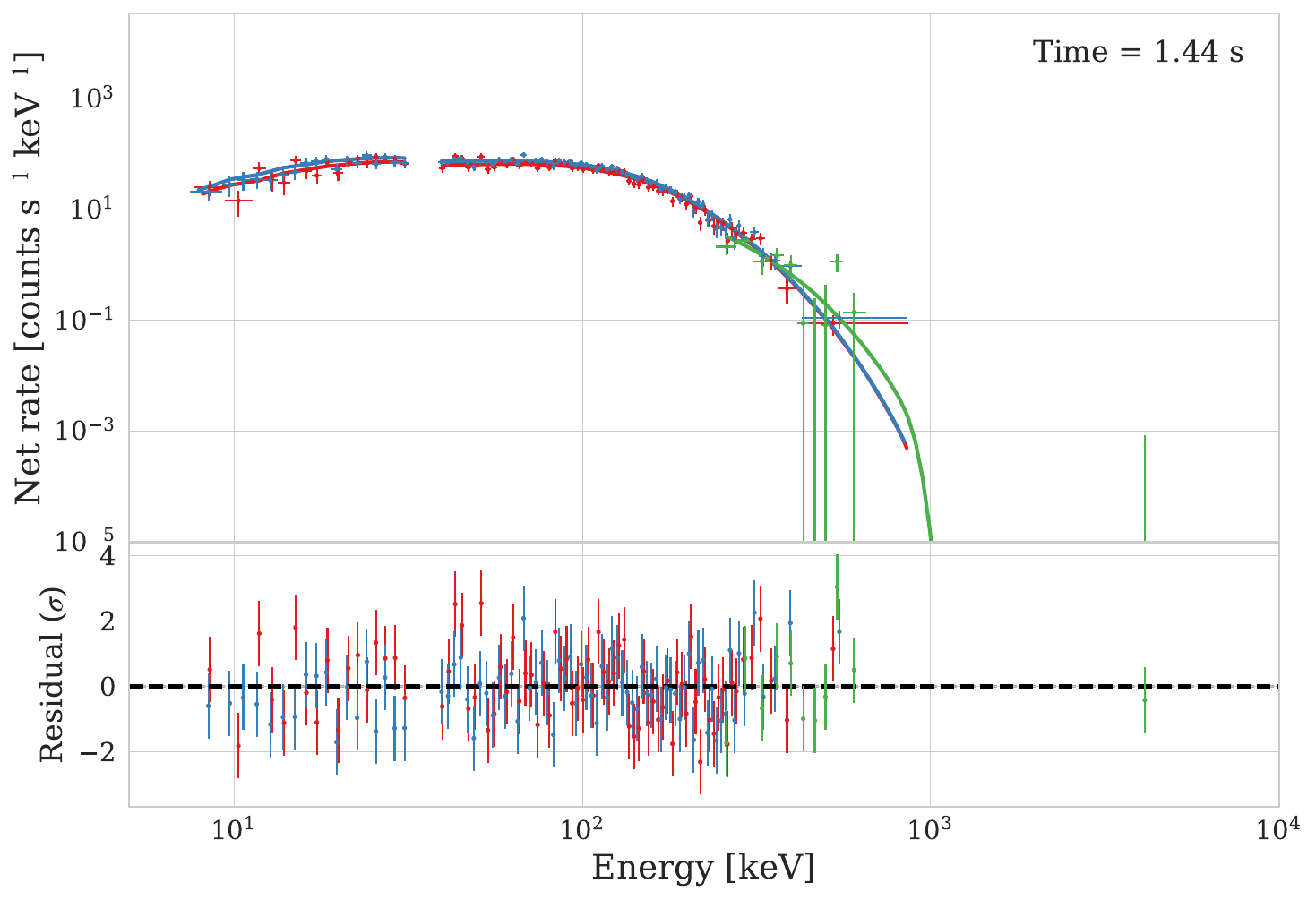}{0.31\textwidth}{(b)}
        \fig{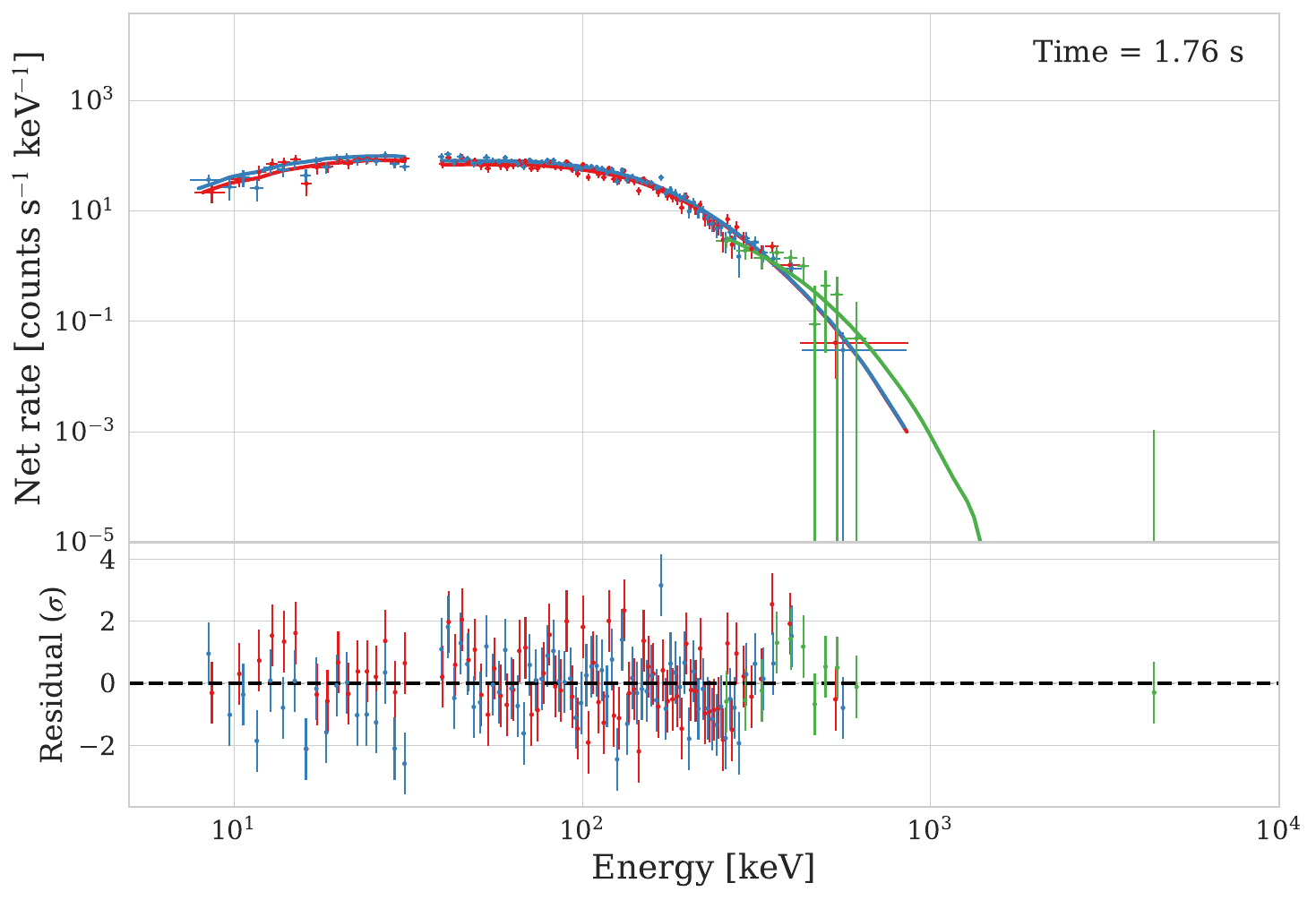}{0.31\textwidth}{(c)}
    }

    \gridline{
        \fig{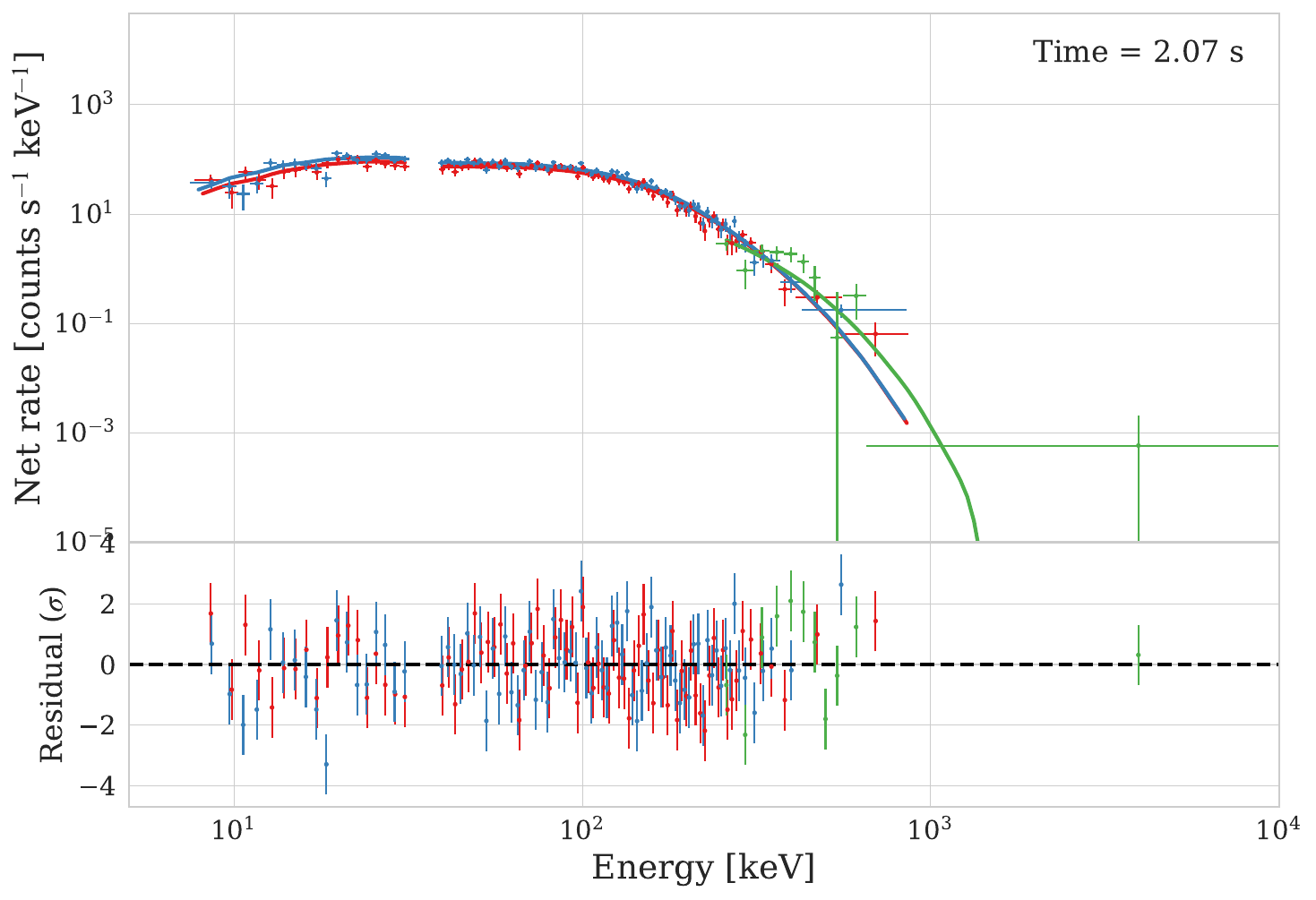}{0.31\textwidth}{(d)}
        \fig{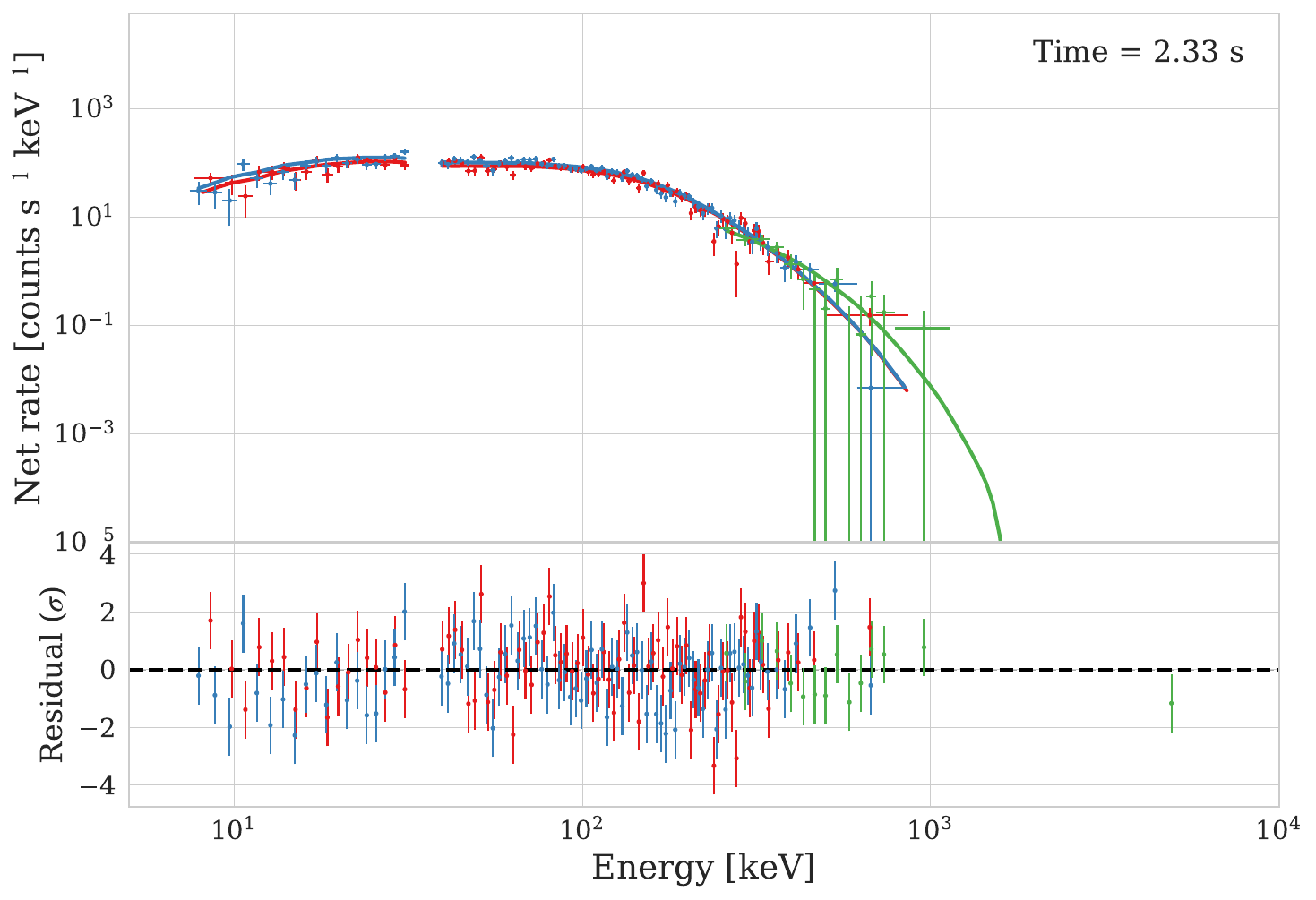}{0.31\textwidth}{(e)}
        \fig{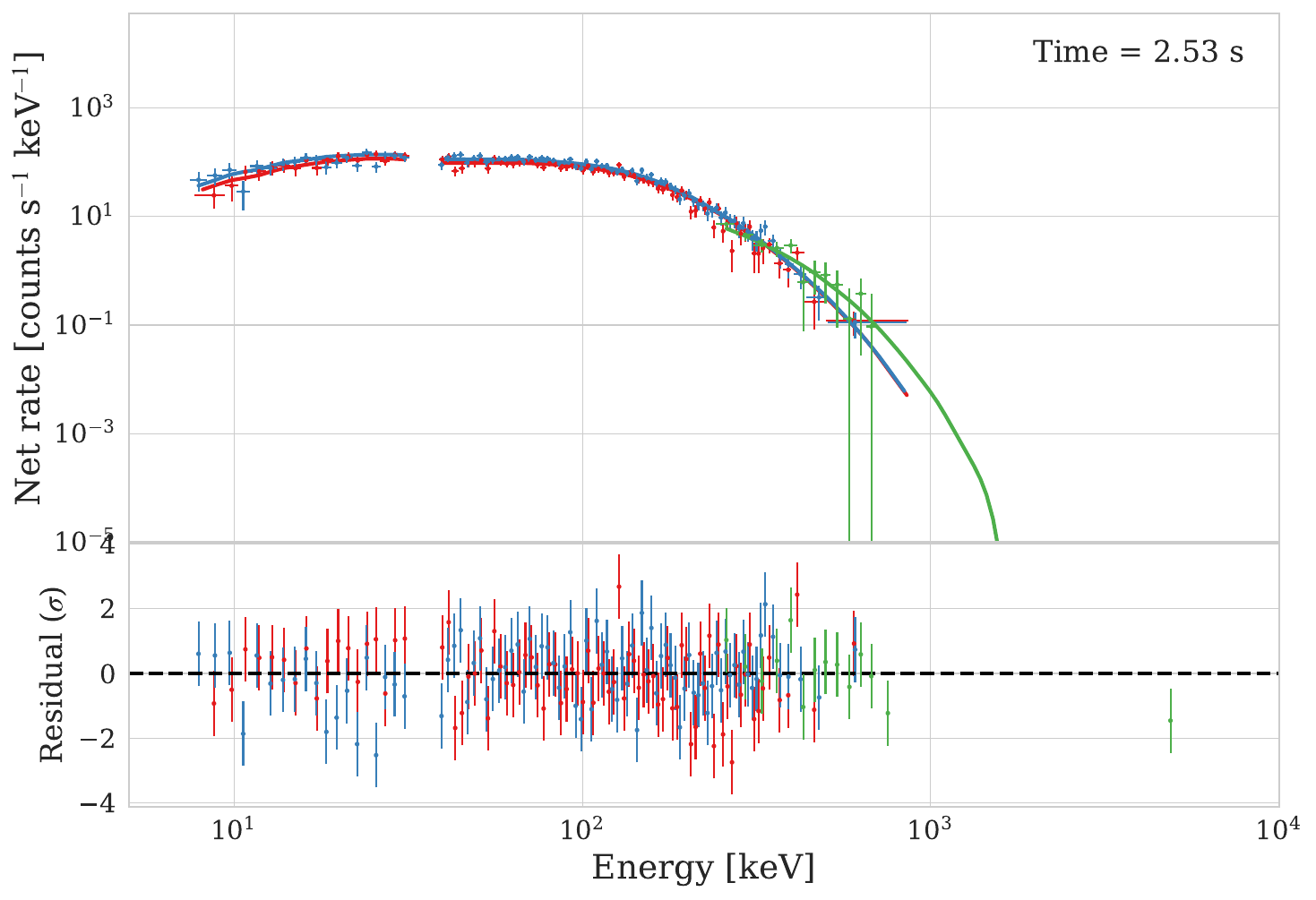}{0.31\textwidth}{(f)}
    }

    \gridline{
        \fig{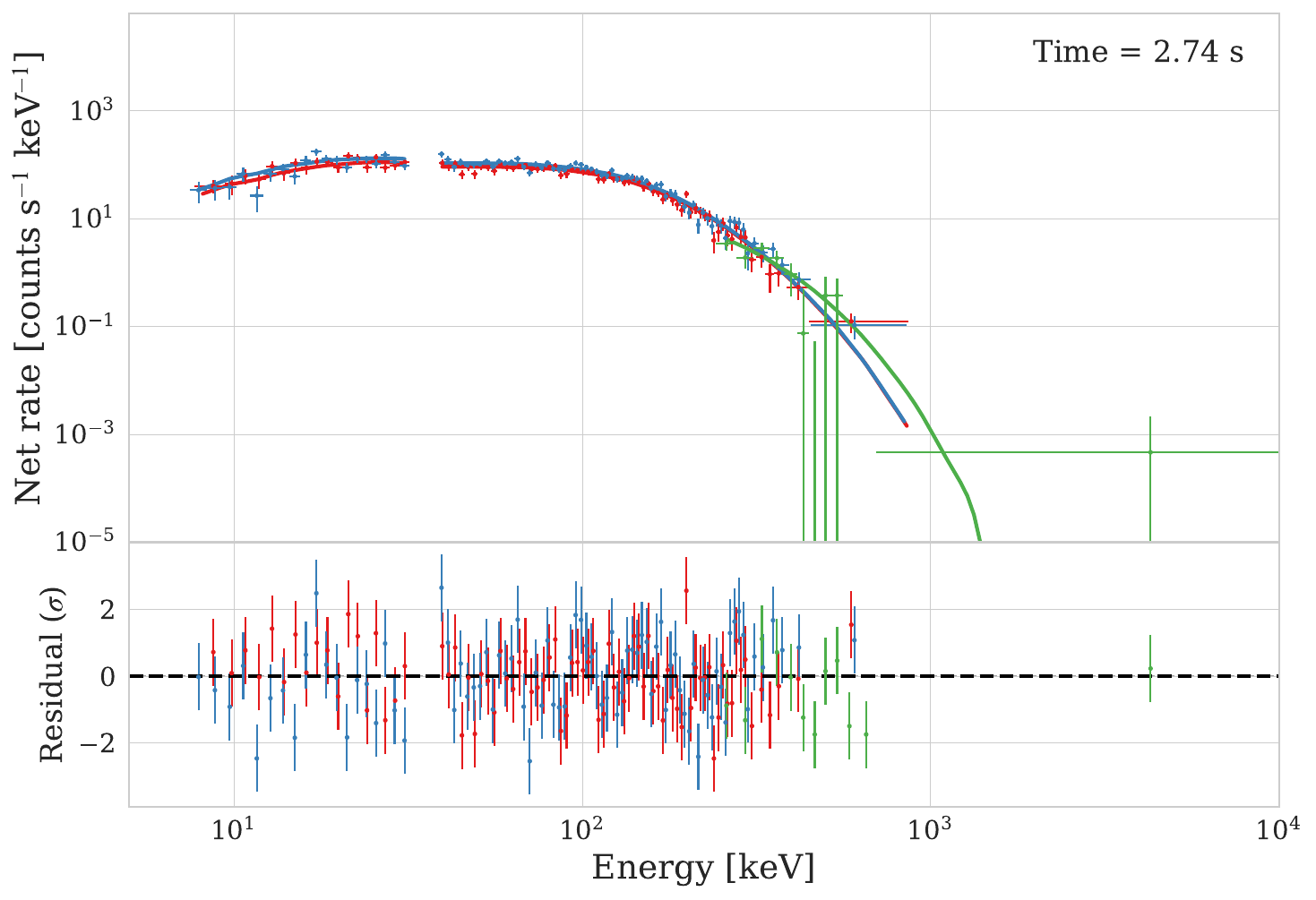}{0.31\textwidth}{(g)}
        \fig{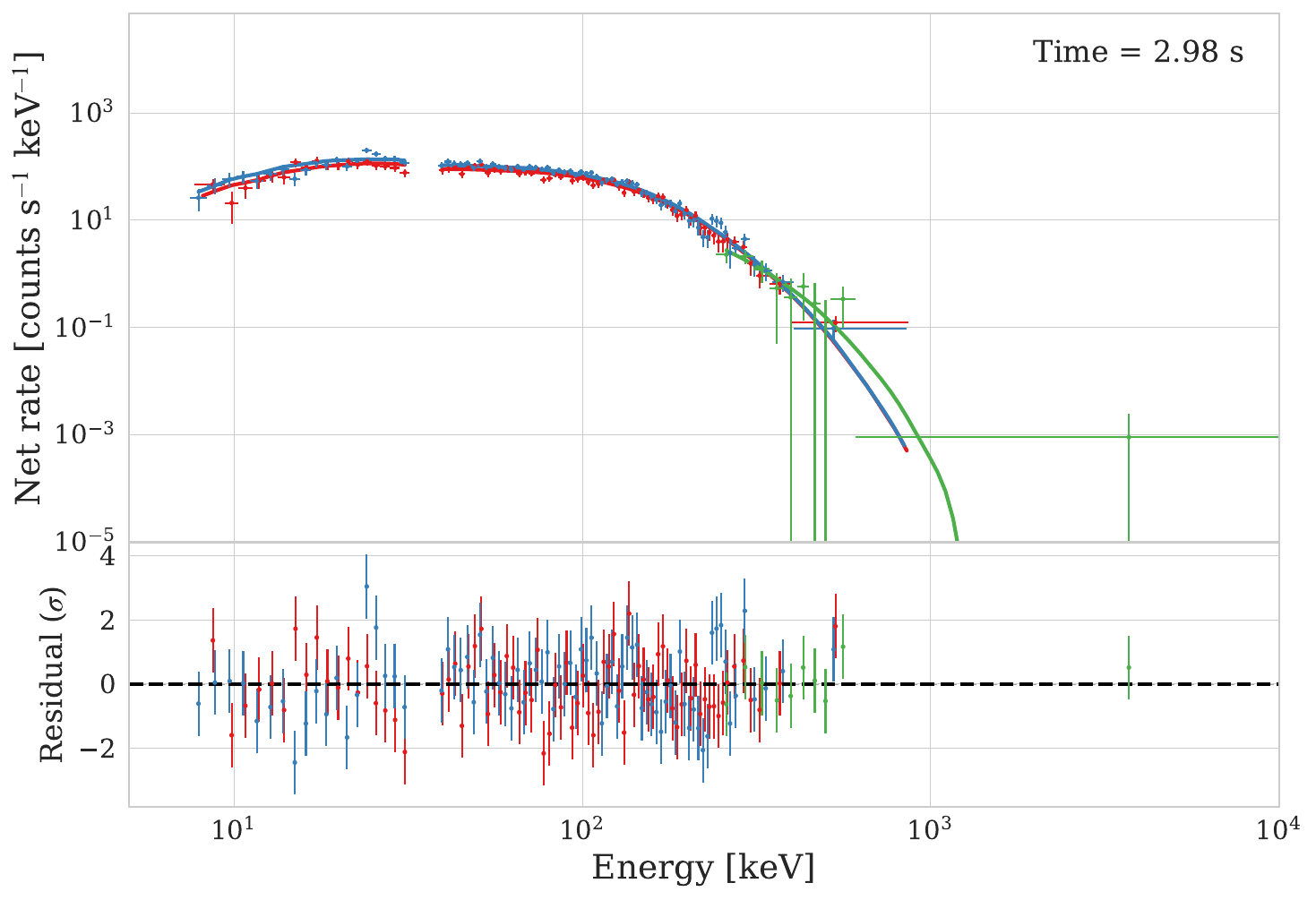}{0.31\textwidth}{(h)}
        \fig{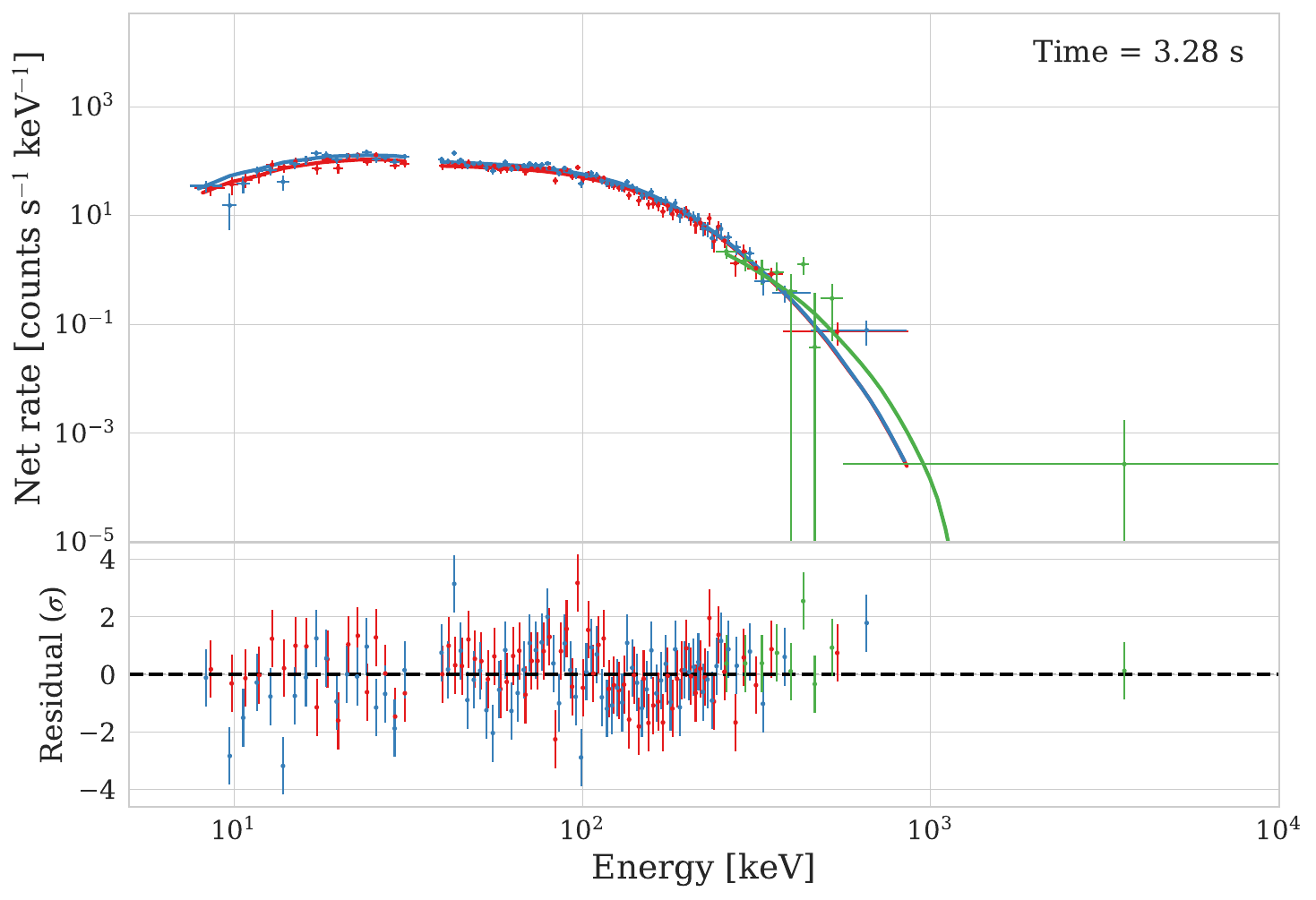}{0.31\textwidth}{(i)}
    }

    \gridline{
        \fig{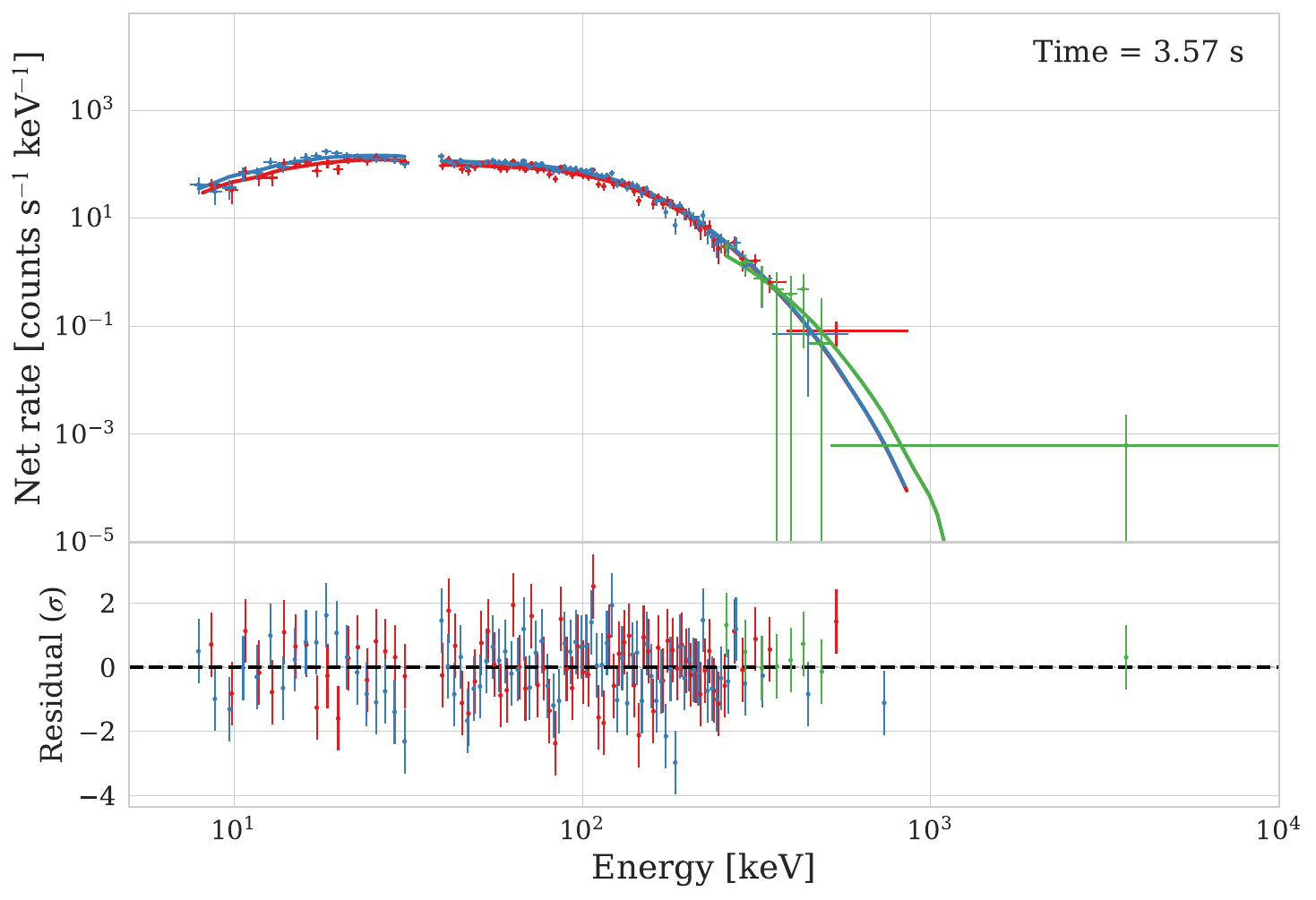}{0.31\textwidth}{(j)}
    }

    \caption{
        Time-resolved count spectra of GRB~220426A and their
        best-fitting spectral models. The upper panel of each
        subfigure shows the background-subtracted net count rate,
        while the lower panel shows the residual significance in
        units of $\sigma$. Panels are ordered chronologically.
        The figure continues on the following page.
    }
    \label{fig:counts_220426285}
\end{figure*}

\clearpage

\begin{figure*}[p]
    \centering

    % Use the same figure number as the preceding page.
    \figurenum{\ref{fig:counts_220426285}}

    \gridline{
        \fig{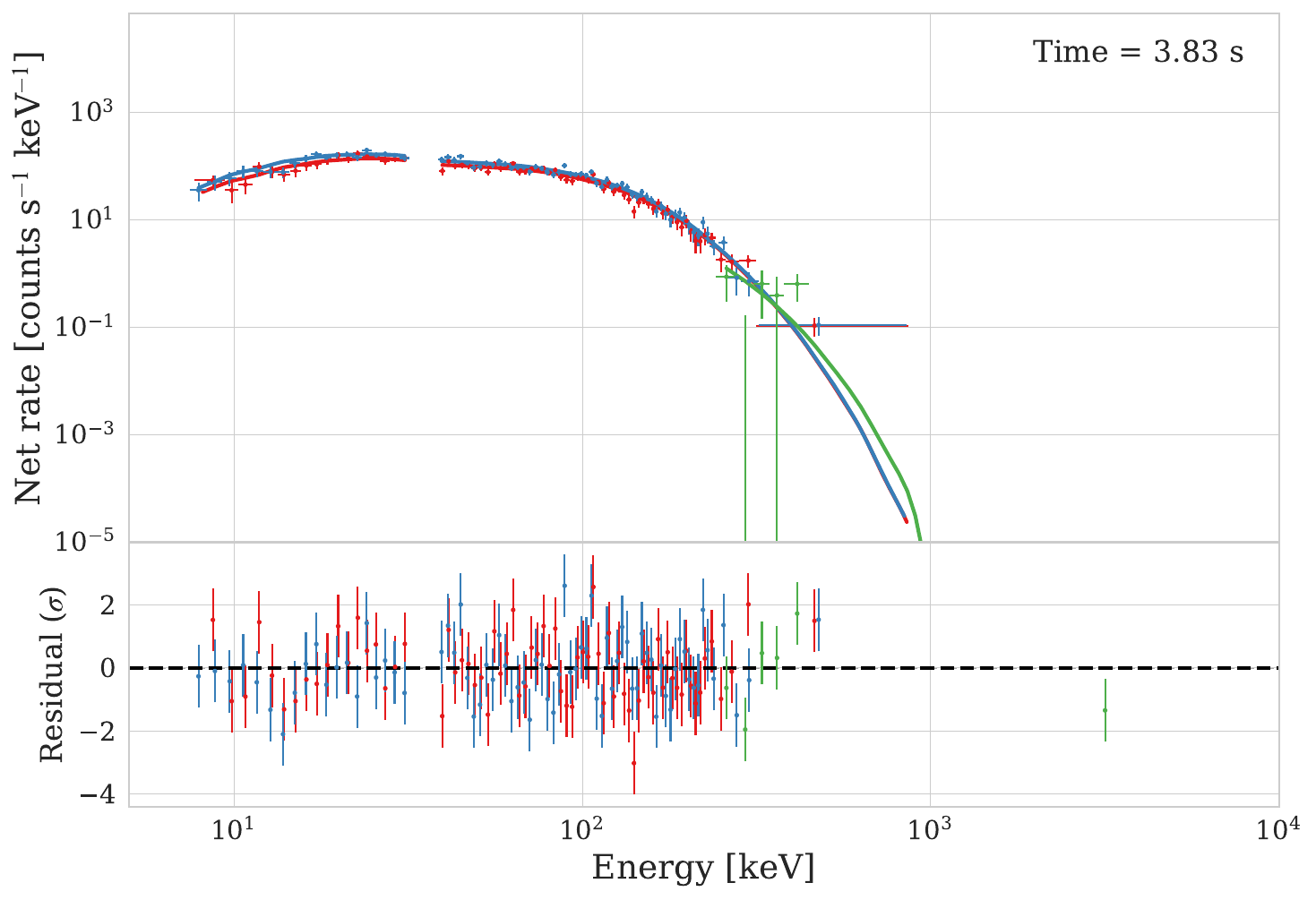}{0.31\textwidth}{(k)}
        \fig{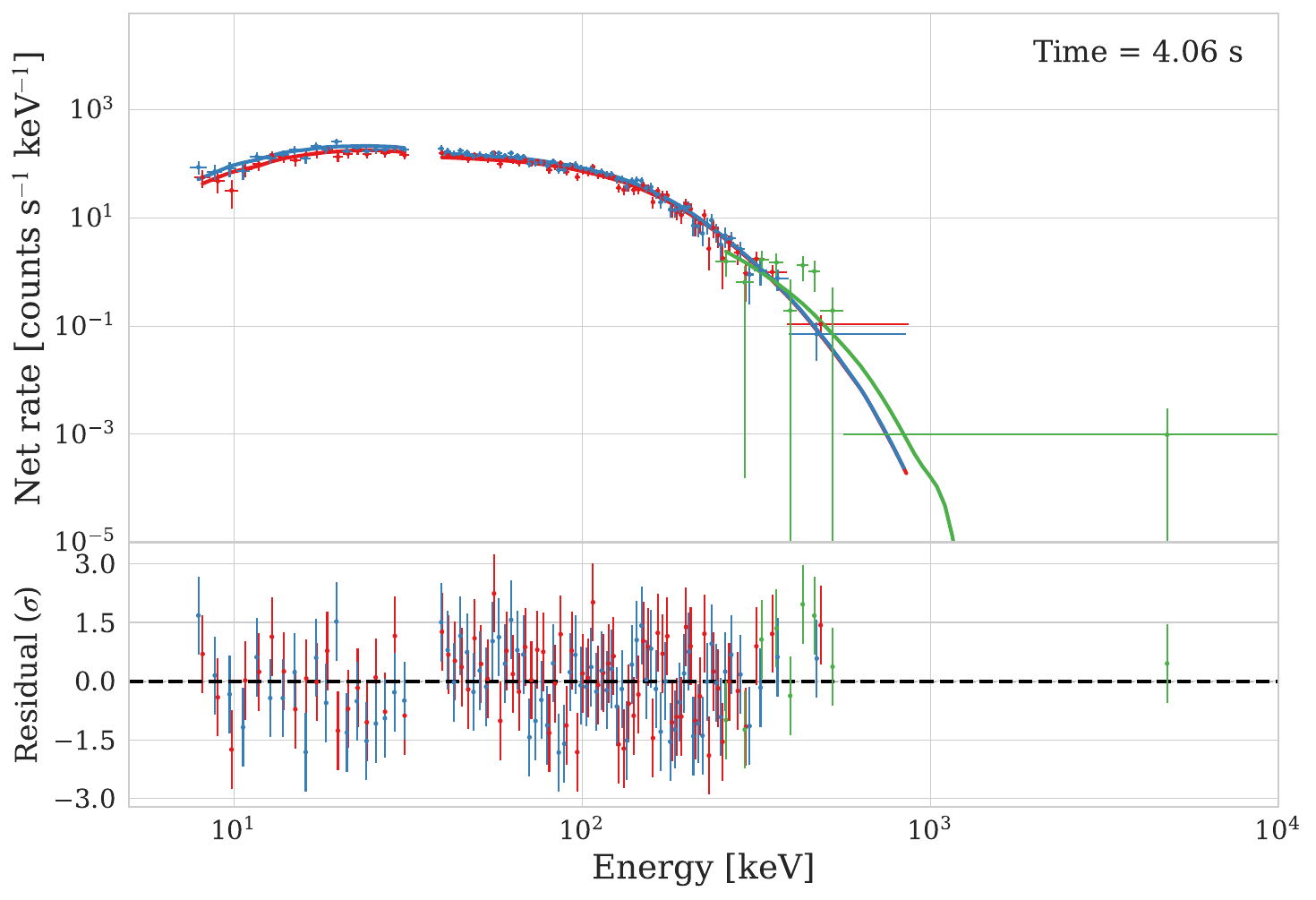}{0.31\textwidth}{(l)}
        \fig{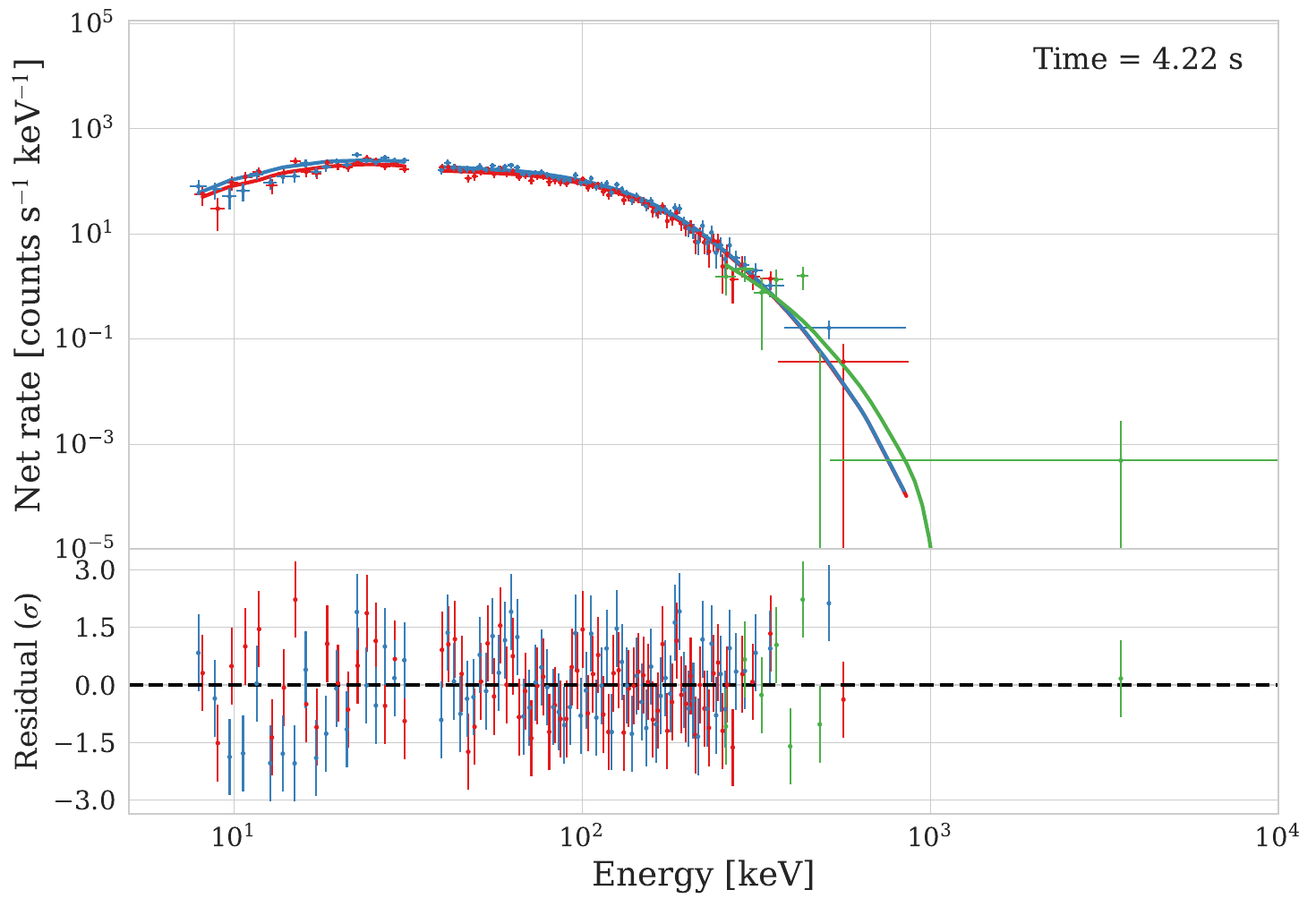}{0.31\textwidth}{(m)}
    }

    \gridline{
        \fig{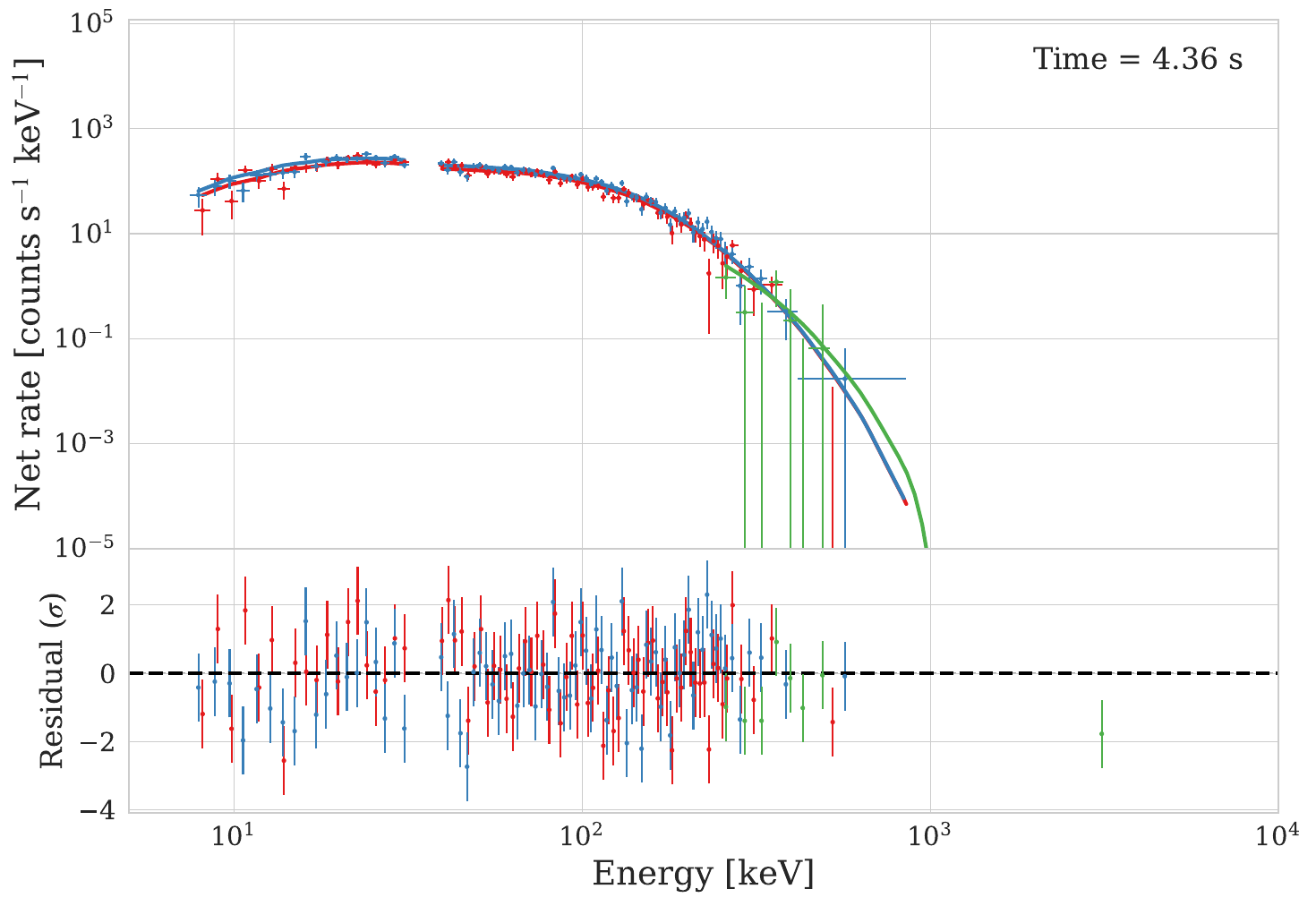}{0.31\textwidth}{(n)}
        \fig{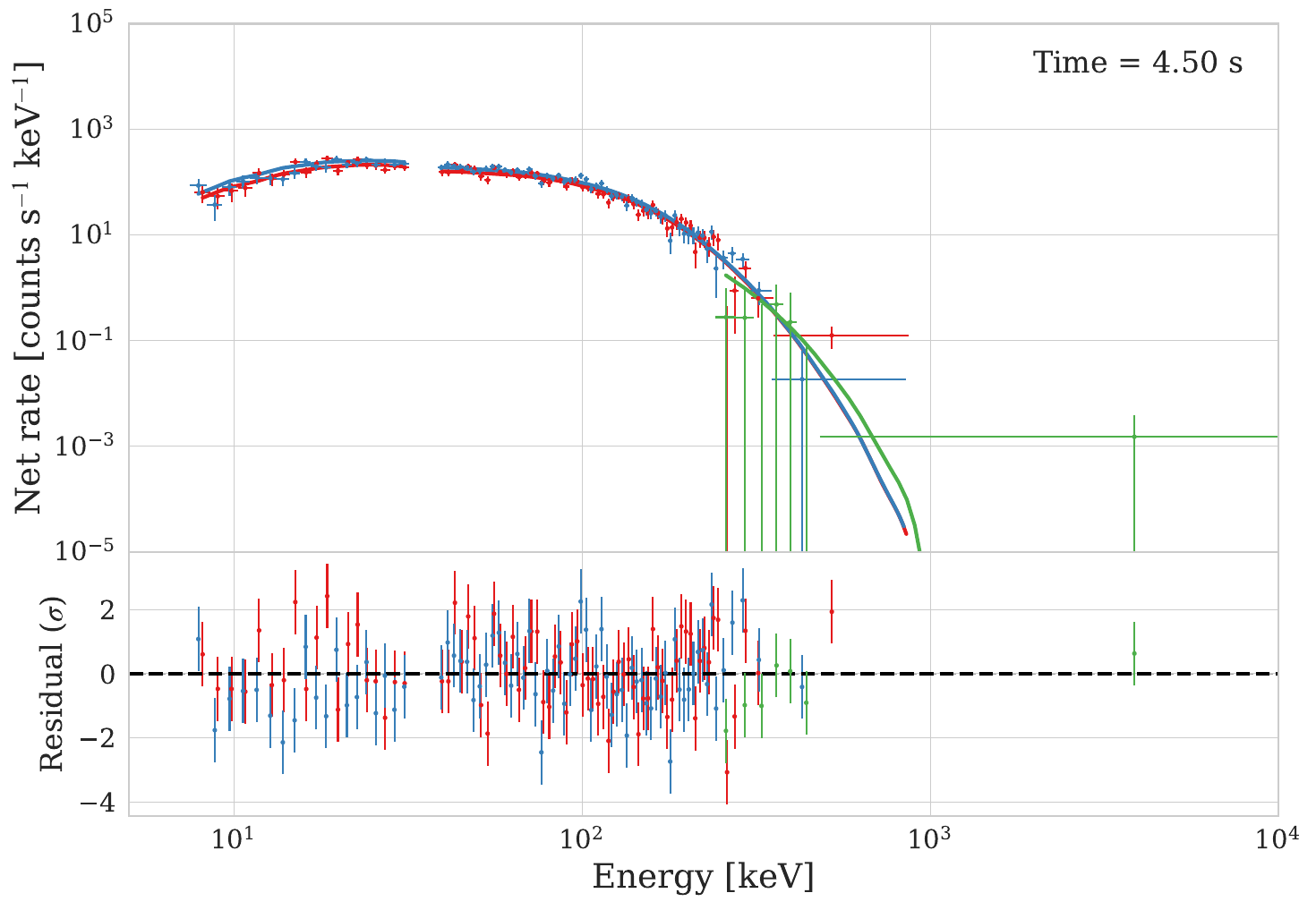}{0.31\textwidth}{(o)}
        \fig{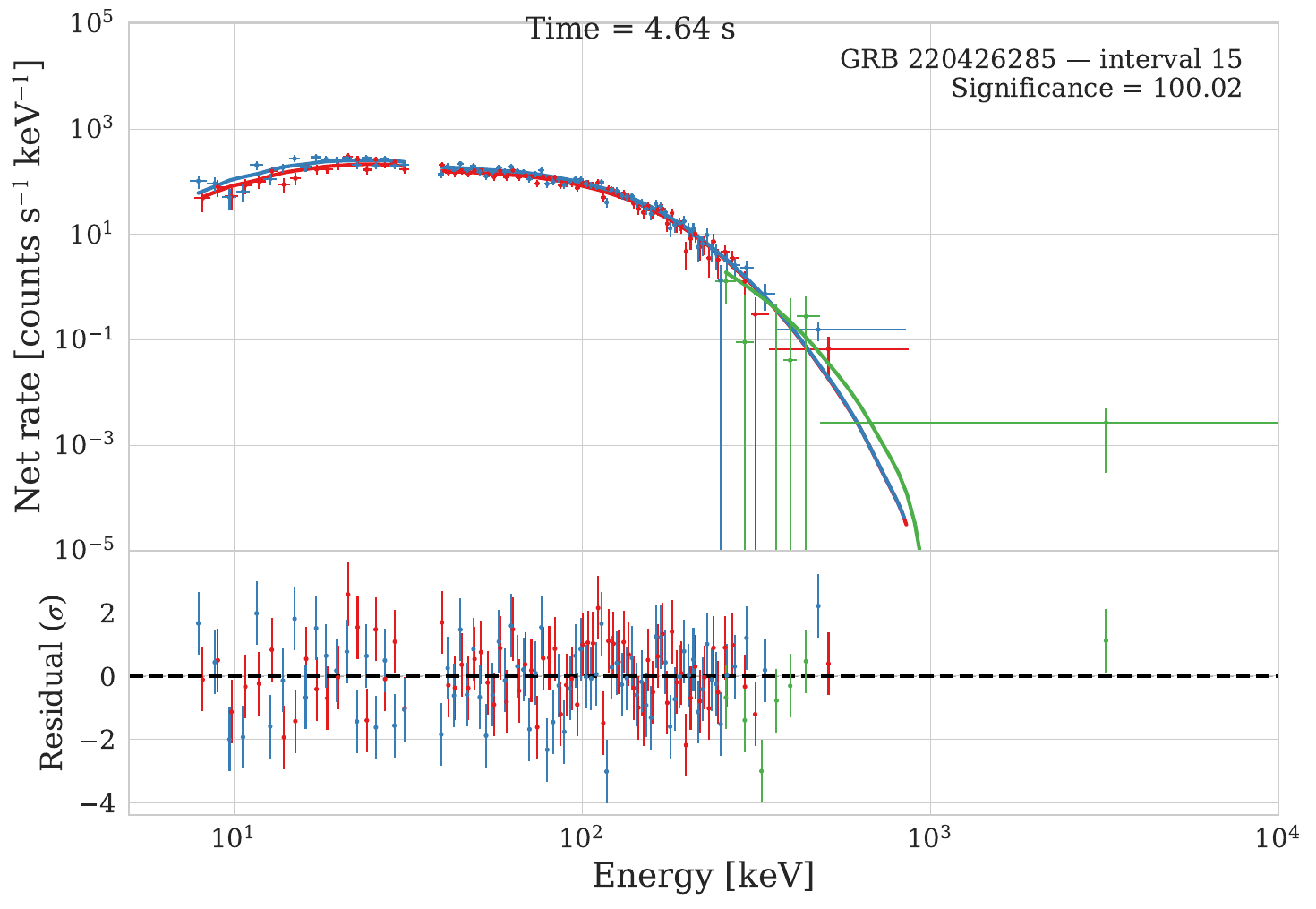}{0.31\textwidth}{(p)}
    }

    \gridline{
        \fig{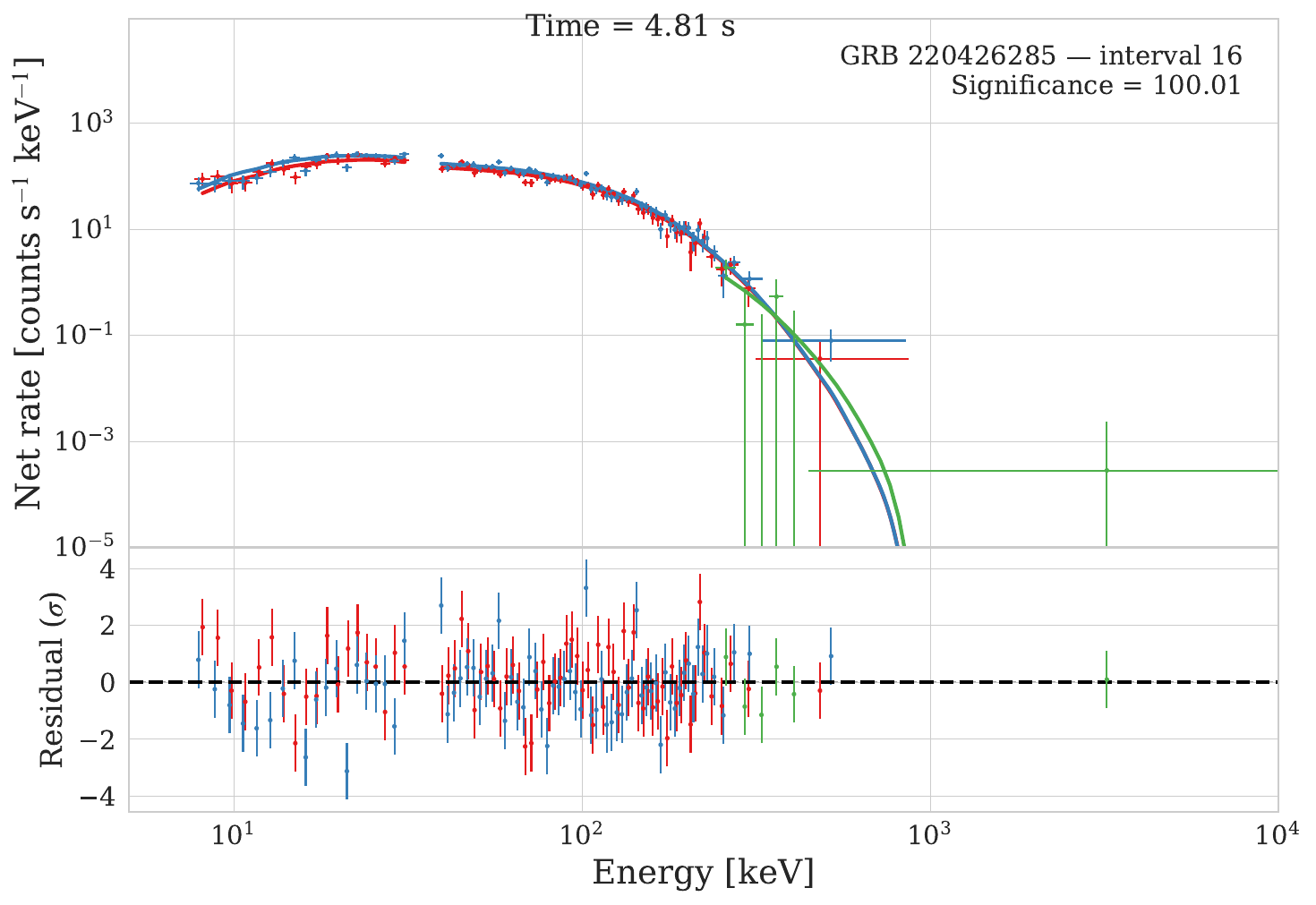}{0.31\textwidth}{(q)}
        \fig{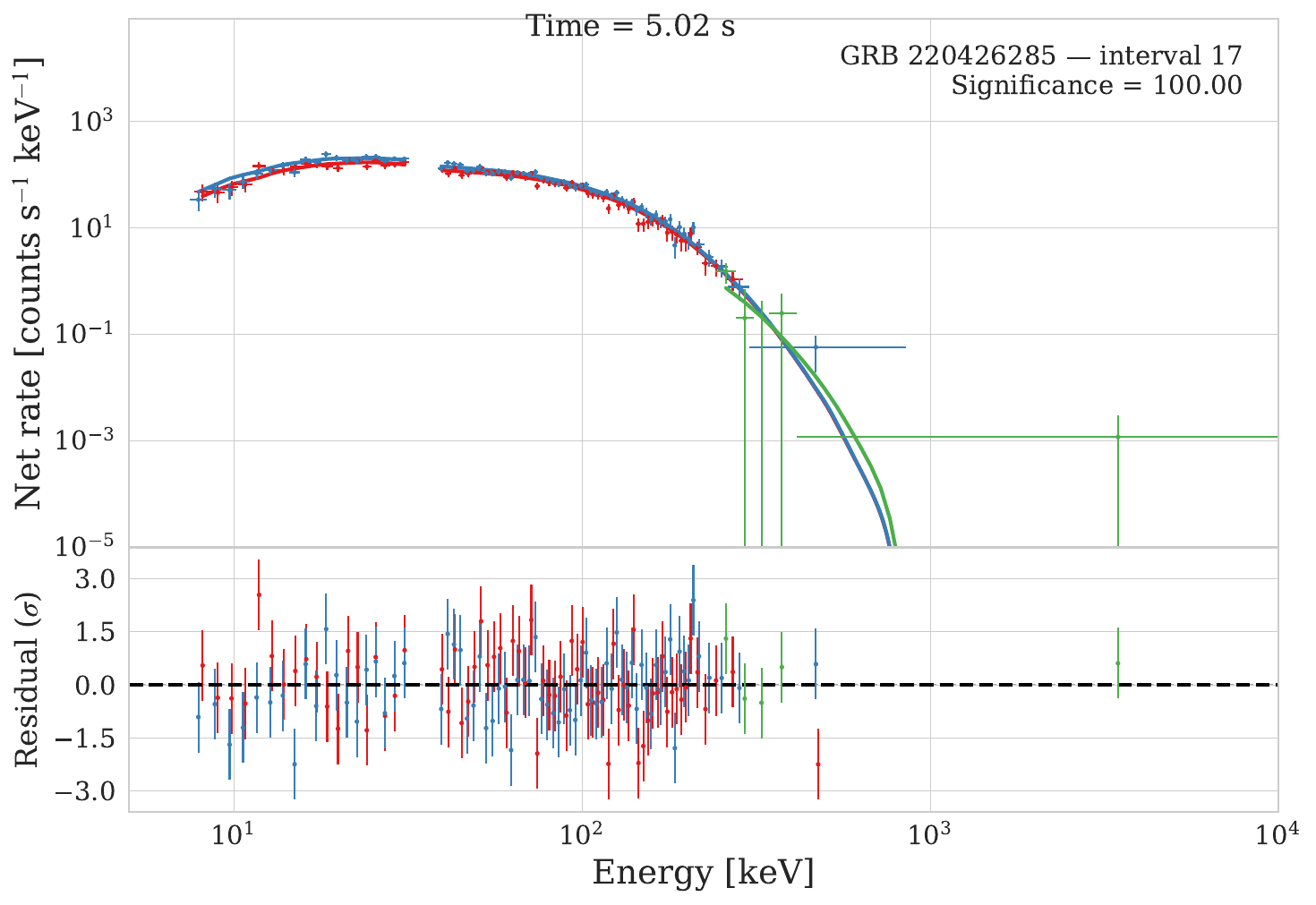}{0.31\textwidth}{(r)}
        \fig{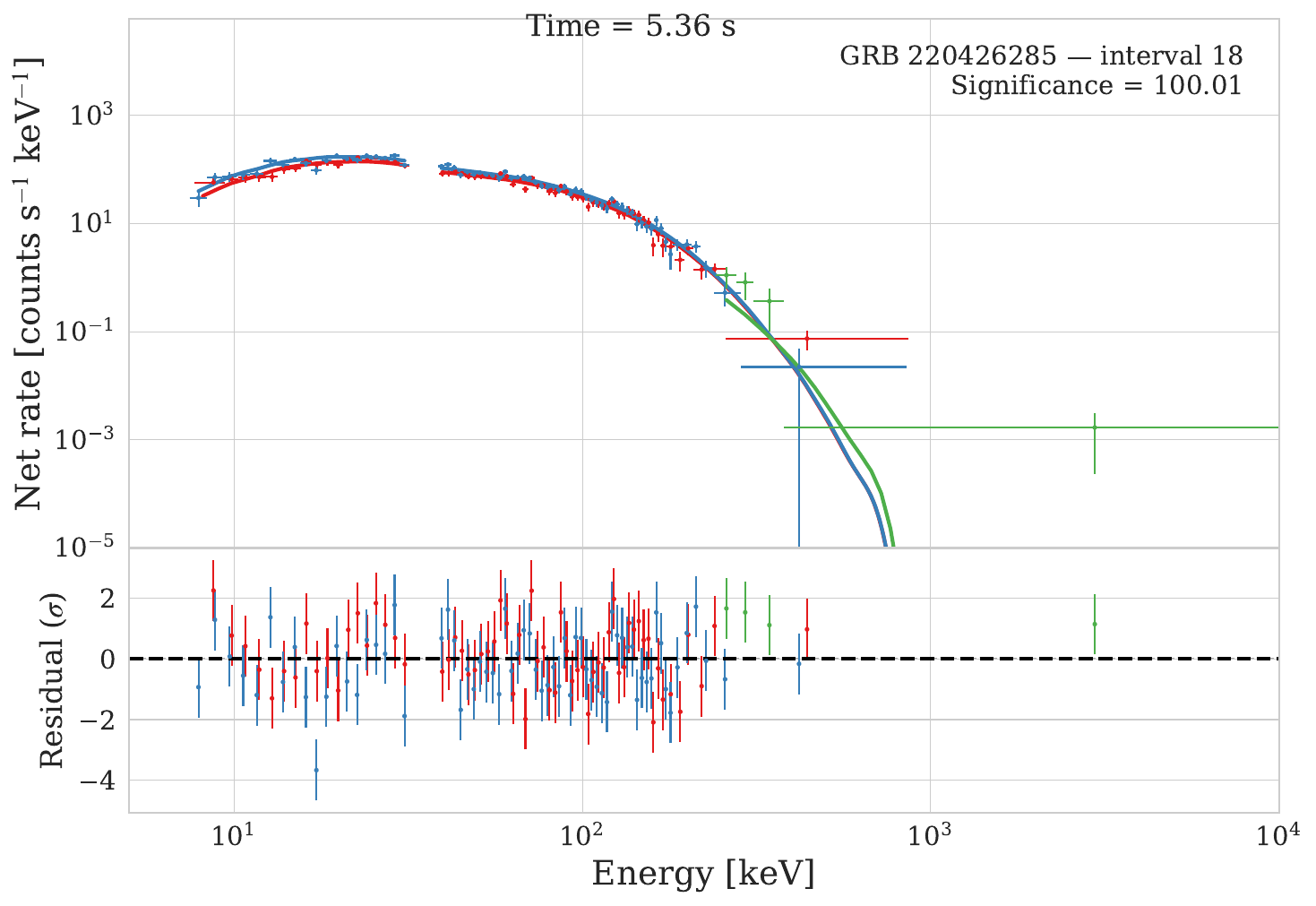}{0.31\textwidth}{(s)}
    }

    \gridline{
        \fig{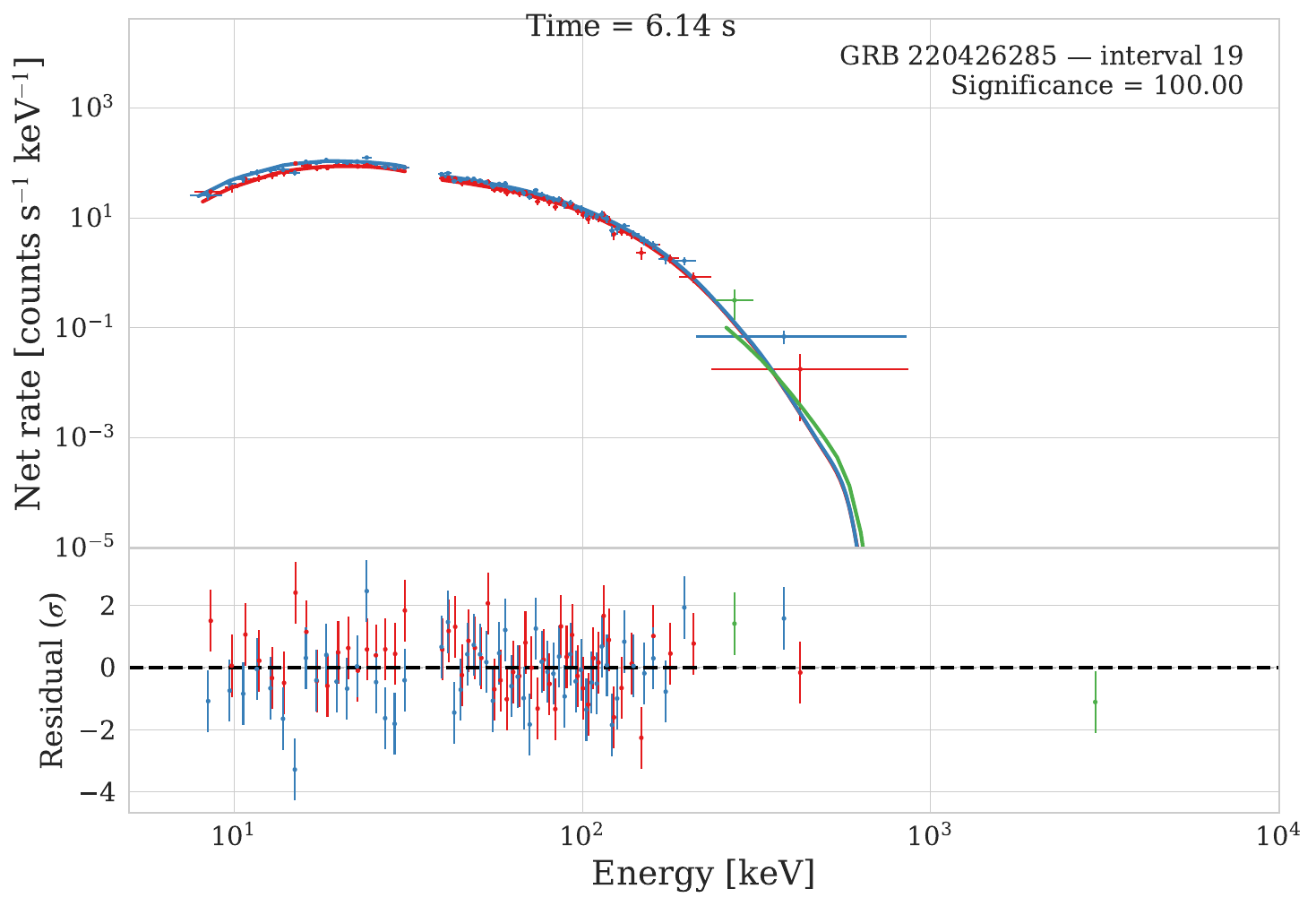}{0.31\textwidth}{(t)}
    }

    \caption{
        Time-resolved count spectra of GRB~220426A
        (Figure~\ref{fig:counts_220426285}, continued).
    }
\end{figure*}

%\bibliography{references.bib}
\bibliography{ref2020.bib}

@ARTICLE{Drenkhahn02,
   author = {{Drenkhahn}, G.},
    title = "{Acceleration of GRB outflows by Poynting flux dissipation}",
  journal = {\aap},
   eprint = {arXiv:astro-ph/0112509},
     year = 2002,
    month = may,
   volume = 387,
    pages = {714-724},
      doi = {10.1051/0004-6361:20020390},
   adsurl = {http://adsabs.harvard.edu/abs/2002A%26A...387..714D}
}

@ARTICLE{Peer17,
   author = {{Pe'er}, A.},
    title = "{Constraining Magnetization of Gamma-Ray Bursts Outflows Using Prompt Emission Fluence}",
  journal = {\apj},
archivePrefix = "arXiv",
   eprint = {1604.06590},
 primaryClass = "astro-ph.HE",
     year = 2017,
    month = dec,
   volume = 850,
      eid = {200},
    pages = {200},
      doi = {10.3847/1538-4357/aa974e},
   adsurl = {http://adsabs.harvard.edu/abs/2017ApJ...850..200P}
}

@ARTICLE{Gehrels+06,
       author = {{Gehrels}, N. and {Norris}, J.~P. and {Barthelmy}, S.~D. and {Granot}, J. and {Kaneko}, Y. and {Kouveliotou}, C. and {Markwardt}, C.~B. and {M{\'e}sz{\'a}ros}, P. and {Nakar}, E. and {Nousek}, J.~A. and {O'Brien}, P.~T. and {Page}, M. and {Palmer}, D.~M. and {Parsons}, A.~M. and {Roming}, P.~W.~A. and {Sakamoto}, T. and {Sarazin}, C.~L. and {Schady}, P. and {Stamatikos}, M. and {Woosley}, S.~E.},
        title = "{A new {\ensuremath{\gamma}}-ray burst classification scheme from GRB060614}",
      journal = {\nat},
         year = 2006,
        month = dec,
       volume = {444},
       number = {7122},
        pages = {1044-1046},
          doi = {10.1038/nature05376},
archivePrefix = {arXiv},
       eprint = {astro-ph/0610635},
 primaryClass = {astro-ph},
       adsurl = {https://ui.adsabs.harvard.edu/abs/2006Natur.444.1044G}
}

@ARTICLE{ZY11,
   author = {{Zhang}, B. and {Yan}, H.},
    title = "{The Internal-collision-induced Magnetic Reconnection and Turbulence (ICMART) Model of Gamma-ray Bursts}",
  journal = {\apj},
archivePrefix = "arXiv",
   eprint = {1011.1197},
 primaryClass = "astro-ph.HE",
     year = 2011,
    month = jan,
   volume = 726,
      eid = {90},
    pages = {90},
      doi = {10.1088/0004-637X/726/2/90},
   adsurl = {http://adsabs.harvard.edu/abs/2011ApJ...726...90Z}
}

@ARTICLE{Lyutikov06,
   author = {{Lyutikov}, M.},
    title = "{The electromagnetic model of gamma-ray bursts}",
  journal = {New Journal of Physics},
   eprint = {astro-ph/0512342},
     year = 2006,
    month = jul,
   volume = 8,
    pages = {119},
      doi = {10.1088/1367-2630/8/7/119},
   adsurl = {http://adsabs.harvard.edu/abs/2006NJPh....8..119L}
}

@ARTICLE{Giannios2008,
       author = {{Giannios}, D.},
        title = "{Prompt GRB emission from gradual energy dissipation}",
      journal = {\aap},
         year = 2008,
        month = mar,
       volume = {480},
       number = {2},
        pages = {305-312},
          doi = {10.1051/0004-6361:20079085},
archivePrefix = {arXiv},
       eprint = {0711.2632},
 primaryClass = {astro-ph},
       adsurl = {https://ui.adsabs.harvard.edu/abs/2008A&A...480..305G}
}

@ARTICLE{Campana2006,
       author = {{Campana}, S. and {Mangano}, V. and {Blustin}, A.~J. and {Brown}, P. and {Burrows}, D.~N. and {Chincarini}, G. and {Cummings}, J.~R. and {Cusumano}, G. and {Della Valle}, M. and {Malesani}, D. and {M{\'e}sz{\'a}ros}, P. and {Nousek}, J.~A. and {Page}, M. and {Sakamoto}, T. and {Waxman}, E. and {Zhang}, B. and {Dai}, Z.~G. and {Gehrels}, N. and {Immler}, S. and {Marshall}, F.~E. and {Mason}, K.~O. and {Moretti}, A. and {O'Brien}, P.~T. and {Osborne}, J.~P. and {Page}, K.~L. and {Romano}, P. and {Roming}, P.~W.~A. and {Tagliaferri}, G. and {Cominsky}, L.~R. and {Giommi}, P. and {Godet}, O. and {Kennea}, J.~A. and {Krimm}, H. and {Angelini}, L. and {Barthelmy}, S.~D. and {Boyd}, P.~T. and {Palmer}, D.~M. and {Wells}, A.~A. and {White}, N.~E.},
        title = "{The association of GRB 060218 with a supernova and the evolution of the shock wave}",
      journal = {\nat},
         year = 2006,
        month = aug,
       volume = {442},
       number = {7106},
        pages = {1008-1010},
          doi = {10.1038/nature04892},
archivePrefix = {arXiv},
       eprint = {astro-ph/0603279},
 primaryClass = {astro-ph},
       adsurl = {https://ui.adsabs.harvard.edu/abs/2006Natur.442.1008C}
}

@ARTICLE{Srinivasaragavan2026,
       author = {{Srinivasaragavan}, Gokul P. and {Li}, Dongyue and {Hall}, Xander J. and {Gottlieb}, Ore and {Schroeder}, Genevieve and {Liu}, Heyang and {O'Connor}, Brendan and {Jin}, Chichuan and {Kasliwal}, Mansi and {Ahumada}, Tom{\'a}s and {Wu}, Qinyu and {Fryer}, Christopher L. and {Niblett}, Annabelle E. and {Xu}, Dong and {Ravasio}, Maria Edvige and {Daja}, Grace and {Li}, Wenxiong and {Anand}, Shreya and {Ho}, Anna Y.~Q. and {Sun}, Hui and {Perley}, Daniel A. and {Yan}, Lin and {Burns}, Eric and {Cenko}, S. Bradley and {Sollerman}, Jesper and {Sarin}, Nikhil and {Piro}, Anthony L. and {Aryan}, Amar and {Miller}, M. Coleman and {An}, Jie and {An}, Tao and {Andrews}, Moira and {Augustin}, Jule and {Bellm}, Eric C. and {Bochenek}, Aleksandra and {Busmann}, Malte and {Chanchaiworawit}, Krittapas and {Cheng}, Huaqing and {Caballero-Garc{\'\i}a}, Maria D. and {Castro-Tirado}, Alberto J. and {Esamdin}, Ali and {Fab{\'a}-Moreno}, Jennifer and {Farah}, Joseph and {Fern{\'a}ndez-Garc{\'\i}a}, Emilio and {Fu}, Shaoyu and {Fynbo}, Johan P.~U. and {Gassert}, Julius and {Padilla Gonzalez}, Estefania and {P{\'e}rez-Garc{\'\i}a}, Ignacio and {Graham}, Matthew and {Gritsevich}, Maria and {Gruen}, Daniel and {Guziy}, Sergiy and {He}, Linbo and {Howell}, D. Andrew and {Hu}, Jingwei and {Hu}, You-Dong and {Iskandar}, Abdusamatjan and {Jaimes}, Joahan Castaneda and {Jiang}, Ji-An and {Jiang}, Ning and {Jiang}, Shuaiqing and {Liang}, Runduo and {Ling}, Zhixing and {Liu}, Jialian and {Liu}, Xing and {Liu}, Yuan and {Masci}, Frank J. and {McCully}, Curtis and {Newsome}, Megan and {Noysena}, Kanthanakorn and {Pandey}, Shashi B. and {Ni}, Kangrui and {Palmese}, Antonella and {Peng}, Han-Long and {Purdum}, Josiah and {Qin}, Yu-Jing and {Rose}, Sam and {Rusholme}, Ben and {S{\'a}nchez-Ram{\'\i}rez}, Rub{\'e}n and {Sevilla}, Cassie and {Smith}, Roger and {Song}, Yujia and {Sravan}, Niharika and {Stein}, Robert and {Tabor}, Constantin and {Terreran}, Giacomo and {Tinyanont}, Samaporn and {Vega}, Pablo and {Wang}, Letian and {Waing}, Tinggui and {Wang}, Xiaofeng and {Wu}, Siyu and {Wu}, Xuefeng and {Wynn}, Kathryn and {Xu}, Yunfei and {Yan}, Shengyu and {Yuan}, Weimin and {Zhang}, Binbin and {Zhang}, Chen and {Zhu}, Zipei and {Zuo}, Xiaoxiong and {Bhullar}, Gursimran},
        title = "{EP250827b/SN 2025wkm: An X-Ray Flash-supernova Powered by a Central Engine and Circumstellar Interaction}",
      journal = {\apjl},
         year = 2026,
        month = aug,
       volume = {1006},
       number = {2},
          eid = {L36},
        pages = {L36},
          doi = {10.3847/2041-8213/ae7a68},
archivePrefix = {arXiv},
       eprint = {2512.10239},
 primaryClass = {astro-ph.HE},
       adsurl = {https://ui.adsabs.harvard.edu/abs/2026ApJ..1006L..36S}
}

@ARTICLE{Nakar2015,
       author = {{Nakar}, Ehud},
        title = "{A Unified Picture for Low-luminosity and Long Gamma-Ray Bursts Based on the Extended Progenitor of llGRB 060218/SN 2006aj}",
      journal = {\apj},
         year = 2015,
        month = jul,
       volume = {807},
       number = {2},
          eid = {172},
        pages = {172},
          doi = {10.1088/0004-637X/807/2/172},
archivePrefix = {arXiv},
       eprint = {1503.00441},
 primaryClass = {astro-ph.HE},
       adsurl = {https://ui.adsabs.harvard.edu/abs/2015ApJ...807..172N}
}

@ARTICLE{SL2018,
       author = {{Sobacchi}, E. and {Lyubarsky}, Y.~E.},
        title = "{External confinement and surface modes in magnetized force-free jets}",
      journal = {\mnras},
         year = 2018,
        month = jan,
       volume = {473},
       number = {2},
        pages = {2813-2821},
          doi = {10.1093/mnras/stx2592},
       adsurl = {https://ui.adsabs.harvard.edu/abs/2018MNRAS.473.2813S}
}

@ARTICLE{Komissarov2009,
       author = {{Komissarov}, Serguei S. and {Vlahakis}, Nektarios and {K{\"o}nigl}, Arieh and {Barkov}, Maxim V.},
        title = "{Magnetic acceleration of ultrarelativistic jets in gamma-ray burst sources}",
      journal = {\mnras},
         year = 2009,
        month = apr,
       volume = {394},
       number = {3},
        pages = {1182-1212},
          doi = {10.1111/j.1365-2966.2009.14410.x},
archivePrefix = {arXiv},
       eprint = {0811.1467},
 primaryClass = {astro-ph},
       adsurl = {https://ui.adsabs.harvard.edu/abs/2009MNRAS.394.1182K}
}

@ARTICLE{Bromberg2014,
       author = {{Bromberg}, Omer and {Granot}, Jonathan and {Lyubarsky}, Yuri and {Piran}, Tsvi},
        title = "{The dynamics of a highly magnetized jet propagating inside a star}",
      journal = {\mnras},
         year = 2014,
        month = sep,
       volume = {443},
       number = {2},
        pages = {1532-1548},
          doi = {10.1093/mnras/stu995},
archivePrefix = {arXiv},
       eprint = {1402.4142},
 primaryClass = {astro-ph.HE},
       adsurl = {https://ui.adsabs.harvard.edu/abs/2014MNRAS.443.1532B}
}

@ARTICLE{Gottlieb2022,
       author = {{Gottlieb}, Ore and {Lalakos}, Aretaios and {Bromberg}, Omer and {Liska}, Matthew and {Tchekhovskoy}, Alexander},
        title = "{Black hole to breakout: 3D GRMHD simulations of collapsar jets reveal a wide range of transients}",
      journal = {\mnras},
         year = 2022,
        month = mar,
       volume = {510},
       number = {4},
        pages = {4962-4975},
          doi = {10.1093/mnras/stab3784},
archivePrefix = {arXiv},
       eprint = {2109.14619},
 primaryClass = {astro-ph.HE},
       adsurl = {https://ui.adsabs.harvard.edu/abs/2022MNRAS.510.4962G}
}

@ARTICLE{Gottlieb2020,
       author = {{Gottlieb}, Ore and {Bromberg}, Omer and {Singh}, Chandra B. and {Nakar}, Ehud},
        title = "{The structure of weakly magnetized {\ensuremath{\gamma}}-ray burst jets}",
      journal = {\mnras},
         year = 2020,
        month = nov,
       volume = {498},
       number = {3},
        pages = {3320-3333},
          doi = {10.1093/mnras/staa2567},
archivePrefix = {arXiv},
       eprint = {2007.11590},
 primaryClass = {astro-ph.HE},
       adsurl = {https://ui.adsabs.harvard.edu/abs/2020MNRAS.498.3320G}
}

@ARTICLE{Guidorzi2015,
       author = {{Guidorzi}, C.},
        title = "{MEPSA: A flexible peak search algorithm designed for uniformly spaced time series}",
      journal = {Astronomy and Computing},
         year = 2015,
        month = apr,
       volume = {10},
        pages = {54-60},
          doi = {10.1016/j.ascom.2015.01.001},
archivePrefix = {arXiv},
       eprint = {1501.01117},
 primaryClass = {astro-ph.IM},
       adsurl = {https://ui.adsabs.harvard.edu/abs/2015A&C....10...54G}
}

@ARTICLE{Marchant20,
       author = {{Marchant}, Pablo and {Moriya}, Takashi J.},
        title = "{The impact of stellar rotation on the black hole mass-gap from pair-instability supernovae}",
      journal = {\aap},
         year = 2020,
        month = aug,
       volume = {640},
          eid = {L18},
        pages = {L18},
          doi = {10.1051/0004-6361/202038902},
archivePrefix = {arXiv},
       eprint = {2007.06220},
 primaryClass = {astro-ph.HE},
       adsurl = {https://ui.adsabs.harvard.edu/abs/2020A&A...640L..18M}
}

@ARTICLE{nakarpiran,
       author = {{Nakar}, Ehud and {Piran}, Tsvi},
        title = "{The Observable Signatures of GRB Cocoons}",
      journal = {\apj},
         year = 2017,
        month = jan,
       volume = {834},
       number = {1},
          eid = {28},
        pages = {28},
          doi = {10.3847/1538-4357/834/1/28},
archivePrefix = {arXiv},
       eprint = {1610.05362},
 primaryClass = {astro-ph.HE},
       adsurl = {https://ui.adsabs.harvard.edu/abs/2017ApJ...834...28N}
}

@ARTICLE{fuller,
       author = {{Fuller}, Jim and {Ro}, Stephen},
        title = "{Pre-supernova outbursts via wave heating in massive stars - II. Hydrogen-poor stars}",
      journal = {\mnras},
         year = 2018,
        month = may,
       volume = {476},
       number = {2},
        pages = {1853-1868},
          doi = {10.1093/mnras/sty369},
archivePrefix = {arXiv},
       eprint = {1710.04251},
 primaryClass = {astro-ph.SR},
       adsurl = {https://ui.adsabs.harvard.edu/abs/2018MNRAS.476.1853F}
}

@ARTICLE{sun,
       author = {{Sun}, Ning-Chen and {Maund}, Justyn R. and {Crowther}, Paul A.},
        title = "{The changing-type SN 2014C may come from an 11-M$_{☉}$ star stripped by binary interaction and violent eruption}",
      journal = {\mnras},
         year = 2020,
        month = oct,
       volume = {497},
       number = {4},
        pages = {5118-5135},
          doi = {10.1093/mnras/staa2277},
archivePrefix = {arXiv},
       eprint = {2003.09325},
 primaryClass = {astro-ph.SR},
       adsurl = {https://ui.adsabs.harvard.edu/abs/2020MNRAS.497.5118S}
}

@ARTICLE{ercolino,
       author = {{Ercolino}, A. and {Jin}, H. and {Langer}, N. and {Dessart}, L.},
        title = "{Mass-transferring binary stars as progenitors of interacting hydrogen-free supernovae}",
      journal = {\aap},
         year = 2025,
        month = apr,
       volume = {696},
          eid = {A103},
        pages = {A103},
          doi = {10.1051/0004-6361/202453426},
archivePrefix = {arXiv},
       eprint = {2412.09893},
 primaryClass = {astro-ph.SR},
       adsurl = {https://ui.adsabs.harvard.edu/abs/2025A&A...696A.103E}
}

@ARTICLE{Matzner2003,
       author = {{Matzner}, Christopher D.},
        title = "{Supernova hosts for gamma-ray burst jets: dynamical constraints}",
      journal = {\mnras},
         year = 2003,
        month = oct,
       volume = {345},
       number = {2},
        pages = {575-589},
          doi = {10.1046/j.1365-8711.2003.06969.x},
archivePrefix = {arXiv},
       eprint = {astro-ph/0203085},
 primaryClass = {astro-ph},
       adsurl = {https://ui.adsabs.harvard.edu/abs/2003MNRAS.345..575M}
}

@ARTICLE{Gupta2026,
       author = {{Gupta}, Soumya and {Sahayanathan}, Sunder and {Shanu}, Saharsh and {Nath}, Rishabh},
        title = "{Are Single-zone Emission Models Sufficient to Explain GRB 220426A and GRB 230812B?}",
      journal = {\apj},
         year = 2026,
        month = jun,
       volume = {1003},
       number = {2},
          eid = {197},
        pages = {197},
          doi = {10.3847/1538-4357/ae6809},
archivePrefix = {arXiv},
       eprint = {2605.11543},
 primaryClass = {astro-ph.HE},
       adsurl = {https://ui.adsabs.harvard.edu/abs/2026ApJ..1003..197G}
}

@ARTICLE{Oganesyan2026,
       author = {{Oganesyan}, Gor and {Sharan Salafia}, Om and {Sobacchi}, Emanuele and {Macera}, Samanta and {Ghirlanda}, Giancarlo and {Nava}, Lara and {Ierardi}, Annarita and {Banerjee}, Biswajit and {Mei}, Alessio and {Ascenzi}, Stefano and {Branchesi}, Marica},
        title = "{MeV absorption in gamma-ray bursts as a probe of their progenitor environments}",
      journal = {\aap},
         year = 2026,
        month = jun,
       volume = {710},
          eid = {L37},
        pages = {L37},
          doi = {10.1051/0004-6361/202659042},
archivePrefix = {arXiv},
       eprint = {2601.14393},
 primaryClass = {astro-ph.HE},
       adsurl = {https://ui.adsabs.harvard.edu/abs/2026A&A...710L..37O}
}

@ARTICLE{Salafia2026,
       author = {{Salafia}, Om Sharan and {Celotti}, Annalisa and {Sobacchi}, Emanuele and {Nava}, Lara and {Oganesyan}, Gor and {Ghirlanda}, Giancarlo and {Boula}, Stella and {Ravasio}, Maria Edvige and {Ghisellini}, Gabriele},
        title = "{A self-consistent explanation of the MeV line in GRB 221009A unveils a dense circumstellar medium}",
      journal = {\aap},
         year = 2026,
        month = aug,
       volume = {712},
          eid = {A97},
        pages = {A97},
          doi = {10.1051/0004-6361/202558639},
archivePrefix = {arXiv},
       eprint = {2601.14257},
 primaryClass = {astro-ph.HE},
       adsurl = {https://ui.adsabs.harvard.edu/abs/2026A&A...712A..97S}
}

@ARTICLE{Norris1986,
       author = {{Norris}, J.~P. and {Share}, G.~H. and {Messina}, D.~C. and {Dennis}, B.~R. and {Desai}, U.~D. and {Cline}, T.~L. and {Matz}, S.~M. and {Chupp}, E.~L.},
        title = "{Spectral Evolution of Pulse Structures in Gamma-Ray Bursts}",
      journal = {\apj},
         year = 1986,
        month = feb,
       volume = {301},
        pages = {213},
          doi = {10.1086/163889},
       adsurl = {https://ui.adsabs.harvard.edu/abs/1986ApJ...301..213N}
}

@ARTICLE{Hamidani2021,
       author = {{Hamidani}, Hamid and {Ioka}, Kunihito},
        title = "{Jet propagation in expanding medium for gamma-ray bursts}",
      journal = {\mnras},
         year = 2021,
        month = jan,
       volume = {500},
       number = {1},
        pages = {627-642},
          doi = {10.1093/mnras/staa3276},
archivePrefix = {arXiv},
       eprint = {2007.10690},
 primaryClass = {astro-ph.HE},
       adsurl = {https://ui.adsabs.harvard.edu/abs/2021MNRAS.500..627H}
}

@ARTICLE{Lane2025,
       author = {{Lane}, Zachary G. and {Ridden-Harper}, Ryan and {Rest}, Sofia and {Rest}, Armin and {Ransome}, Conor L. and {Wang}, Qinan and {Montilla}, Clarinda and {Steed}, Micaela and {Andreoni}, Igor and {Armstrong}, Patrick and {Brown}, Peter J. and {Cooke}, Jeffrey and {Coulter}, David A. and {Fox}, Ori and {Freeburn}, James and {Galoppo}, Marco and {Gal-Yam}, Avishay and {Goldberg}, Jared A. and {Harvey-Hawes}, Christopher and {Hiramatsu}, Daichi and {Hounsell}, Rebekah and {Leicester}, Brayden and {Lelkes}, Kl{\'a}ra and {Linial}, Itai and {Moln{\'a}r}, L{\'a}szlo and {Moore}, Thomas and {Mourier}, Pierre and {Nugent}, Anya E. and {O'Neill}, David and {Roxburgh}, Hugh and {Shukawa}, Koji and {Smartt}, Stephen J. and {Smith}, Nathan and {Smith}, Ken W. and {Vergara Carrasco}, Sebastian and {Villar}, V. Ashley and {Vink{\'o}}, J{\'o}zsef and {Wasserman}, Tal and {Yossef}, Zenati and {Zimmerman}, Erez},
        title = "{SN 2019vxm: A Shocking Coincidence between Fermi and TESS}",
      journal = {arXiv e-prints},
         year = 2025,
        month = nov,
          eid = {arXiv:2511.15975},
        pages = {arXiv:2511.15975},
          doi = {10.48550/arXiv.2511.15975},
archivePrefix = {arXiv},
       eprint = {2511.15975},
 primaryClass = {astro-ph.HE},
       adsurl = {https://ui.adsabs.harvard.edu/abs/2025arXiv251115975L}
}

@ARTICLE{Renzo2020,
       author = {{Renzo}, M. and {Farmer}, R. and {Justham}, S. and {G{\"o}tberg}, Y. and {de Mink}, S.~E. and {Zapartas}, E. and {Marchant}, P. and {Smith}, N.},
        title = "{Predictions for the hydrogen-free ejecta of pulsational pair-instability supernovae}",
      journal = {\aap},
         year = 2020,
        month = aug,
       volume = {640},
          eid = {A56},
        pages = {A56},
          doi = {10.1051/0004-6361/202037710},
archivePrefix = {arXiv},
       eprint = {2002.05077},
 primaryClass = {astro-ph.SR},
       adsurl = {https://ui.adsabs.harvard.edu/abs/2020A&A...640A..56R}
}

@ARTICLE{Leung2019,
       author = {{Leung}, Shing-Chi and {Nomoto}, Ken'ichi and {Blinnikov}, Sergei},
        title = "{Pulsational Pair-instability Supernovae. I. Pre-collapse Evolution and Pulsational Mass Ejection}",
      journal = {\apj},
         year = 2019,
        month = dec,
       volume = {887},
       number = {1},
          eid = {72},
        pages = {72},
          doi = {10.3847/1538-4357/ab4fe5},
archivePrefix = {arXiv},
       eprint = {1901.11136},
 primaryClass = {astro-ph.HE},
       adsurl = {https://ui.adsabs.harvard.edu/abs/2019ApJ...887...72L}
}

@ARTICLE{Camisasca,
       author = {{Camisasca}, A.~E. and {Guidorzi}, C. and {Amati}, L. and {Frontera}, F. and {Song}, X.~Y. and {Xiao}, S. and {Xiong}, S.~L. and {Zhang}, S.~N. and {Margutti}, R. and {Kobayashi}, S. and {Mundell}, C.~G. and {Ge}, M.~Y. and {Gomboc}, A. and {Jia}, S.~M. and {Jordana-Mitjans}, N. and {Li}, C.~K. and {Li}, X.~B. and {Maccary}, R. and {Shrestha}, M. and {Xue}, W.~C. and {Zhang}, S.},
        title = "{GRB minimum variability timescale with Insight-HXMT and Swift. Implications for progenitor models, dissipation physics, and GRB classifications}",
      journal = {\aap},
         year = 2023,
        month = mar,
       volume = {671},
          eid = {A112},
        pages = {A112},
          doi = {10.1051/0004-6361/202245657},
archivePrefix = {arXiv},
       eprint = {2301.01176},
 primaryClass = {astro-ph.HE},
       adsurl = {https://ui.adsabs.harvard.edu/abs/2023A&A...671A.112C}
}

@ARTICLE{Maccary,
       author = {{Maccary}, R. and {Guidorzi}, C. and {Camisasca}, A.~E. and {Maistrello}, M. and {Kobayashi}, S. and {Amati}, L. and {Bazzanini}, L. and {Bulla}, M. and {Ferro}, L. and {Frontera}, F. and {Tsvetkova}, A.},
        title = "{Gamma-ray burst minimum variability timescales with Fermi/GBM}",
      journal = {\aap},
         year = 2025,
        month = oct,
       volume = {702},
          eid = {A95},
        pages = {A95},
          doi = {10.1051/0004-6361/202555418},
archivePrefix = {arXiv},
       eprint = {2508.08995},
 primaryClass = {astro-ph.HE},
       adsurl = {https://ui.adsabs.harvard.edu/abs/2025A&A...702A..95M}
}

@ARTICLE{Malacaria2022,
       author = {{Malacaria}, C. and {Meegan}, C. and {Fermi GBM Team}},
        title = "{GRB 220426A: Fermi GBM detection}",
      journal = {GRB Coordinates Network},
         year = 2022,
        month = apr,
       volume = {31955},
        pages = {1},
       adsurl = {https://ui.adsabs.harvard.edu/abs/2022GCN.31955....1M}
}

@ARTICLE{Sander2019,
       author = {{Sander}, A.~A.~C. and {Hamann}, W.-R. and {Todt}, H. and {Hainich}, R. and {Shenar}, T. and {Ramachandran}, V. and {Oskinova}, L.~M.},
        title = "{The Galactic WC and WO stars. The impact of revised distances from Gaia DR2 and their role as massive black hole progenitors}",
      journal = {\aap},
         year = 2019,
        month = jan,
       volume = {621},
          eid = {A92},
        pages = {A92},
          doi = {10.1051/0004-6361/201833712},
archivePrefix = {arXiv},
       eprint = {1807.04293},
 primaryClass = {astro-ph.GA},
       adsurl = {https://ui.adsabs.harvard.edu/abs/2019A&A...621A..92S}
}

@ARTICLE{Tchekhovskoy2015,
       author = {{Tchekhovskoy}, Alexander and {Giannios}, Dimitrios},
        title = "{Magnetic flux of progenitor stars sets gamma-ray burst luminosity and variability}",
      journal = {\mnras},
         year = 2015,
        month = feb,
       volume = {447},
       number = {1},
        pages = {327-344},
          doi = {10.1093/mnras/stu2229},
archivePrefix = {arXiv},
       eprint = {1409.4414},
 primaryClass = {astro-ph.HE},
       adsurl = {https://ui.adsabs.harvard.edu/abs/2015MNRAS.447..327T}
}

@ARTICLE{BrombergTchekhovskoy2016,
       author = {{Bromberg}, Omer and {Tchekhovskoy}, Alexander},
        title = "{Relativistic MHD simulations of core-collapse GRB jets: 3D instabilities and magnetic dissipation}",
      journal = {\mnras},
         year = 2016,
        month = feb,
       volume = {456},
       number = {2},
        pages = {1739-1760},
          doi = {10.1093/mnras/stv2591},
archivePrefix = {arXiv},
       eprint = {1508.02721},
 primaryClass = {astro-ph.HE},
       adsurl = {https://ui.adsabs.harvard.edu/abs/2016MNRAS.456.1739B}
}

@INCOLLECTION{Smith17,
       author = {{Smith}, Nathan},
        title = "{Interacting Supernovae: Types IIn and Ibn}",
    booktitle = {Handbook of Supernovae},
         year = 2017,
       editor = {{Alsabti}, Athem W. and {Murdin}, Paul},
        pages = {403},
          doi = {10.1007/978-3-319-21846-5_38},
       adsurl = {https://ui.adsabs.harvard.edu/abs/2017hsn..book..403S}
}

@ARTICLE{Bromberg2016,
       author = {{Bromberg}, Omer and {Tchekhovskoy}, Alexander},
        title = "{Relativistic MHD simulations of core-collapse GRB jets: 3D instabilities and magnetic dissipation}",
      journal = {\mnras},
         year = 2016,
        month = feb,
       volume = {456},
       number = {2},
        pages = {1739-1760},
          doi = {10.1093/mnras/stv2591},
archivePrefix = {arXiv},
       eprint = {1508.02721},
 primaryClass = {astro-ph.HE},
       adsurl = {https://ui.adsabs.harvard.edu/abs/2016MNRAS.456.1739B}
}

@ARTICLE{Ryde2022,
       author = {{Ryde}, Felix and {Iyyani}, Shabnam and {Ahlgren}, Bj{\"o}rn and {Pe'er}, Asaf and {Sharma}, Vidushi and {Lundman}, Christoffer and {Axelsson}, Magnus},
        title = "{Onset of Particle Acceleration during the Prompt Phase in Gamma-Ray Bursts as Revealed by Synchrotron Emission in GRB 160821A}",
      journal = {\apjl},
         year = 2022,
        month = jun,
       volume = {932},
       number = {2},
          eid = {L15},
        pages = {L15},
          doi = {10.3847/2041-8213/ac73fe},
archivePrefix = {arXiv},
       eprint = {2206.00680},
 primaryClass = {astro-ph.HE},
       adsurl = {https://ui.adsabs.harvard.edu/abs/2022ApJ...932L..15R}
}

@ARTICLE{Samuelsson2023,
       author = {{Samuelsson}, Filip and {Ryde}, Felix},
        title = "{Observational Characteristics of Radiation-mediated Shocks in Photospheric Gamma-Ray Burst Emission}",
      journal = {\apj},
         year = 2023,
        month = oct,
       volume = {956},
       number = {1},
          eid = {42},
        pages = {42},
          doi = {10.3847/1538-4357/ace441},
archivePrefix = {arXiv},
       eprint = {2206.11701},
 primaryClass = {astro-ph.HE},
       adsurl = {https://ui.adsabs.harvard.edu/abs/2023ApJ...956...42S}
}

@ARTICLE{Nugis2000,
       author = {{Nugis}, T. and {Lamers}, H.~J.~G.~L.~M.},
        title = "{Mass-loss rates of Wolf-Rayet stars as a function of stellar parameters}",
      journal = {\aap},
         year = 2000,
        month = aug,
       volume = {360},
        pages = {227-244},
       adsurl = {https://ui.adsabs.harvard.edu/abs/2000A&A...360..227N}
}

@ARTICLE{Srinivasaragavan2025,
       author = {{Srinivasaragavan}, Gokul P. and {Hamidani}, Hamid and {Schroeder}, Genevieve and {Sarin}, Nikhil and {Ho}, Anna Y.~Q. and {Piro}, Anthony L. and {Cenko}, S. Bradley and {Anand}, Shreya and {Sollerman}, Jesper and {Perley}, Daniel A. and {Maeda}, Keiichi and {O'Connor}, Brendan and {Kuncarayakti}, Hanindyo and {Miller}, M. Coleman and {Ahumada}, Tom{\'a}s and {Vail}, Jada L. and {Duffell}, Paul and {Dastidar}, Ranadeep and {Andreoni}, Igor and {Bochenek}, Aleksandra and {Brennan}, Se{\'a}n. J. and {Carney}, Jonathan and {Chen}, Ping and {Freeburn}, James and {Gal-Yam}, Avishay and {Jacobson-Gal{\'a}n}, Wynn and {Kasliwal}, Mansi M. and {Li}, Jiaxuan and {Li}, Maggie L. and {Sravan}, Niharika and {Warshofsky}, Daniel E.},
        title = "{EP250108a/SN 2025kg: A Jet-driven Stellar Explosion Interacting with Circumstellar Material}",
      journal = {\apjl},
         year = 2025,
        month = aug,
       volume = {988},
       number = {2},
          eid = {L60},
        pages = {L60},
          doi = {10.3847/2041-8213/ade870},
archivePrefix = {arXiv},
       eprint = {2504.17516},
 primaryClass = {astro-ph.HE},
       adsurl = {https://ui.adsabs.harvard.edu/abs/2025ApJ...988L..60S}
}

@ARTICLE{Perley2022,
       author = {{Perley}, Daniel A. and {Sollerman}, Jesper and {Schulze}, Steve and {Yao}, Yuhan and {Fremling}, Christoffer and {Gal-Yam}, Avishay and {Ho}, Anna Y.~Q. and {Yang}, Yi and {Kool}, Erik C. and {Irani}, Ido and {Yan}, Lin and {Andreoni}, Igor and {Baade}, Dietrich and {Bellm}, Eric C. and {Brink}, Thomas G. and {Chen}, Ting-Wan and {Cikota}, Aleksandar and {Coughlin}, Michael W. and {Dahiwale}, Aishwarya and {Dekany}, Richard and {Duev}, Dmitry A. and {Filippenko}, Alexei V. and {Hoeflich}, Peter and {Kasliwal}, Mansi M. and {Kulkarni}, S.~R. and {Lunnan}, Ragnhild and {Masci}, Frank J. and {Maund}, Justyn R. and {Medford}, Michael S. and {Riddle}, Reed and {Rosnet}, Philippe and {Shupe}, David L. and {Strotjohann}, Nora Linn and {Tzanidakis}, Anastasios and {Zheng}, WeiKang},
        title = "{The Type Icn SN 2021csp: Implications for the Origins of the Fastest Supernovae and the Fates of Wolf-Rayet Stars}",
      journal = {\apj},
         year = 2022,
        month = mar,
       volume = {927},
       number = {2},
          eid = {180},
        pages = {180},
          doi = {10.3847/1538-4357/ac478e},
archivePrefix = {arXiv},
       eprint = {2111.12110},
 primaryClass = {astro-ph.HE},
       adsurl = {https://ui.adsabs.harvard.edu/abs/2022ApJ...927..180P}
}

@ARTICLE{Irani2024,
       author = {{Irani}, Ido and {Chen}, Ping and {Morag}, Jonathan and {Schulze}, Steve and {Gal-Yam}, Avishay and {Strotjohann}, Nora L. and {Yaron}, Ofer and {Zimmerman}, Erez A. and {Sharon}, Amir and {Perley}, Daniel A. and {Sollerman}, J. and {Tohuvavohu}, Aaron and {Das}, Kaustav K. and {Kasliwal}, Mansi M. and {Bruch}, Rachel and {Brink}, Thomas G. and {Zheng}, WeiKang and {Filippenko}, Alexei V. and {Patra}, Kishore C. and {Vasylyev}, Sergiy S. and {Yang}, Yi and {Graham}, Matthew J. and {Bloom}, Joshua S. and {Mazzali}, Paolo and {Purdum}, Josiah and {Laher}, Russ R. and {Wold}, Avery and {Sharma}, Yashvi and {Lacroix}, Leander and {Medford}, Michael S.},
        title = "{SN 2022oqm─A Ca-rich Explosion of a Compact Progenitor Embedded in C/O Circumstellar Material}",
      journal = {\apj},
         year = 2024,
        month = feb,
       volume = {962},
       number = {2},
          eid = {109},
        pages = {109},
          doi = {10.3847/1538-4357/ad04d7},
archivePrefix = {arXiv},
       eprint = {2210.02554},
 primaryClass = {astro-ph.HE},
       adsurl = {https://ui.adsabs.harvard.edu/abs/2024ApJ...962..109I}
}

@ARTICLE{PeerRyde2024,
       author = {{Pe'er}, Asaf and {Ryde}, Felix},
        title = "{Gamma-Ray Burst Interaction with the Circumburst Medium: The CBM Phase Following the Prompt Phase in GRBs}",
      journal = {\apj},
         year = 2024,
        month = nov,
       volume = {976},
       number = {1},
          eid = {55},
        pages = {55},
          doi = {10.3847/1538-4357/ad82ed},
archivePrefix = {arXiv},
       eprint = {2406.03841},
 primaryClass = {astro-ph.HE},
       adsurl = {https://ui.adsabs.harvard.edu/abs/2024ApJ...976...55P}
}

@ARTICLE{Lloyd2020,
       author = {{Lloyd-Ronning}, Nicole and {Hurtado}, Valeria U. and {Aykutalp}, Aycin and {Johnson}, Jarrett and {Ceccobello}, Chiara},
        title = "{The evolution of gamma-ray burst jet opening angle through cosmic time}",
      journal = {\mnras},
         year = 2020,
        month = may,
       volume = {494},
       number = {3},
        pages = {4371-4381},
          doi = {10.1093/mnras/staa1057},
archivePrefix = {arXiv},
       eprint = {1912.00057},
 primaryClass = {astro-ph.HE},
       adsurl = {https://ui.adsabs.harvard.edu/abs/2020MNRAS.494.4371L}
}

@ARTICLE{Hamidani2025,
       author = {{Hamidani}, Hamid and {Ioka}, Kunihito and {Kashiyama}, Kazumi and {Tanaka}, Masaomi},
        title = "{Gamma-Ray Burst Jets in Circumstellar Material: Dynamics, Breakout, and Diversity of Transients}",
      journal = {\apj},
         year = 2025,
        month = jul,
       volume = {988},
       number = {1},
          eid = {30},
        pages = {30},
          doi = {10.3847/1538-4357/addd13},
archivePrefix = {arXiv},
       eprint = {2503.16242},
 primaryClass = {astro-ph.HE},
       adsurl = {https://ui.adsabs.harvard.edu/abs/2025ApJ...988...30H}
}

@ARTICLE{DuffellHo2020,
       author = {{Duffell}, Paul C. and {Ho}, Anna Y.~Q.},
        title = "{How Dense of a Circumstellar Medium Is Sufficient to Choke a Jet?}",
      journal = {\apj},
         year = 2020,
        month = sep,
       volume = {900},
       number = {2},
          eid = {193},
        pages = {193},
          doi = {10.3847/1538-4357/aba90a},
       adsurl = {https://ui.adsabs.harvard.edu/abs/2020ApJ...900..193D}
}

@ARTICLE{SuzukiMaeda2024,
       author = {{Suzuki}, Akihiro and {Irwin}, Christopher M. and {Maeda}, Keiichi},
        title = "{Dynamical properties of mildly relativistic ejecta produced by the mass-loading of gamma-ray burst jets in dense ambient media}",
      journal = {\pasj},
         year = 2024,
        month = aug,
       volume = {76},
       number = {4},
        pages = {863-879},
          doi = {10.1093/pasj/psae055},
archivePrefix = {arXiv},
       eprint = {2406.06939},
 primaryClass = {astro-ph.HE},
       adsurl = {https://ui.adsabs.harvard.edu/abs/2024PASJ...76..863S}
}

@ARTICLE{Meszaros1993,
       author = {{Meszaros}, P. and {Rees}, M.~J.},
        title = "{Relativistic Fireballs and Their Impact on External Matter: Models for Cosmological Gamma-Ray Bursts}",
      journal = {\apj},
         year = 1993,
        month = mar,
       volume = {405},
        pages = {278},
          doi = {10.1086/172360},
       adsurl = {https://ui.adsabs.harvard.edu/abs/1993ApJ...405..278M}
}

@INPROCEEDINGS{Battelino2007,
       author = {{Battelino}, Milan and {Ryde}, Felix and {Omodei}, Nicola and {Band}, David L.},
        title = "{Simulation of prompt emission from GRBs with a photospheric component and its detectability by GLAST}",
    booktitle = {The First GLAST Symposium},
         year = 2007,
       editor = {{Ritz}, Steven and {Michelson}, Peter and {Meegan}, Charles A.},
       series = {American Institute of Physics Conference Series},
       volume = {921},
        month = jul,
    publisher = {AIP},
        pages = {478-479},
          doi = {10.1063/1.2757410},
       adsurl = {https://ui.adsabs.harvard.edu/abs/2007AIPC..921..478B}
}

@ARTICLE{Deng2022,
       author = {{Deng}, Li-Tao and {Lin}, Da-Bin and {Zhou}, Li and {Wang}, Kai and {Yang}, Xing and {Hou}, Shu-Jin and {Li}, Jing and {Wang}, Xiang-Gao and {Lu}, Rui-Jing and {Liang}, En-Wei},
        title = "{Spectral Analysis of GRB 220426A: Another Case of a Thermally Dominated Burst}",
      journal = {\apjl},
         year = 2022,
        month = aug,
       volume = {934},
       number = {2},
          eid = {L22},
        pages = {L22},
          doi = {10.3847/2041-8213/ac8169},
       adsurl = {https://ui.adsabs.harvard.edu/abs/2022ApJ...934L..22D}
}

@ARTICLE{Song2022,
       author = {{Song}, Xin-Ying and {Zhang}, Shuang-Nan and {Ge}, Ming-Yu and {Zhang}, Shu},
        title = "{The origin of the photospheric emission of GRB 220426A}",
      journal = {\mnras},
         year = 2022,
        month = dec,
       volume = {517},
       number = {2},
        pages = {2088-2102},
          doi = {10.1093/mnras/stac2764},
archivePrefix = {arXiv},
       eprint = {2209.10832},
 primaryClass = {astro-ph.HE},
       adsurl = {https://ui.adsabs.harvard.edu/abs/2022MNRAS.517.2088S}
}

@ARTICLE{Wang2022,
       author = {{Wang}, Yun and {Zheng}, Tian-Ci and {Jin}, Zhi-Ping},
        title = "{GRB 220426A: A Thermal Radiation-Dominated Gamma-Ray Burst}",
      journal = {\apj},
         year = 2022,
        month = dec,
       volume = {940},
       number = {2},
          eid = {142},
        pages = {142},
          doi = {10.3847/1538-4357/aca017},
archivePrefix = {arXiv},
       eprint = {2205.08427},
 primaryClass = {astro-ph.HE},
       adsurl = {https://ui.adsabs.harvard.edu/abs/2022ApJ...940..142W}
}

@ARTICLE{Wistemar2026,
       author = {{Wistemar}, Oscar and {Alamaa}, Filip and {Ryde}, Felix},
        title = "{Photospheric emission from GRB 211211A altered by a strong radiation-mediated shock}",
      journal = {\mnras},
         year = 2025,
        month = dec,
       volume = {544},
       number = {4},
        pages = {3683-3695},
          doi = {10.1093/mnras/staf1757},
archivePrefix = {arXiv},
       eprint = {2506.08122},
 primaryClass = {astro-ph.HE},
       adsurl = {https://ui.adsabs.harvard.edu/abs/2025MNRAS.544.3683W}
}

@ARTICLE{Wistemar2025,
       author = {{Wistemar}, Oscar and {Ryde}, Felix and {Alamaa}, Filip},
        title = "{A Generalized Method to Measure the Lorentz Factor from Gamma-Ray Burst Photospheric Emission}",
      journal = {\apj},
         year = 2025,
        month = jun,
       volume = {986},
       number = {2},
          eid = {118},
        pages = {118},
          doi = {10.3847/1538-4357/add52d},
archivePrefix = {arXiv},
       eprint = {2504.00092},
 primaryClass = {astro-ph.HE},
       adsurl = {https://ui.adsabs.harvard.edu/abs/2025ApJ...986..118W}
}

@ARTICLE{Das2024,
       author = {{Das}, Kaustav K. and {Kasliwal}, Mansi M. and {Sollerman}, Jesper and {Fremling}, Christoffer and {Irani}, I. and {Leung}, Shing-Chi and {Yang}, Sheng and {Wu}, Samantha and {Fuller}, Jim and {Anand}, Shreya and {Andreoni}, Igor and {Barbarino}, C. and {Brink}, Thomas G. and {De}, Kishalay and {Dugas}, Alison and {Groom}, Steven L. and {Helou}, George and {Hinds}, K. -Ryan and {Ho}, Anna Y.~Q. and {Karambelkar}, Viraj and {Kulkarni}, S.~R. and {Perley}, Daniel A. and {Purdum}, Josiah and {Regnault}, Nicolas and {Schulze}, Steve and {Sharma}, Yashvi and {Sit}, Tawny and {Sravan}, Niharika and {Srinivasaragavan}, Gokul P. and {Stein}, Robert and {Taggart}, Kirsty and {Tartaglia}, Leonardo and {Tzanidakis}, Anastasios and {Wold}, Avery and {Yan}, Lin and {Yao}, Yuhan and {Zolkower}, Jeffry},
        title = "{Probing Presupernova Mass Loss in Double-peaked Type Ibc Supernovae from the Zwicky Transient Facility}",
      journal = {\apj},
         year = 2024,
        month = sep,
       volume = {972},
       number = {1},
          eid = {91},
        pages = {91},
          doi = {10.3847/1538-4357/ad595f},
archivePrefix = {arXiv},
       eprint = {2306.04698},
 primaryClass = {astro-ph.HE},
       adsurl = {https://ui.adsabs.harvard.edu/abs/2024ApJ...972...91D}
}

@ARTICLE{Ioka11,
       author = {{Ioka}, K. and {Ohira}, Y. and {Kawanaka}, N. and {Mizuta}, A.},
        title = "{Gamma-Ray Burst without Baryonic and Magnetic Load?}",
      journal = {Progress of Theoretical Physics},
         year = 2011,
        month = sep,
       volume = {126},
       number = {3},
        pages = {555-564},
          doi = {10.1143/PTP.126.555},
archivePrefix = {arXiv},
       eprint = {1103.5746},
 primaryClass = {astro-ph.HE},
       adsurl = {https://ui.adsabs.harvard.edu/abs/2011PThPh.126..555I}
}

@ARTICLE{Sullivan24,
       author = {{Sullivan}, Andrew G. and {Blandford}, Roger D. and {Begelman}, Mitchell C. and {Birkinshaw}, Mark and {Readhead}, Anthony C.~S.},
        title = "{Small-scale radio jets and tidal disruption events: a theory of high-luminosity compact symmetric objects}",
      journal = {\mnras},
         year = 2024,
        month = mar,
       volume = {528},
       number = {4},
        pages = {6302-6311},
          doi = {10.1093/mnras/stae322},
archivePrefix = {arXiv},
       eprint = {2401.14399},
 primaryClass = {astro-ph.HE},
       adsurl = {https://ui.adsabs.harvard.edu/abs/2024MNRAS.528.6302S}
}

@ARTICLE{Mizuta09,
       author = {{Mizuta}, Akira and {Aloy}, Miguel A.},
        title = "{Angular Energy Distribution of Collapsar-Jets}",
      journal = {\apj},
         year = 2009,
        month = jul,
       volume = {699},
       number = {2},
        pages = {1261-1273},
          doi = {10.1088/0004-637X/699/2/1261},
archivePrefix = {arXiv},
       eprint = {0812.4813},
 primaryClass = {astro-ph},
       adsurl = {https://ui.adsabs.harvard.edu/abs/2009ApJ...699.1261M}
}

@ARTICLE{Harrison18,
       author = {{Harrison}, Richard and {Gottlieb}, Ore and {Nakar}, Ehud},
        title = "{Numerically calibrated model for propagation of a relativistic unmagnetized jet in dense media}",
      journal = {\mnras},
         year = 2018,
        month = jun,
       volume = {477},
       number = {2},
        pages = {2128-2140},
          doi = {10.1093/mnras/sty760},
archivePrefix = {arXiv},
       eprint = {1707.06234},
 primaryClass = {astro-ph.HE},
       adsurl = {https://ui.adsabs.harvard.edu/abs/2018MNRAS.477.2128H}
}

@ARTICLE{NalewajkoSikora09,
       author = {{Nalewajko}, Krzysztof and {Sikora}, Marek},
        title = "{A structure and energy dissipation efficiency of relativistic reconfinement shocks}",
      journal = {\mnras},
         year = 2009,
        month = jan,
       volume = {392},
       number = {3},
        pages = {1205-1210},
          doi = {10.1111/j.1365-2966.2008.14123.x},
archivePrefix = {arXiv},
       eprint = {0810.3912},
 primaryClass = {astro-ph},
       adsurl = {https://ui.adsabs.harvard.edu/abs/2009MNRAS.392.1205N}
}

@ARTICLE{Bromberg07,
       author = {{Bromberg}, Omer and {Levinson}, Amir},
        title = "{Hydrodynamic Collimation of Relativistic Outflows: Semianalytic Solutions and Application to Gamma-Ray Bursts}",
      journal = {\apj},
         year = 2007,
        month = dec,
       volume = {671},
       number = {1},
        pages = {678-688},
          doi = {10.1086/522668},
archivePrefix = {arXiv},
       eprint = {0705.2040},
 primaryClass = {astro-ph},
       adsurl = {https://ui.adsabs.harvard.edu/abs/2007ApJ...671..678B}
}

@ARTICLE{KomissarovFalle97,
       author = {{Komissarov}, S.~S. and {Falle}, S.~A.~E.~G.},
        title = "{Simulations of Superluminal Radio Sources}",
      journal = {\mnras},
         year = 1997,
        month = jul,
       volume = {288},
       number = {4},
        pages = {833-848},
          doi = {10.1093/mnras/288.4.833},
       adsurl = {https://ui.adsabs.harvard.edu/abs/1997MNRAS.288..833K}
}

@ARTICLE{Salafia2020,
       author = {{Salafia}, O.~S. and {Barbieri}, C. and {Ascenzi}, S. and {Toffano}, M.},
        title = "{Gamma-ray burst jet propagation, development of angular structure, and the luminosity function}",
      journal = {\aap},
         year = 2020,
        month = apr,
       volume = {636},
          eid = {A105},
        pages = {A105},
          doi = {10.1051/0004-6361/201936335},
archivePrefix = {arXiv},
       eprint = {1907.07599},
 primaryClass = {astro-ph.HE},
       adsurl = {https://ui.adsabs.harvard.edu/abs/2020A&A...636A.105S}
}

@ARTICLE{Bromberg11,
       author = {{Bromberg}, Omer and {Nakar}, Ehud and {Piran}, Tsvi and {Sari}, Re'em},
        title = "{The Propagation of Relativistic Jets in External Media}",
      journal = {\apj},
         year = 2011,
        month = oct,
       volume = {740},
       number = {2},
          eid = {100},
        pages = {100},
          doi = {10.1088/0004-637X/740/2/100},
archivePrefix = {arXiv},
       eprint = {1107.1326},
 primaryClass = {astro-ph.HE},
       adsurl = {https://ui.adsabs.harvard.edu/abs/2011ApJ...740..100B}
}

@ARTICLE{BegelmanCioffi89,
       author = {{Begelman}, Mitchell C. and {Cioffi}, Denis F.},
        title = "{Overpressured Cocoons in Extragalactic Radio Sources}",
      journal = {\apjl},
         year = 1989,
        month = oct,
       volume = {345},
        pages = {L21},
          doi = {10.1086/185542},
       adsurl = {https://ui.adsabs.harvard.edu/abs/1989ApJ...345L..21B}
}

@ARTICLE{Crowther2007,
       author = {{Crowther}, Paul A.},
        title = "{Physical Properties of Wolf-Rayet Stars}",
      journal = {\araa},
         year = 2007,
        month = sep,
       volume = {45},
       number = {1},
        pages = {177-219},
          doi = {10.1146/annurev.astro.45.051806.110615},
archivePrefix = {arXiv},
       eprint = {astro-ph/0610356},
 primaryClass = {astro-ph},
       adsurl = {https://ui.adsabs.harvard.edu/abs/2007ARA&A..45..177C}
}

@ARTICLE{Samulesson2022,
       author = {{Samuelsson}, Filip and {Lundman}, Christoffer and {Ryde}, Felix},
        title = "{An Efficient Method for Fitting Radiation-mediated Shocks to Gamma-Ray Burst Data: The Kompaneets RMS Approximation}",
      journal = {\apj},
         year = 2022,
        month = jan,
       volume = {925},
       number = {1},
          eid = {65},
        pages = {65},
          doi = {10.3847/1538-4357/ac332a},
archivePrefix = {arXiv},
       eprint = {2111.01810},
 primaryClass = {astro-ph.HE},
       adsurl = {https://ui.adsabs.harvard.edu/abs/2022ApJ...925...65S}
}

@INPROCEEDINGS{Vianello2015,
       author = {{Vianello}, G. and {Lauer}, R. and {Younk}, P. and {Tibaldo}, L. and {Burgess}, J.~M. and {Ayala Solares}, H. and {Harding}, J.~P. and {Hui}, C.~M. and {Omodei}, N. and {Zhou}, H.},
        title = "{The Multi-Mission Maximum Likelihood framework}",
    booktitle = {34th International Cosmic Ray Conference (ICRC2015)},
         year = 2015,
       series = {International Cosmic Ray Conference},
       volume = {34},
        month = jul,
          eid = {1042},
        pages = {1042},
       adsurl = {https://ui.adsabs.harvard.edu/abs/2015ICRC...34.1042V}
}

@ARTICLE{Vianello2018,
       author = {{Vianello}, Giacomo},
        title = "{The Significance of an Excess in a Counting Experiment: Assessing the Impact of Systematic Uncertainties and the Case with a Gaussian Background}",
      journal = {\apjs},
         year = "2018",
        month = "May",
       volume = {236},
       number = {1},
          eid = {17},
        pages = {17},
          doi = {10.3847/1538-4365/aab780},
archivePrefix = {arXiv},
       eprint = {1712.00118},
 primaryClass = {physics.data-an},
       adsurl = {https://ui.adsabs.harvard.edu/abs/2018ApJS..236...17V}
}

@ARTICLE{Acuner2019,
       author = {{Acuner}, Zeynep and {Ryde}, Felix and {Yu}, Hoi-Fung},
        title = "{Non-dissipative photospheres in GRBs: spectral appearance in the Fermi/GBM catalogue}",
      journal = {Monthly Notices of the Royal Astronomical Society},
         year = "2019",
        month = "Aug",
       volume = {487},
       number = {4},
        pages = {5508-5519},
          doi = {10.1093/mnras/stz1356},
archivePrefix = {arXiv},
       eprint = {1906.01318},
 primaryClass = {astro-ph.HE},
       adsurl = {https://ui.adsabs.harvard.edu/abs/2019MNRAS.487.5508A}
}

@ARTICLE{Gottlieb2019,
   author = {{Gottlieb}, O. and {Levinson}, A. and {Nakar}, E.},
    title = "{High efficiency photospheric emission in gamma-ray bursts}",
  journal = {arXiv e-prints},
archivePrefix = "arXiv",
   eprint = {1904.07244},
 primaryClass = "astro-ph.HE",
     year = 2019,
    month = apr,
   adsurl = {http://adsabs.harvard.edu/abs/2019arXiv190407244G}
}

@ARTICLE{Yu2018,
   author = {{Yu}, H.-F. and {Dereli-B{\'e}gu{\'e}}, H. and {Ryde}, F.},
    title = "{Bayesian Time-Resolved Spectra of GRB Pulses}",
  journal = {arXiv e-prints},
archivePrefix = "arXiv",
   eprint = {1810.07313},
 primaryClass = "astro-ph.HE",
     year = 2018,
    month = oct,
   adsurl = {http://adsabs.harvard.edu/abs/2018arXiv181007313Y}
}

@ARTICLE{Ryde2019,
   author = {{Ryde}, F. and {Yu}, H.-F. and {Dereli-B{\'e}gu{\'e}}, H. and 
	{Lundman}, C. and {Pe'er}, A. and {Li}, L.},
    title = "{On the {$\alpha$}-Intensity Correlation in Gamma-Ray Bursts: Subphotospheric Heating with Varying Entropy}",
  journal = {\mnras},
archivePrefix = "arXiv",
   eprint = {1901.01775},
 primaryClass = "astro-ph.HE",
     year = 2019,
    month = jan,
      doi = {10.1093/mnras/stz083},
   adsurl = {http://adsabs.harvard.edu/abs/2019MNRAS.tmp...65R}
}

@ARTICLE{Ryde2017,
   author = {{Ryde}, F. and {Lundman}, C. and {Acuner}, Z.},
    title = "{Emission from accelerating jets in gamma-ray bursts: radiation-dominated flows with increasing mass outflow rates}",
  journal = {\mnras},
     year = 2017,
    month = dec,
   volume = 472,
    pages = {1897-1906},
      doi = {10.1093/mnras/stx2019},
   adsurl = {http://adsabs.harvard.edu/abs/2017MNRAS.472.1897R}
}

@ARTICLE{Yu2016,
   author = {{Yu}, H.-F. and {Preece}, R.~D. and {Greiner}, J. and {Narayana Bhat}, P. and 
	{Bissaldi}, E. and {Briggs}, M.~S. and {Cleveland}, W.~H. and 
	{Connaughton}, V. and {Goldstein}, A. and {von Kienlin}, A. and 
	{Kouveliotou}, C. and {Mailyan}, B. and {Meegan}, C.~A. and 
	{Paciesas}, W.~S. and {Rau}, A. and {Roberts}, O.~J. and {Veres}, P. and 
	{Wilson-Hodge}, C. and {Zhang}, B.-B. and {van Eerten}, H.~J.
	},
    title = "{The Fermi GBM gamma-ray burst time-resolved spectral catalog: brightest bursts in the first four years}",
  journal = {Astron.Astrophys.},
     year = 2016,
    month = apr,
   volume = 588,
      eid = {A135},
    pages = {A135},
   adsurl = {http://adsabs.harvard.edu/abs/2016A%26A...588A.135Y}
}

@ARTICLE{Ryde1999,
   author = {{Ryde}, F. and {Svensson}, R.},
    title = "{On the Shape of Pulse Spectra in Gamma-Ray Bursts}",
  journal = {\apj},
   eprint = {astro-ph/9808213},
     year = 1999,
    month = feb,
   volume = 512,
    pages = {693-698},
      doi = {10.1086/306818},
   adsurl = {http://adsabs.harvard.edu/abs/1999ApJ...512..693R}
}

@ARTICLE{peer2015,
   author = {{Pe'er}, A. and {Barlow}, H. and {O'Mahony}, S. and {Margutti}, R. and 
	{Ryde}, F. and {Larsson}, J. and {Lazzati}, D. and {Livio}, M.
	},
    title = "{Hydrodynamic Properties of Gamma-Ray Burst Outflows Deduced from the Thermal Component}",
  journal = {\apj},
archivePrefix = "arXiv",
   eprint = {1507.00873},
 primaryClass = "astro-ph.HE",
     year = 2015,
    month = nov,
   volume = 813,
      eid = {127},
    pages = {127},
      doi = {10.1088/0004-637X/813/2/127},
   adsurl = {http://adsabs.harvard.edu/abs/2015ApJ...813..127P}
}

@ARTICLE{Peer2008,
   author = {{Pe'er}, A.},
    title = "{Temporal Evolution of Thermal Emission from Relativistically Expanding Plasma}",
  journal = {\apj},
archivePrefix = "arXiv",
   eprint = {0802.0725},
     year = 2008,
    month = jul,
   volume = 682,
      eid = {463-473},
    pages = {463-473},
      doi = {10.1086/588136},
   adsurl = {http://adsabs.harvard.edu/abs/2008ApJ...682..463P}
}

@article{Nakar2017,
author = {Nakar, Ehud and Piran, Tsvi},
title = {{THE OBSERVABLE SIGNATURES OF GRB COCOONS}},
journal = {Astrophysical Journal},
year = {2017},
volume = {834},
number = {1},
pages = {28},
month = jan
}

@ARTICLE{Ito2013,
   author = {{Ito}, H. and {Nagataki}, S. and {Ono}, M. and {Lee}, S.-H. and 
	{Mao}, J. and {Yamada}, S. and {Pe'er}, A. and {Mizuta}, A. and 
	{Harikae}, S.},
    title = "{Photospheric Emission from Stratified Jets}",
  journal = {\apj},
archivePrefix = "arXiv",
   eprint = {1306.4822},
 primaryClass = "astro-ph.HE",
     year = 2013,
    month = nov,
   volume = 777,
      eid = {62},
    pages = {62},
      doi = {10.1088/0004-637X/777/1/62},
   adsurl = {http://adsabs.harvard.edu/abs/2013ApJ...777...62I}
}

@ARTICLE{Meszaros2006,
       author = {{M{\'e}sz{\'a}ros}, P.},
        title = "{Gamma-ray bursts}",
      journal = {Reports on Progress in Physics},
         year = 2006,
        month = aug,
       volume = {69},
       number = {8},
        pages = {2259-2321},
archivePrefix = {arXiv},
 primaryClass = {astro-ph},
       adsurl = {https://ui.adsabs.harvard.edu/abs/2006RPPh...69.2259M}
}

@article{Band1993,
	Adsurl = {http://adsabs.harvard.edu/abs/1993ApJ...413..281B},
	Author = {{Band}, D. and {Matteson}, J. and et al.},
	Doi = {10.1086/172995},
	Journal = {ApJ},
	Month = aug,
	Pages = {281-292},
	Title = {{BATSE observations of gamma-ray burst spectra. I - Spectral diversity}},
	Volume = 413,
	Year = 1993}

@ARTICLE{Ryde2004,
       author = {{Ryde}, Felix},
        title = "{The Cooling Behavior of Thermal Pulses in Gamma-Ray Bursts}",
      journal = {Astron.J.},
         year = 2004,
        month = oct,
       volume = {614},
       number = {2},
        pages = {827-846},
archivePrefix = {arXiv},
 primaryClass = {astro-ph},
       adsurl = {https://ui.adsabs.harvard.edu/abs/2004ApJ...614..827R}
}

@ARTICLE{Ryde2005,
       author = {{Ryde}, Felix},
        title = "{Is Thermal Emission in Gamma-Ray Bursts Ubiquitous?}",
      journal = {Astrophys.J.Lett.},
         year = 2005,
        month = jun,
       volume = {625},
       number = {2},
        pages = {L95-L98},
archivePrefix = {arXiv},
 primaryClass = {astro-ph},
       adsurl = {https://ui.adsabs.harvard.edu/abs/2005ApJ...625L..95R}
}

@ARTICLE{RydePeer2009,
       author = {{Ryde}, Felix and {Pe'er}, Asaf},
        title = "{Quasi-blackbody Component and Radiative Efficiency of the Prompt Emission of Gamma-ray Bursts}",
      journal = {Astron.J.},
         year = 2009,
        month = sep,
       volume = {702},
       number = {2},
        pages = {1211-1229},
archivePrefix = {arXiv},
 primaryClass = {astro-ph},
       adsurl = {https://ui.adsabs.harvard.edu/abs/2009ApJ...702.1211R}
}

@article{Peer2007,
	Adsurl = {http://adsabs.harvard.edu/abs/2007ApJ...664L...1P},
	Author = {{Pe'er}, A. and {Ryde}, F. and et al.},
	Doi = {10.1086/520534},
	Eprint = {arXiv:astro-ph/0703734},
	Journal = {ApJL},
	Month = jul,
	Pages = {L1-L4},
	Title = {{A New Method of Determining the Initial Size and Lorentz Factor of Gamma-Ray Burst Fireballs Using a Thermal Emission Component}},
	Volume = 664,
	Year = 2007}

@article{Guiriec2011,
	Adsurl = {http://adsabs.harvard.edu/abs/2011ApJ...727L..33G},
	Archiveprefix = {arXiv},
	Author = {{Guiriec}, S. and {Connaughton}, V. and et al.},
	Doi = {10.1088/2041-8205/727/2/L33},
	Eid = {L33},
	Eprint = {1010.4601},
	Journal = {ApJL},
	Month = feb,
	Pages = {L33},
	Primaryclass = {astro-ph.HE},
	Title = {{Detection of a Thermal Spectral Component in the Prompt Emission of GRB 100724B}},
	Volume = 727,
	Year = 2011}

@ARTICLE{Ryde2010,
       author = {{Ryde}, F. and {Axelsson}, M. and {Zhang}, B.~B. and {McGlynn}, S. and {Pe'er}, A. and {Lundman}, C. and {Larsson}, S. and {Battelino}, M. and {Zhang}, B. and {Bissaldi}, E. and {Bregeon}, J. and {Briggs}, M.~S. and {Chiang}, J. and {de Palma}, F. and {Guiriec}, S. and {Larsson}, J. and {Longo}, F. and {McBreen}, S. and {Omodei}, N. and {Petrosian}, V. and {Preece}, R. and {van der Horst}, A.~J.},
        title = "{Identification and Properties of the Photospheric Emission in GRB090902B}",
      journal = {Astrophys. J. Lett.},
         year = 2010,
        month = feb,
       volume = {709},
       number = {2},
        pages = {L172-L177},
archivePrefix = {arXiv},
 primaryClass = {astro-ph.HE},
       adsurl = {https://ui.adsabs.harvard.edu/abs/2010ApJ...709L.172R}
}

@article{PeerRyde2011,
	Adsurl = {http://adsabs.harvard.edu/abs/2011ApJ...732...49P},
	Author = {{Pe'er}, A. and {Ryde}, F.},
	Doi = {10.1088/0004-637X/732/1/49},
	Eid = {49},
	Journal = {ApJ},
	Month = may,
	Pages = {49},
	Title = {{A Theory of Multicolor Blackbody Emission from Relativistically Expanding Plasmas}},
	Volume = 732,
	Year = 2011}

@article{Hascoet2013,
	Adsurl = {http://adsabs.harvard.edu/abs/2013A%26A...551A.124H},
	Archiveprefix = {arXiv},
	Author = {{Hasco{\"e}t}, R. and {Daigne}, F. and {Mochkovitch}, R.},
	Doi = {10.1051/0004-6361/201220023},
	Eid = {A124},
	Eprint = {1302.0235},
	Journal = {A$\&$A},
	Month = mar,
	Pages = {A124},
	Primaryclass = {astro-ph.HE},
	Title = {{Prompt thermal emission in gamma-ray bursts}},
	Volume = 551,
	Year = 2013}

@article{Thompson2007,
	Adsurl = {http://adsabs.harvard.edu/abs/2007ApJ...666.1012T},
	Author = {{Thompson}, C. and {M{\'e}sz{\'a}ros}, P. and {Rees}, M.~J.},
	Doi = {10.1086/518551},
	Eprint = {arXiv:astro-ph/0608282},
	Journal = {ApJ},
	Month = sep,
	Pages = {1012-1023},
	Title = {{Thermalization in Relativistic Outflows and the Correlation between Spectral Hardness and Apparent Luminosity in Gamma-Ray Bursts}},
	Volume = 666,
	Year = 2007}

@article{Lundman2013,
	Adsurl = {http://adsabs.harvard.edu/abs/2013MNRAS.428.2430L},
	Archiveprefix = {arXiv},
	Author = {{Lundman}, C. and {Pe'er}, A. and {Ryde}, F.},
	Doi = {10.1093/mnras/sts219},
	Eprint = {1208.2965},
	Journal = {MNRAS},
	Month = jan,
	Pages = {2430-2442},
	Primaryclass = {astro-ph.HE},
	Title = {{A theory of photospheric emission from relativistic, collimated outflows}},
	Volume = 428,
	Year = 2013}

@article{Ghirlanda2013,
	Adsurl = {http://adsabs.harvard.edu/abs/2013MNRAS.428.1410G},
	Archiveprefix = {arXiv},
	Author = {{Ghirlanda}, G. and {Ghisellini}, G. and et al.},
	Doi = {10.1093/mnras/sts128},
	Eprint = {1210.1215},
	Journal = {MNRAS},
	Month = jan,
	Pages = {1410-1423},
	Primaryclass = {astro-ph.HE},
	Title = {{The faster the narrower: characteristic bulk velocities and jet opening angles of gamma-ray bursts}},
	Volume = 428,
	Year = 2013}

@article{Rees&Meszaros2005,
	Adsurl = {http://adsabs.harvard.edu/abs/2005ApJ...628..847R},
	Author = {{Rees}, M.~J. and {M{\'e}sz{\'a}ros}, P.},
	Doi = {10.1086/430818},
	Eprint = {arXiv:astro-ph/0412702},
	Journal = {ApJ},
	Month = aug,
	Pages = {847-852},
	Title = {{Dissipative Photosphere Models of Gamma-Ray Bursts and X-Ray Flashes}},
	Volume = 628,
	Year = 2005}

@article{Beloborodov2011,
	Adsurl = {http://adsabs.harvard.edu/abs/2011ApJ...737...68B},
	Archiveprefix = {arXiv},
	Author = {{Beloborodov}, A.~M.},
	Doi = {10.1088/0004-637X/737/2/68},
	Eid = {68},
	Eprint = {1011.6005},
	Journal = {ApJ},
	Month = aug,
	Pages = {68},
	Primaryclass = {astro-ph.HE},
	Title = {{Radiative Transfer in Ultrarelativistic Outflows}},
	Volume = 737,
	Year = 2011}

@article{Meszaros&Rees2000,
	Adsurl = {http://adsabs.harvard.edu/abs/2000ApJ...530..292M},
	Author = {{M{\'e}sz{\'a}ros}, P. and {Rees}, M.~J.},
	Doi = {10.1086/308371},
	Eprint = {arXiv:astro-ph/9908126},
	Journal = {ApJ},
	Month = feb,
	Pages = {292-298},
	Title = {{Steep Slopes and Preferred Breaks in Gamma-Ray Burst Spectra: The Role of Photospheres and Comptonization}},
	Volume = 530,
	Year = 2000}

@article{Iyyani2013,
	Adsurl = {http://adsabs.harvard.edu/abs/2013MNRAS.433.2739I},
	Archiveprefix = {arXiv},
	Author = {{Iyyani}, S. and {Ryde}, F. and {Axelsson}, M. and {Burgess}, J.~M. and et al.},
	Doi = {10.1093/mnras/stt863},
	Eprint = {1305.3611},
	Journal = {MNRAS},
	Month = aug,
	Pages = {2739-2748},
	Primaryclass = {astro-ph.HE},
	Title = {{Variable jet properties in GRB 110721A: time resolved observations of the jet photosphere}},
	Volume = 433,
	Year = 2013}

@article{Meegan2009,
	Adsurl = {http://adsabs.harvard.edu/abs/2009ApJ...702..791M},
	Archiveprefix = {arXiv},
	Author = {{Meegan}, C. and {Lichti}, G. and {Bhat}, P.~N. and et al.},
	Doi = {10.1088/0004-637X/702/1/791},
	Eid = {791},
	Eprint = {0908.0450},
	Journal = {ApJ},
	Month = sep,
	Pages = {791-804},
	Primaryclass = {astro-ph.IM},
	Title = {{The Fermi Gamma-ray Burst Monitor}},
	Volume = 702,
	Year = 2009}

@ARTICLE{Goldstein2012,
   author = {{Goldstein}, A. and {Burgess}, J.~M. and {Preece}, R.~D. and 
	{Briggs}, M.~S. and {Guiriec}, S. and {van der Horst}, A.~J. and 
	{Connaughton}, V. and {Wilson-Hodge}, C.~A. and {Paciesas}, W.~S. and 
	{Meegan}, C.~A. and {von Kienlin}, A. and {Bhat}, P.~N. and 
	{Bissaldi}, E. and {Chaplin}, V. and {Diehl}, R. and {Fishman}, G.~J. and 
	{Fitzpatrick}, G. and {Foley}, S. and {Gibby}, M. and {Giles}, M. and 
	{Greiner}, J. and {Gruber}, D. and {Kippen}, R.~M. and {Kouveliotou}, C. and 
	{McBreen}, S. and {McGlynn}, S. and {Rau}, A. and {Tierney}, D.
	},
    title = "{The Fermi GBM Gamma-Ray Burst Spectral Catalog: The First Two Years}",
  journal = {\apjs},
archivePrefix = "arXiv",
   eprint = {1201.2981},
 primaryClass = "astro-ph.HE",
     year = 2012,
    month = mar,
   volume = 199,
      eid = {19},
    pages = {19},
      doi = {10.1088/0067-0049/199/1/19},
   adsurl = {http://adsabs.harvard.edu/abs/2012ApJS..199...19G}
}

@Manual{R,
    title = {R: A Language and Environment for Statistical Computing},
    author = {{R Core Team}},
    organization = {R Foundation for Statistical Computing},
    address = {Vienna, Austria},
    year = {2013},
    note = {{ISBN} 3-900051-07-0},
    url = {http://www.R-project.org/},
  }

@ARTICLE{Kobayashi_etal_1997,
       author = {{Kobayashi}, Shiho and {Piran}, Tsvi and {Sari}, Re'em},
        title = "{Can Internal Shocks Produce the Variability in Gamma-Ray Bursts?}",
      journal = {Astrophys.J.},
         year = 1997,
        month = nov,
       volume = {490},
        pages = {92},
archivePrefix = {arXiv},
 primaryClass = {astro-ph},
       adsurl = {https://ui.adsabs.harvard.edu/abs/1997ApJ...490...92K}
}

@ARTICLE{Acuner2020,
       author = {{Acuner}, Zeynep and {Ryde}, Felix and {Pe'er}, Asaf and {Mortlock}, Daniel and {Ahlgren}, Bj{\"o}rn},
        title = "{The Fraction of Gamma-Ray Bursts with an Observed Photospheric Emission Episode}",
      journal = {Astrophys.J.},
         year = 2020,
        month = apr,
       volume = {893},
       number = {2},
          eid = {128},
        pages = {128},
archivePrefix = {arXiv},
 primaryClass = {astro-ph.HE},
       adsurl = {https://ui.adsabs.harvard.edu/abs/2020ApJ...893..128A}
}

\end{document}